\documentclass[10pt]{jfm}
\usepackage{graphicx}
\usepackage{float}

\usepackage{multirow}
\newcommand{\beq}{\begin{equation}}
\newcommand{\eeq}{\end{equation}}
\newcommand{\bseq}{\begin{subequations}}
\newcommand{\eseq}{\end{subequations}}
\newcommand{\beqn}{\begin{eqnarray}}
\newcommand{\eeqn}{\end{eqnarray}}
\newcommand{\ba}{\begin{array}}
\newcommand{\ea}{\end{array}}
\newcommand{\bct}{\begin{center}}
\newcommand{\ect}{\end{center}}
\newcommand{\btmz}{\begin{itemize}}
\newcommand{\etmz}{\end{itemize}}
\newcommand{\benum}{\begin{enumerate}}
\newcommand{\eenum}{\end{enumerate}}

\newcommand{\matbegin}{
        \left[
}
\newcommand{\matend}{
        \right]
}

\newcommand{\thbth}[9]{
 \matbegin \begin{array}{ccc}
                #1 & #2 & #3 \\
                #4 & #5 & #6 \\
                #7 & #8 & #9
                \end{array}\matend}

\newcommand{\be}{\begin{equation}}
\newcommand{\ee}{\end{equation}}

\newcommand{\cplxs}{ C\kern -.35em \rule{0.03 em}{.7 ex}~   }

\def\complex{\hbox{C\kern -.45em \rule{0.03 em}{1.5 ex}}~}

\newcommand{\bi}{\begin{itemize}}
\newcommand{\ei}{\end{itemize}}

\usepackage{amsfonts,amssymb,latexsym,color,amsmath,pifont,epsfig,mathdots,mathtools,natbib}
\usepackage{setspace}
\usepackage{subfigure}
\usepackage{url}
\usepackage{rotating}

\newcommand{\cS}{{\cal S}}
\newcommand{\cM}{{\cal M}}
\newcommand{\cN}{{\cal N}}

\newcommand{\cO}{{\cal O}}

\newcommand{\non}{\nonumber}
\newcommand{\ds}{\displaystyle}

\newcommand{\mrd}{\mathrm{d}}
\newcommand{\mre}{\mathrm{e}}
\newcommand{\mri}{\mathrm{i}}

\newcommand{\fvec}{{\bf f}}

\newcommand{\bu}{{\bf u}}
\newcommand{\bU}{{\bf U}}

\newcommand{\blambda}{\mbox{\boldmath$\lambda$}}
\newcommand{\bphi}{\mbox{\boldmath$\phi$}}
\newcommand{\bpsi}{\mbox{\boldmath$\psi$}}

\newcommand{\p}{\partial}

\providecommand{\abs}[1]{\lvert#1\rvert}

\newcommand{\tc}{\textcolor}

\title{A foundation for analytical developments in the logarithmic region of turbulent channels}

\shorttitle{Analytical developments in the logarithmic region of turbulent channels}

\author{Rashad Moarref$^{1}$, Ati S.\ Sharma$^{2}$, \\[0.15cm] Joel A.\ Tropp$^{3}$ \and Beverley J.\ McKeon$^{1}$}

\affiliation{
$^{1}$Graduate Aerospace Laboratories, California Institute of Technology, CA 91125, USA
\\[0.1cm]
$^{2}$Engineering and the environment, University of Southampton, SO17 1BJ, UK
\\[0.1cm]
$^{3}$Computing \& Mathematical Sciences, California Institute of Technology, CA 91125, USA
}

\shortauthor{R.\ Moarref, A.\ S.\ Sharma, J.\ A.\ Tropp \& B.\ J.\ McKeon}

\begin{document}

\maketitle

%\tableofcontents

    \begin{abstract}

An analytical framework for studying the logarithmic region of turbulent channels is formulated. We build on recent findings (Moarref~\emph{et al.}, J. Fluid Mech., {\bf 734}, 2013) that the velocity fluctuations in the logarithmic region can be decomposed into a weighted sum of geometrically self-similar resolvent modes. The resolvent modes and the weights represent the linear amplification mechanisms and the scaling influence of the nonlinear interactions in the Navier-Stokes equations (NSE), respectively (McKeon \& Sharma, J. Fluid Mech., {\bf 658}, 2010). Originating from the NSE, this framework provides an analytical support for Townsend's attached-eddy model. Our main result is that self-similarity enables order reduction in modeling the logarithmic region by establishing a quantitative link between the self-similar structures and the velocity spectra. Specifically, the energy intensities, the Reynolds stresses, and the energy budget are expressed in terms of the resolvent modes with speeds corresponding to the top of the logarithmic region. The weights of the triad modes -the modes that directly interact via the quadratic nonlinearity in the NSE- are coupled via the interaction coefficients that depend solely on the resolvent modes (McKeon~\emph{et al.}, Phys. Fluids, {\bf 25}, 2013). We use the hierarchies of self-similar modes in the logarithmic region to extend the notion of triad modes to triad hierarchies. It is shown that the interaction coefficients for the triad modes that belong to a triad hierarchy follow an exponential function. The combination of these findings can be used to better understand the dynamics and interaction of flow structures in the logarithmic region. The compatibility of the proposed model with theoretical and experimental results is further discussed.

    \end{abstract}
    
\section{Introduction}
\label{sec.introduction}

A better understanding of wall-bounded turbulent flows at high Reynolds numbers is essential to modeling, controlling, and optimizing engineering systems such as air and water vehicles. Notwithstanding developments in high-Reynolds-number experiments and direct numerical simulations (DNS), several important aspects of the scaling and interaction of turbulent flow structures remain unknown, see e.g.~\cite{smimckmar11}. Over the past $50$ years, significant effort has been devoted to understanding the presence of an inertial sublayer in wall turbulence where the characteristic length scale is the distance from the wall. One of the main features of the inertial sublayer is the presence of a logarithmic mean velocity that is supported by an overwhelming body of experimental evidence, see e.g.~\cite{marmonhulsmi13}, and theoretical explanations that go back to Prandtl, K{\'a}rm{\'a}n, and Millikan, see e.g.~\cite{col56}. Using experiments at friction-Reynolds-numbers $2\times10^4 \leq Re_\tau \leq 6\times 10^5$,~\cite{marmonhulsmi13} showed that the mean velocity and the streamwise energy intensity are logarithmic in the interval $3 \sqrt{Re_\tau} \leq y^+ \leq 0.15 Re_\tau$ where $y^+$ is the wall-normal coordinate normalized with viscous length-scale. In particular, they showed that the mean velocity $U$ normalized by the friction velocity is given by
       \be
	U(y^+)
	\; = \;
	B
	\, + \,
	(1/\kappa)
	\,
	\ln (y^+),
	\non
	\ee
where $B = 4.3$ and the K{\'a}rm{\'a}n constant $\kappa = 0.39$ optimally match the measured mean velocity in the mean-square sense for the above wall-normal interval. 

One of the most successful structural descriptions of the inertial sublayer is based on the attached-eddy hypothesis~\citep{tow76}. It emerged from the analysis of the equilibrium layers where the rates of turbulent energy production and dissipation are large and in balance~\citep{tow61}. Townsend hypothesized that the equilibrium layer is populated by a forest of geometrically self-similar attached eddies, i.e. eddies whose size scale with their height and whose height is proportional to the distance of their centers from the wall, see~figure~\ref{fig.perry-chong-fig14-hierarchies-attached-eddies-JFM1982} in~\S~\ref{sec.hierarchies}.

\cite{tow76} and~\cite{percho82} used the attached-eddy model to systematically predict that the mean velocity and wall-parallel energy intensities are logarithmic and the wall-normal energy intensity and the streamwise/wall-normal Reynolds stress are constant in the inertial sublayer for high Reynolds numbers. In addition, the attached-eddy model was used to predict a region where the one-dimensional streamwise spectrum follows an inverse power-law~\citep{perhencho86,perli90}.~\cite{permarli94} used the attached-eddy model to propose a wall-turbulence closure model suitable for incorporating coherent structure concepts. The attached-eddy model was extended to the outer and inner regions of the turbulent flow by~\cite{maruddper97} and~\cite{markun03}, respectively. The predictions of the attached-eddy model have been validated by high-Reynolds-number experiments~\citep{kunmar06,nicmarhafhutcho07,marmonhulsmi13} over the past decade. 

A model-based study of the self-similar scales was performed by~\cite{hwacos10} where a turbulent viscosity was included in the Navier-Stokes equations (NSE) linearized around the turbulent mean velocity. It was shown that the optimal perturbations that yield the largest transient growth are geometrically similar for an intermediate range of spanwise wavenumbers that correspond to the logarithmic region. The self-sustaining processes for these motions were studied using large-eddy simulations~\citep{hwacos11}. 

\subsection{Paper outline}
\label{sec.contributions}

We propose a framework for analytical developments in the logarithmic region of wall turbulence. This framework is based on the resolvent-mode decomposition proposed by~\cite{mcksha10} where the resolvent represent the linear amplification mechanisms in the NSE and the weights of the resolvent modes represent the role of nonlinearity, see~\S~\ref{sec.resolvent} for an overview. We build on our recent effort~\citep{moashatromckJFM13} where the Reynolds-number scaling and geometric self-similarity of the resolvent modes were studied. As reviewed in~\S~\ref{sec.hierarchies}, the logarithmic part of the mean velocity induces self-similar scalings on a subset of resolvent modes and yields hierarchies of self-similar velocity fluctuations. Our development establishes a formal connection between the resolvent modes and the Townsend's attached-eddy model.

In~\S~\ref{sec.log-region}, we use the self-similar scalings of the resolvent modes to derive analytical results for the energy spectra and co-spectra, the energy intensities and Reynolds stresses, and the energy budget in the logarithmic region. A significant reduction in the number of modes is achieved by formulating the above quantities in terms of the largest resolvent modes in the hierarchies. In this formulation, the weights of smaller resolvent modes appear as weights on the wall-normal shape of the largest modes in the corresponding hierarchies. The implications of the self-similar scalings for the rank-1 approximation to the resolvent-mode decomposition is discussed in~\S~\ref{sec.rank-1}. 

The nonlinear interaction of the self-similar resolvent modes is studied in~\S~\ref{sec.scaling-weights}. We identify the scaling of the interaction coefficient associated with the resolvent modes that directly interact in the logarithmic region through the quadratic nonlinearity in the NSE. In~\S~\ref{sec.exp-scaling}, we outline several constraints on the weights of the resolvent modes by requiring that the model's predictions are compatible with experimental observations in the logarithmic region. A summary of main results is provided in~\S~\ref{sec.conclusion}. The proposed framework is expected to furnish a better understanding of the self-sustaining mechanisms in the logarithmic region and enable development of dynamically significant reduced-order models therein. 

\subsection{Resolvent-mode decomposition}
\label{sec.resolvent}

\begin{figure}
        \begin{center}
        \includegraphics[height=2.8cm]
                {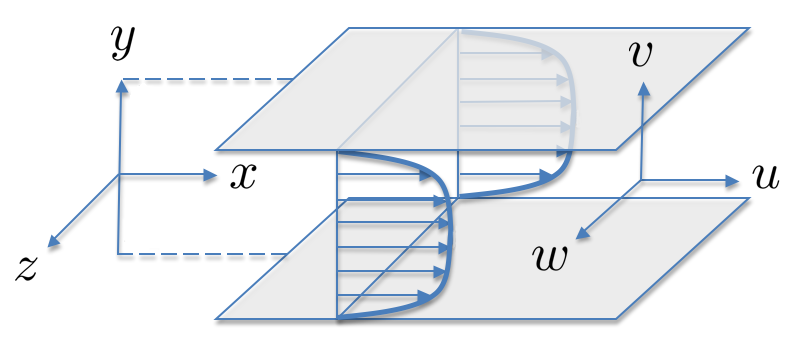}
        \end{center}
        \caption{Pressure-driven channel flow.}
        \label{fig.channel}
      \end{figure}

The pressure-driven flow of an incompressible Newtonian fluid in a channel with geometry shown in figure~\ref{fig.channel} is governed by the nondimensional Navier-Stokes equations (NSE)
	\be
	\ba{l}
	\bu_t
	\, + \,
	(\bu \cdot \nabla) \bu
	\, + \,
	\nabla P
	\; = \;
	(1/Re_\tau) \Delta \bu,
	\\[0.15cm]
	\nabla \cdot \bu
	\; = \;
	0,
	\ea
	\label{eq.NS}
	\ee
where $\bu (x,y,z,t) = [\,u~v~w\,]^T$ is the velocity vector, $P (x,y,z,t)$ is the pressure, $\nabla$ is the gradient, and $\Delta = \nabla \cdot \nabla$ is the Laplacian. The streamwise, wall-normal, and spanwise directions are denoted by $x \in (-\infty,\infty)$, $y \in [0,2]$, and $z \in (-\infty,\infty)$, and $t$ denotes time. The subscript $t$ represents temporal derivative, e.g. $\bu_t = \p \bu/ \p t$. The Reynolds number $Re_\tau = u_\tau h/\nu$ is defined based on the channel half-height $h$, kinematic viscosity $\nu$, and friction velocity $u_\tau = \sqrt{\tau_w/\rho}$, where $\tau_w$ is the shear stress at the wall, and $\rho$ is the density. Unless explicitly indicated, velocity is normalized by $u_\tau$, spatial variables by $h$, time by $h/u_\tau$, and pressure by $\rho u_\tau^2$. The spatial variables are denoted by $^+$ when normalized by the viscous length scale $\nu/u_\tau$, e.g. $y^+ = Re_\tau y$.

Following~\cite{mcksha10}, the velocity is represented by a weighted sum of resolvent modes. A discussion of the resolvent-mode decomposition is provided in this section. For details regarding the numerical computation of the resolvent modes in channel flows, see~\cite{moashatromckJFM13}. 

The Fourier decomposition of the velocity filed in the homogeneous directions $x$, $z$, and $t$ yields 
	\be
	\bu (x,y,z,t)
	\; = \;
	\ds{
	\iiint_{-\infty}^{\infty}
	}
	\,
	\hat{\bu} (y, \lambda_x, \lambda_z, \omega)\,
	\mre^{\mri 
	(
	2\pi x/\lambda_x
	\, + \, 
	2\pi z/\lambda_z 
	\, - \,
	\omega t
	)
	}
	\mrd \lambda_x \,
	\mrd \lambda_z \,
	\mrd \omega,
	\label{eq.fourier}
	\ee
where $\lambda_x$, $\lambda_z$, and $\omega$ denote the streamwise and spanwise wavelengths and the temporal frequency. The Fourier coefficients, denoted by the hat, are three-dimensional three-component propagating waves with streamwise and spanwise wavenumbers $\kappa_x = 2\pi/\lambda_x$ and $\kappa_z = 2\pi/\lambda_z$ and streamwise speed $c = \omega/\kappa_x$. The velocity fluctuations around the turbulent mean velocity $\bU = [\,U(y)~0~0\,]^T = \hat{\bu}(y,0,0,0)$ satisfy
	\be
	-\mri \omega \hat{\bu}
	\, + \,
	(\bU \cdot \nabla) \hat{\bu}
	\, + \,
	(\hat{\bu} \cdot \nabla) \bU
	\, + \,
	\nabla \hat{p}
	\, - \,
	(1/{Re}_\tau) \Delta \hat{\bu}
	\; = \;
	\hat{\fvec},
	~~
	\nabla \cdot \hat{\bu}
	\; = \;
	0,
	\label{eq.NS-lin}
	\ee
where $\fvec = [\,f_1~f_2~f_3\,]^T = -(\bu \cdot \nabla) \bu$ is considered as a forcing term that drives the fluctuations, $p$ is pressure fluctuations, $\nabla = [\,\mri \kappa_x~\p_y~\mri \kappa_z\,]^T$, and $\Delta = \p_{yy} - \kappa^2$ where $\kappa^2 = \kappa_x^2 + \kappa_z^2$. The input-output relationship between the nonlinear forcing and the velocity is described by
	\be
	\hat{\bu} (y, \blambda, c)
	\; = \;
	H (\blambda, c) \, \hat{\fvec} (y, \blambda, c),
	\non
	\ee
where $H$ is the resolvent operator and $\blambda = [\,\lambda_x~\lambda_z\,]$ is the wavelength vector. In the above equation and the rest of this paper, the variables are parameterized with $c$ instead of $\omega$ since $c$ plays an integral role in determining the scalings of the resolvent modes~\citep{moashatromckJFM13}. Notice that for given $\kappa_x$, knowledge of either $c$ or $\omega$ yields the other parameter.

For any $\blambda$ and $c$, the Schmidt (singular value) decomposition of $H$ in the non-homogeneous direction $y$ yields an orthonormal set of forcing modes $\hat{\bphi}_j = [\,\hat{f}_{1j}~\hat{f}_{2j}~\hat{f}_{3j}\,]^T$ and an orthonormal set of response (resolvent) modes $\hat{\bpsi}_j = [\,\hat{u}_j~\hat{v}_j~\hat{w}_j\,]^T$ that are ordered by the corresponding gains $\sigma_1 \geq \sigma_2 \geq \cdots \geq 0\,$ such that $H \hat{\bphi}_j = \sigma_j \hat{\bpsi}_j$. Therefore, if the nonlinear forcing is approximated by a weighted sum of the first $N$ forcing modes,
	\be
	\ba{rcl}
	\hat{\fvec} (y, \blambda, c)
	&\!\! = \!\!&
	\ds{\sum_{j = 1}^{N}}
	\;
	\chi_j (\blambda, c)\,
	\,
	\hat{\bphi}_j (y, \blambda, c),
	\ea
	\label{eq.f}
	\ee
the velocity is determined by a weighted sum of the first $N$ resolvent modes,
	\be
	\ba{rcl}
	\hat{\bu} (y, \blambda, c)
	&\!\! = \!\!&
	\ds{\sum_{j = 1}^{N}}
	\;
	\chi_j (\blambda, c)\, \sigma_j (\blambda, c)\,
	\,
	\hat{\bpsi}_j (y, \blambda, c).
	\ea
	\label{eq.u}
	\ee 
The weights $\chi_j$ represent the scaling influence of the nonlinear interaction of the resolvent modes and can be obtained by projecting the nonlinear forcing onto the forcing modes,
	\be
	\ba{rcl}
	\chi_j (\blambda, c)
	&\!\! = \!\!&
	\ds{\int_{0}^{2}}
	\hat{\bphi}_j^* (y, \blambda, c)
	\,
	\hat{\fvec} (y, \blambda, c)
	\,
	\mrd y,
	\ea
	\label{eq.f-phi}
	\ee	
where the star denotes the complex conjugate.

         \begin{table}
         \centering
         \begin{tabular}{lcccccc}
            Class
            \hspace{-0.05cm}  
            & 
            \hspace{-0.05cm}  
            range of $c$
            \hspace{0.15cm}  
            & 
            \hspace{0.15cm}  
            $\lambda_x$
            \hspace{0.1cm}  
            & 
            \hspace{0.1cm}  
            $y, \lambda_z$
            \hspace{0.1cm}  
            & 
            \hspace{0.1cm}  
            $\sigma_j$
            \hspace{0.1cm}  
            & 
            \hspace{-0.1cm}  
            $\hat{u}_j, \hat{f}_{2j}, \hat{f}_{3j}$
            \hspace{-0.1cm}  
            & 
            \hspace{-0.1cm}  
            $\hat{v}_j, \hat{w}_j, \hat{f}_{1j}$
            \hspace{-0.1cm}  
            \\[0.2cm] %\hline
            Inner
            \hspace{-0.05cm}  
            & 
            \hspace{-0.05cm}  
            $0 \leq c \leq 16$
            \hspace{0.15cm}  
            & 
            \hspace{0.15cm}  
            $Re_\tau^{-1}$
            \hspace{0.1cm}  
            & 
            \hspace{0.1cm}  
            $Re_\tau^{-1}$
            \hspace{0.1cm}  
            & 
            \hspace{0.1cm}  
            $Re_\tau^{-1}$
            \hspace{0.1cm}  
            & 
            \hspace{-0.1cm}  
            $Re_\tau^{1/2}$
            \hspace{-0.1cm}  
            & 
            \hspace{-0.1cm}  
            $Re_\tau^{1/2}$
            \hspace{-0.1cm}  
            \\[0.2cm]
            Self-similar
            \hspace{-0.05cm}  
            & 
            \hspace{-0.05cm}  
            $16 \leq c \leq U_{cl} - 6.15$
            \hspace{0.15cm}  
            & 
            \hspace{0.15cm}  
            $y_c^+ y_c$
            \hspace{0.1cm}  
            & 
            \hspace{0.1cm}  
            $y_c$
            \hspace{0.1cm}  
            & 
            \hspace{0.1cm}  
            $(y_c^+)^2 y_c$
            \hspace{0.1cm}  
            & 
            \hspace{-0.1cm}  
            $y_c^{-1/2}$
            \hspace{-0.1cm}  
            & 
            \hspace{-0.1cm}  
            $(y_c^+)^{-1} y_c^{-1/2}$
            \hspace{-0.1cm}  
            \\[0.2cm]
            Outer
            \hspace{-0.05cm}  
            & 
            \hspace{-0.05cm}  
            $0 \leq U_{cl} - c \leq 6.15$
            \hspace{0.15cm}  
            & 
            \hspace{0.15cm}  
            $Re_\tau$
            \hspace{0.1cm}  
            & 
            \hspace{0.1cm}  
            $1$
            \hspace{0.1cm}  
            & 
            \hspace{0.1cm}  
            $Re_\tau^2$
            \hspace{0.1cm}  
            & 
            \hspace{-0.1cm}  
            $1$
            \hspace{-0.1cm}  
            & 
            \hspace{-0.1cm}  
            $Re_\tau^{-1}$
            \hspace{-0.1cm}  
         \end{tabular}
         \caption{Scalings of the inner, outer, and self-similar classes of the resolvent modes~\citep{moashatromckJFM13}. The range of mode speeds that distinguish these classes and the growth/decay rates (with respect to $Re_\tau$ or $y_c$) of the wall-parallel wavelengths, height, gain, and forcing and response modes are shown. The self-similar and outer scales are valid for the modes with aspect ratio $\lambda_x/\lambda_z \geq \gamma$, where a conservative value for $\gamma$ is $\sqrt{3}$ for the self-similar class and $\sqrt{3} Re_\tau$ for the outer-scaled class. The critical wall-normal location corresponding to the mode speed is denoted by $y_c$, i.e. $c = U (y_c)$.}
         \label{table.scalings}
         \end{table}  
         		
\cite{moashatromckJFM13} showed that the resolvent operator admits three classes of scalings such that the appropriately-scaled resolvent modes are either independent of $Re_\tau$ or geometrically self-similar. These scalings primarily depend on the mode speed and the different regions of the turbulent mean velocity. As summarized in table~\ref{table.scalings}, the modes with $0 \leq c \leq U(y^+ = 100) = 16$ scale in inner units, the modes with $0 \leq U_{cl} - c \leq U_{cl} - U(y = 0.1) =  6.15$ scale in outer units ($U_{cl} = U(y = 1)$ denotes the centerline velocity), and the modes with $16 \leq c \leq U_{cl} - 6.15$ are geometrically self-similar, i.e. they scale with their distance from the wall. Notice that an aspect-ratio constraint $\lambda_x/\lambda_z \geq \gamma$ must be satisfied by the modes in the self-similar and outer-scaled classes, where a conservative value for $\gamma$ is $\sqrt{3}$ for the self-similar class and $\sqrt{3} Re_\tau$ for the outer-scaled class. The scalings of the streamwise response modes were previously given in~\cite{moashatromckJFM13}. Here, we also report the scalings of the wall-normal and spanwise response modes as well as the forcing modes. In addition, we pay special attention to the scalings of the self-similar class and show how they can be used to develop analytical results in the overlap region of the mean velocity. 
                    
\subsection{Hierarchies of geometrically self-similar resolvent modes}
\label{sec.hierarchies}

\cite{moashatromckJFM13} showed that the resolvent operator for the modes with $c$ in the logarithmic region of the mean velocity and $\lambda_x/\lambda_z > \gamma$ scales as
	\be
	H
	\, = \,
	\thbth
	{\big(y_c^+ y_c\big) {H}_{11}}{\big(y_c^+\big)^2 (y_c) {H}_{12}}{\big(y_c^+\big)^2 (y_c) {H}_{13}} 
	{(y_c) {H}_{21}}{\big(y_c^+ y_c\big) {H}_{22}}{\big(y_c^+ y_c\big) {H}_{23}} 
	{(y_c) {H}_{31}}{\big(y_c^+ y_c\big) {H}_{32}}{\big(y_c^+ y_c\big) {H}_{33}}.
	\non
	\ee
Here, $y_c$ is the critical wall-normal location where the mode speed equals the local mean velocity, $c = U(y_c)$. In addition, the operators ${H}_{ij}$ are parameterized by $\lambda_x/(y_c^+ y_c)$ and $\lambda_z/y_c$ and act on functions of $y/y_c$. This yields hierarchies of geometrically self-similar resolvent modes that are parameterized by $y_c$ as summarized in table~\ref{table.scalings}. Since the resolvent modes peak close to the critical layer~\citep{mcksha10}, the hierarchies are equivalently parameterized by the wall-normal location of their centers.

\begin{figure}
    \begin{center}
    \includegraphics[width=0.65\columnwidth]
    {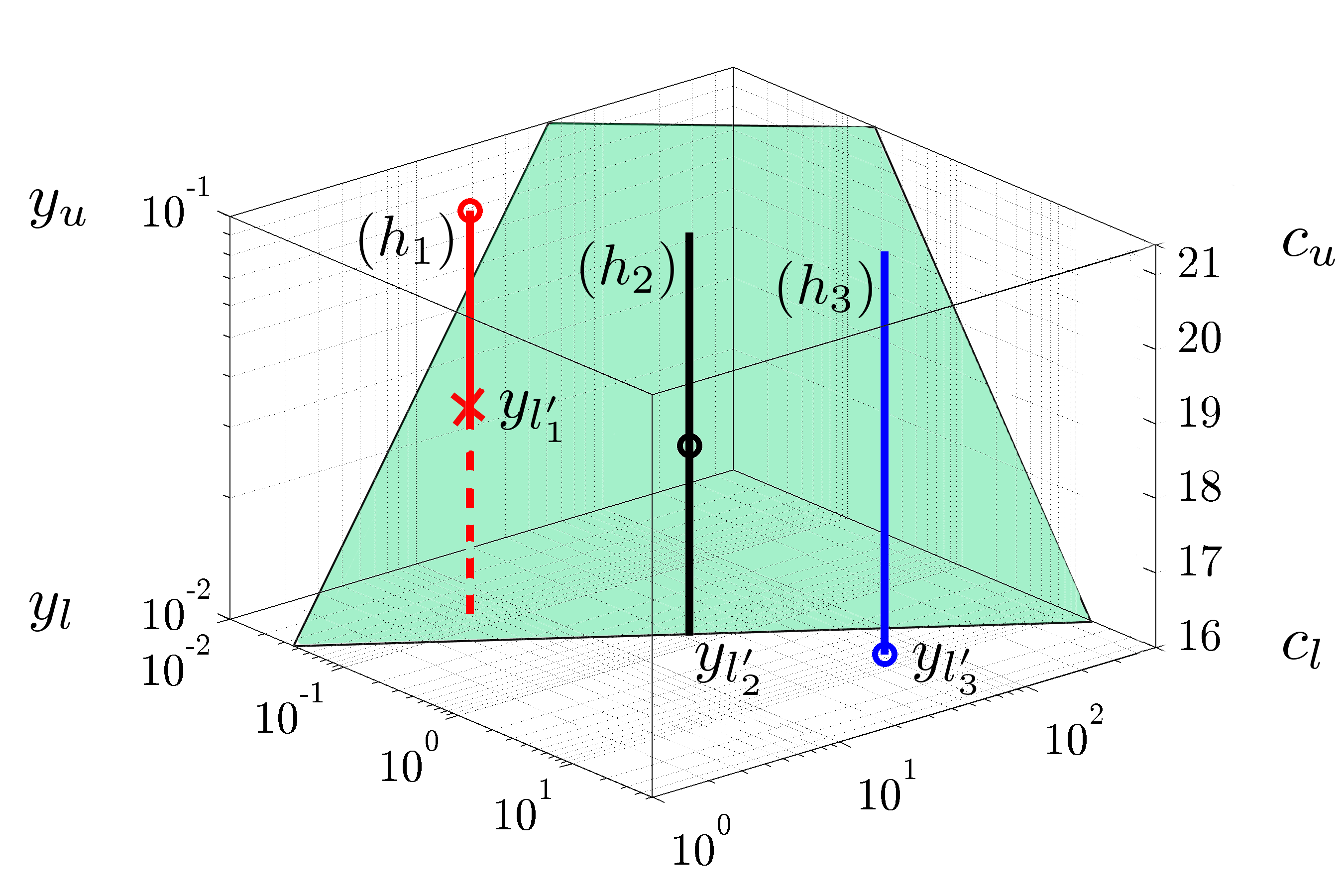}
    \begin{tabular}{c}
    \\[-4.5cm]
    \begin{tabular}{c}
    \hskip-9.cm
    \begin{turn}{90}
    \tc{black}{$~~~~~y_c$}
    \end{turn}
    \hskip8.cm
    \begin{turn}{90}
    \tc{black}{$c = U(y_c)$}
    \end{turn}
    \end{tabular}
    \\[2.1cm]
    \begin{tabular}{c}
    \hskip-8.9cm
    \tc{black}{$\lambda_x/(y_c^+ y_c)$}
    \hskip5.1cm
    \tc{black}{$\lambda_z/y_c$}
    \end{tabular}
    \end{tabular}
    \end{center}
    \caption{
    Illustration of hierarchies of self-similar modes reproduced from~\cite{moashatromckJFM13}. Any point in the box represents a resolvent mode with a unique speed in the logarithmic region of the mean velocity and unique wall-parallel wavelengths. The shaded threshold plane corresponds to the modes with aspect ratio $\lambda_x/\lambda_z = \gamma$. Any vertical line that lies above this plane represents the locus of a hierarchy of self-similar modes. The numbers correspond to the case where a logarithmic mean velocity between $y_l^+ = 100$ and $y_u = 0.1$ is considered for $Re_\tau = 10^4$. For example, hierarchies $h_1$ to $h_3$ include modes (open circles) with $\kappa_x = 1$, $\kappa_z = 10$, and $c = U_{cl} - 6.15$ ($h_1$, red), $(2/3) U_{cl}$ ($h_2$, black), and $16$ ($h_3$, blue). The hierarchy $h_1$ corresponds to the representative VLSM mode. The centers of the smallest and largest modes in any hierarchy take place at $y_{l'}$ and $y_u$, respectively. Notice that $y_{l'}$ is constrained by the lower bound on the logarithmic region of the mean velocity in $h_2$ and $h_3$ and by the aspect-ratio threshold in $h_1$.
    }
    \label{fig.hierarchies}
    \end{figure}

We preserve generality by considering a logarithmic mean velocity for $y_l \leq y \leq y_u$ where $y_l$ and $y_u$ can admit different scalings with $Re_\tau$, and we denote $c_l = U(y_l)$ and $c_u = U(y_u)$. The concept of hierarchies is illustrated in figure~\ref{fig.hierarchies}, reproduced from~\cite{moashatromckJFM13}. Any point in this box represents a resolvent mode with speed $c_l \leq c \leq c_u$ and wall-parallel wavelengths $\lambda_x$ and $\lambda_z$ normalized according to the scalings in table~\ref{table.scalings}. The aspect-ratio threshold $\lambda_x/\lambda_z = \gamma$ is shown by the shaded plane. Any vertical line that lies above the threshold plane represents the locus of a hierarchy of self-similar resolvent modes. In addition, each mode belongs to one and only one hierarchy.

As $y_c$ or $c$ increases from $y_l$ to $y_u$, the modes become larger. It follows from the scalings of the wall-parallel wavelengths that the aspect ratio grows with $y_c^+$, cf.~table~\ref{table.scalings}, within a hierarchy. The wavelengths and speed of the largest mode in a hierarchy are denoted by $\blambda_u$ and $c_u$, and we have $\lambda_{x,u}/\lambda_{z,u} \geq \gamma$. The wavelengths and speed of the smallest mode in a hierarchy are denoted by $\blambda_{l'}$ and $c_{l'}$ where $\lambda_{x,l'} = (y_{l'}^+ y_{l'} / y_u^+ y_u) \lambda_{x,u}$, $\lambda_{z,l'} = (y_{l'} / y_u) \lambda_{z,u}$, and $c_{l'} = U(y_{l'})$. Here, $y_{l'}^+$ is the larger of the lower edge of the logarithmic region $y_l^+$ and the height $\gamma y_u^+ (\lambda_{z,u}/\lambda_{x,u})$ of the smallest mode that satisfies the aspect-ratio constraint. Therefore, the range of scales in a hierarchy depends on the ratio between $y_u$ and $y_{l'}$.

A hierarchy is formally defined as a subset $\cS (\blambda_u)$ of all mode parameters $\cS$ and is characterized by the wavelengths of the largest mode in that hierarchy, see also figure~\ref{fig.hierarchies}, 
	\be
	\ba{l}
	\cS (\blambda_u)
	\, = \,
	\Bigg\{
	(\blambda,c)
	\,|\,
	\lambda_x = \lambda_{x,u} \Big(\dfrac{y_c^+ y_c}{y_u^+ y_u}\Big),
	\lambda_z = \lambda_{z,u} \Big(\dfrac{y_c}{y_u}\Big),
	c = c_u + \dfrac{1}{\kappa} \ln \Big( \dfrac{y_c^+}{y_u^+} \Big),
	\\[0.3cm]
	\hskip2.85cm
	y_c^+ \geq y_{l'}^+ = \mbox{max} \{y_l^+, \gamma y_u^+ (\lambda_{z,u}/\lambda_{x,u})\},
	~y_c \leq y_u
	\Bigg\}.
	\ea
	\label{eq.S_h}
	\ee
We note that the inner-scaled variables $y_c^+$ and $y_u^+$ can be defined in terms of different Reynolds numbers. Specifically, we denote 
	\be
	y_c^+ 
	\, = \,
	Re_\tau y_c,~~
	y_u^+ 
	\, = \, Re_{\tau,u} y_u,
	\label{eq.ycp-yup}
	\ee
where $Re_{\tau,u}$ is the reference Reynolds number for which the largest resolvent mode is computed and $Re_\tau$ is the Reynolds number for the modes that will be determined based on similarity to the largest resolvent mode. Therefore, the hierarchies at arbitrary values of the Reynolds number are determined from the resolvent modes computed at a reference $Re_{\tau,u}$, cf.~(\ref{eq.S_h}).

The vertical lines in figure~\ref{fig.hierarchies} show three hierarchies for the case where a logarithmic mean velocity is considered between $y_l^+ = 100$ and $y_u = 0.1$ for $Re_\tau = 10^4$. The hierarchy $h_2$ passes through the mode with wall-parallel wavelengths $\lambda_x = 6$ and $\lambda_z = 0.6$ and speed $c = (2/3) U_{cl}$. This mode is representative of the very large-scale motions (VLSMs) in turbulent flows~\citep{mcksha10}. The hierarchies $h_1$ and $h_3$ pass through the modes with the same wavelengths as the representative VLSM mode, but their speeds correspond to the upper and lower bounds on the logarithmic region, i.e. $c = U (y_l^+) = 16$ in $h_1$ and $c = U(y_u) = U_{cl}- 6.15$ in $h_3$.

\begin{figure}
    \begin{center}
    \begin{tabular}{cc}
    \subfigure{\includegraphics[width=0.52\columnwidth]
    {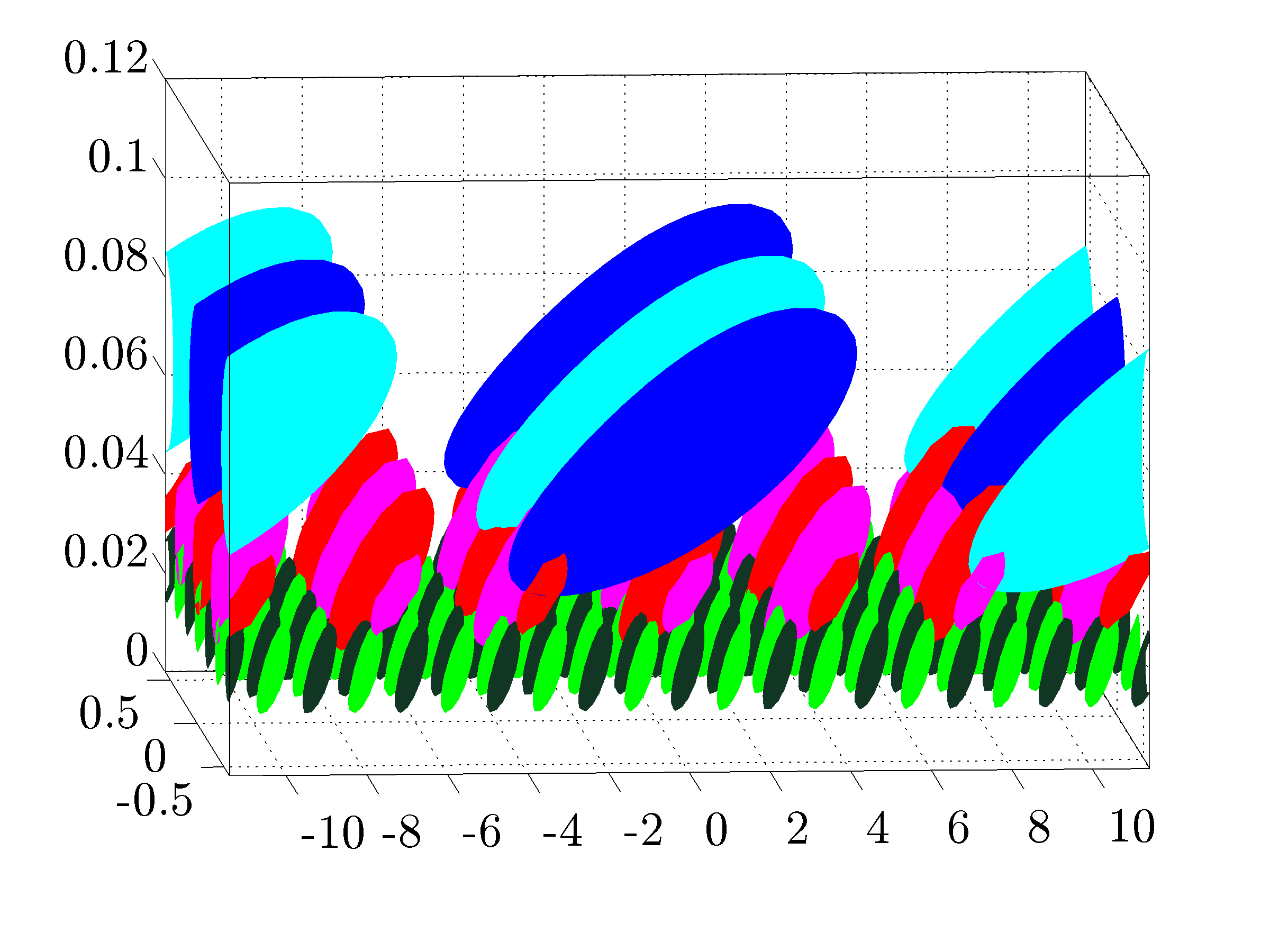}
    \label{fig.log-region-hierarchy-u-3D-R1e4-kx1-kz10-yc-0p0178-0p0316-0p0562}}
    &
    \hskip0cm
    \begin{tabular}{c}
    \\[-5.9cm]
    \subfigure{\includegraphics[width=0.42\columnwidth]
    {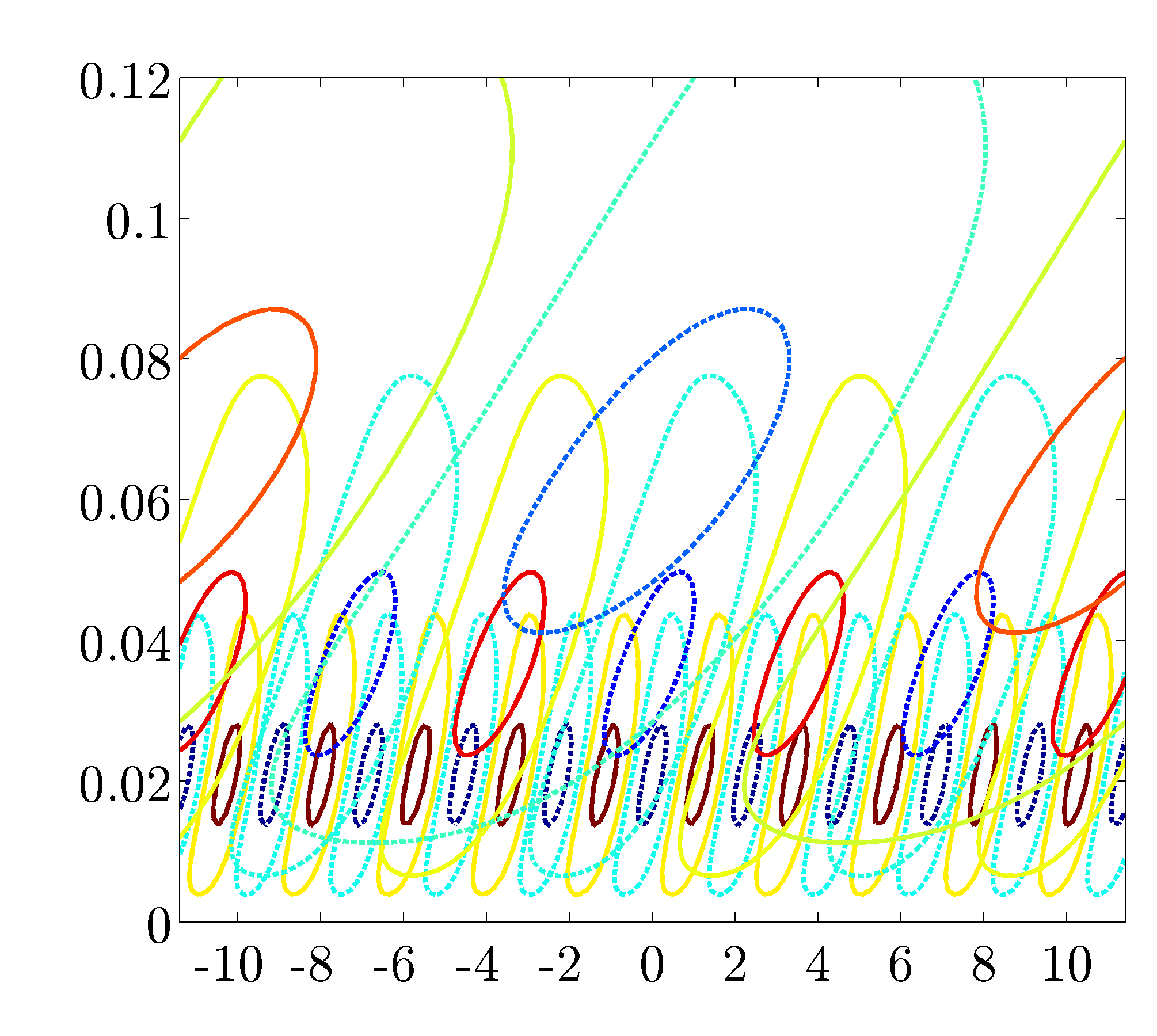}
    \label{fig.log-region-hierarchy-contour-u-vs-lx-y-R1e4-kx1-kz10-yc-0p0178-0p0316-0p0562}}
    \end{tabular}
    \\[-0.1cm]
    $(a)$
    &
    $(b)$
    \end{tabular}
    \begin{tabular}{c}
    \\[-4.7cm]
    \begin{tabular}{c}
    \hskip-6.2cm
    \begin{turn}{90}
    \tc{black}{$y$}
    \end{turn}
    \hskip7.2cm
    \begin{turn}{90}
    \tc{black}{$y$}
    \end{turn}
    \end{tabular}
    \\[2cm]
    \hskip-13cm
    \tc{black}{$z$}
    \\[0.3cm]
    \begin{tabular}{c}
    \hskip1.3cm
    \tc{black}{$x$}
    \hskip6.4cm
    \tc{black}{$x$}
    \end{tabular}
    \end{tabular}
    \\[-0.2cm]
    \hskip0.6cm
    \begin{tabular}{cc}
    \subfigure{\includegraphics[width=0.42\columnwidth]
    {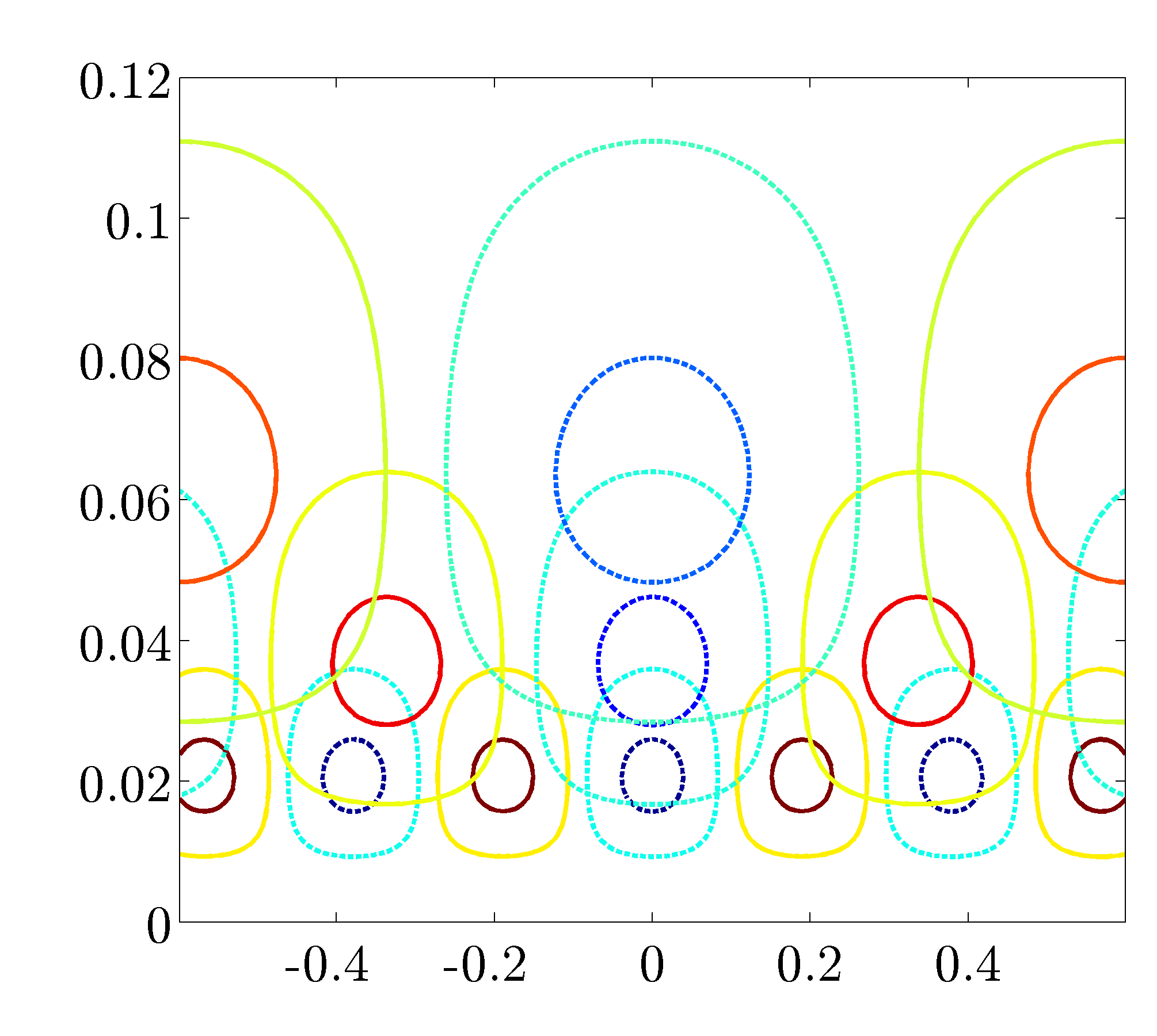}
    \label{fig.log-region-hierarchy-contour-u-vs-lz-y-R1e4-kx1-kz10-yc-0p0178-0p0316-0p0562}}
    &
    \hskip1.1cm
    \begin{tabular}{c}
    \\[-5.65cm]
    \subfigure{\includegraphics[width=0.39\columnwidth]
    {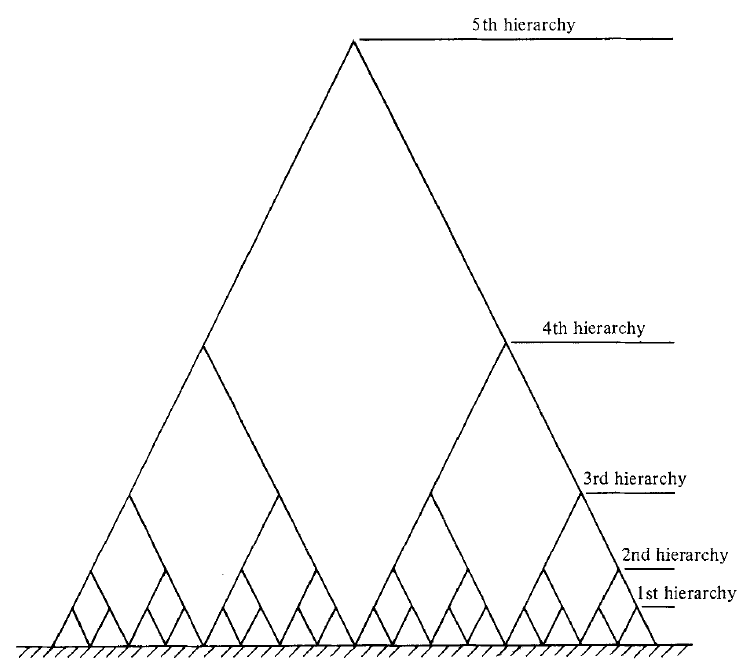}
    \label{fig.perry-chong-fig14-hierarchies-attached-eddies-JFM1982}}
    \end{tabular}
    \\[0.2cm]
    \hskip0.3cm
    $(c)$
    &
    \hskip0.8cm
    $(d)$
    \end{tabular}
    \begin{tabular}{c}
    \\[-4.2cm]
    \begin{tabular}{c}
    \hskip-6.4cm
    \begin{turn}{90}
    \tc{black}{$y$}
    \end{turn}
    \hskip5.8cm
    \end{tabular}
    \\[2.3cm]
    \begin{tabular}{c}
    \hskip-3cm
    \tc{black}{$z$}
    \end{tabular}
    \end{tabular}
    \end{center}
    \caption{Illustration of the geometrically self-similar resolvent modes. (a) The isosurfaces of the principal streamwise velocities for three modes with $(\blambda, c) = (2.3,0.38,17.35)$; blue, $(7.2,0.67,18.70)$; red, and $(23,1.2,20.05)$; green, that belong to the hierarchy $h_2$ in figure~\ref{fig.hierarchies}. The dark and light colors show $\pm70 \%$ of the maximum velocity.
    The contours in (b) and (c) show cross-sections of (a) for $x = 0$ and $z = 0$ at $\pm20 \%$ and $\pm80 \%$ of the maximum velocity. (d) Schematic of the geometrically self-similar hierarchies reproduced from~\cite{percho82}.}
    \label{fig.self-similar}
    \end{figure}

The isosurfaces of streamwise velocity associated with three modes that belong to hierarchy $h_2$ are shown in figure~\ref{fig.log-region-hierarchy-u-3D-R1e4-kx1-kz10-yc-0p0178-0p0316-0p0562}. The larger modes propagate faster and lean more towards the wall since the length of the modes grows quadratically with the height. The cross-sections of the streamwise velocity at $x = 0$ and $z = 0$ are shown in figures~\ref{fig.log-region-hierarchy-contour-u-vs-lx-y-R1e4-kx1-kz10-yc-0p0178-0p0316-0p0562} and~\ref{fig.log-region-hierarchy-contour-u-vs-lz-y-R1e4-kx1-kz10-yc-0p0178-0p0316-0p0562}. As $c$ increases, the modes become larger and their centers move away from the wall. Therefore, the modes are attached and self-similar in the sense of Townsend, i.e. their height is proportional to the distance of their centers from the wall and their size scale with their height; compare figure~\ref{fig.log-region-hierarchy-contour-u-vs-lz-y-R1e4-kx1-kz10-yc-0p0178-0p0316-0p0562} with~\ref{fig.perry-chong-fig14-hierarchies-attached-eddies-JFM1982}, reproduced from~\cite{percho82}.

Notice that the streamwise scaling of the self-similar resolvent modes differs from the streamwise scaling of the hypothesized attached eddies. In the equilibrium layer,~\cite{tow76} argued that ``It is difficult to imagine how the presence of the wall could impose a dissipation length-scale proportional to distance from it unless the main eddies of the flow have diameters proportional to distance of their centers from the wall because their motion is directly influenced by its presence." As a result, he assumed that the velocity field associated with the self-similar eddies is given by
	\be
	\bu (x,y,z)
	\; = \;
	s_1 \big(
	(x-x_a)/y_a, (y-y_a)/y_a, (z-z_a)/y_a
	\big),
	\non
	\ee
where $s_1$ is the velocity in terms of the normalized location of the eddy center $x_a$, $y_a$, and $z_a$. On the other hand, the self-similar resolvent modes have the following form
	\be
	\bu (x,y,z,t)
	\; = \;
	s_2 \big(
	(x - c\,t)/(y_c^+ y_c), (y-y_c)/y_c, z/y_c
	\big),
	\non
	\ee	
where $s_2$ is the velocity in terms of the parameters that position the mode centers at the wall-parallel origin in a moving frame with streamwise speed $c$. The critical location $y_c$ in the present study is equivalent to Townsend's eddy center $y_a$. In agreement with scaling of Townsend's eddies, the spanwise and wall-normal extents of the resolvent modes scale with $y_c$. On the other hand, the streamwise extent of the resolvent modes scales with $y_c^+ y_c$. Notice that this difference does not contradict Townsend's original hypothesis because the dissipation length-scale for the case where $\lambda_x$ and $\lambda_z$ are respectively proportional to $y_c^+ y_c$ and $y_c$ is dominated by the dissipation in the spanwise direction, and hence, proportional to the mode height. 

\cite{tow76} considered a range of eddy scales between $l_0$ and $L_0$ that, respectively, correspond to $y_{l'}$ and $y_u$ in the present study, cf.~(\ref{eq.S_h}). Even though Townsend did not specify the Reynolds-number scalings of the smallest and largest eddies, their dependence on the Reynolds number is not incompatible with the attached-eddy hypothesis~\citep{marmonhulsmi13}. For example,~\cite{perhencho86} considered the case where $l_0$ scales in inner units and $L_0$ scales in outer units. Our framework exhibits the same generality of Townsend's since it allows for different Reynolds-number scalings for $y_{l'}$ and $y_u$. We next examine the implications of the self-similar scalings for developing analytical results in the logarithmic region.
		     
\section{Analytical developments using the self-similar scalings}
\label{sec.log-region}

We derive analytical expressions for the energy spectra and co-spectra, the energy intensities and Reynolds stresses, and the energy budget in the logarithmic region. The linear and non-linear effects are distinguished by expressing each quantity in terms of an inner product of two matrices that depend on the resolvent modes and the resolvent weights, respectively. For example, the `energy spectrum' is written in terms of the inner product of the `energy density matrix' that only depends on the resolvent modes and the `weight matrix' that only depends on the resolvent weights, see definitions in~(\ref{eq.Euu-u})-(\ref{eq.X-ij}) later. It is important to notice the distinction between these terminologies in this paper.

For simplicity, we consider the case where the self-similar scalings hold for the modes with speeds throughout the interval $c_{l'} \leq c \leq c_u$. The edge effects as $c$ approaches the boundaries of this region are discussed in Appendix~\ref{sec.edge}.
	    	        
\subsection{Energy spectra and co-spectra}
\label{sec.analytical-spectra}

The pre-multiplied three-dimensional streamwise energy spectrum is defined as
	\be
	\ba{rcl}
	E_{uu} (y, \blambda, c)
	&\!\! = \!\!&
	(2\pi/\lambda_x)^2 \, (2\pi/\lambda_z) \;
	\hat{u} (y, \blambda, c)
	\;
	\overline{\hat{u}} (y, \blambda, c),
	\ea
	\label{eq.Euu-u}
	\ee
where the overline denotes the complex conjugate of a number. The additional factor of $2\pi/\lambda_x$ facilitates computation of temporal averages by integration over $c$ instead of $\omega$. For example, the pre-multiplied two-dimensional streamwise energy spectrum is obtained from
	\be
	\ba{rcl}
	E_{uu} (y,\blambda)
	&\!\! = \!\!&
	\ds{
	\int_{0}^{U_{cl}}
	}
	\,
	E_{uu} (y, \blambda, c)
	\,
	\mrd c.
	\ea
	\label{eq.Euu}
	\ee
The contribution of the first $N$ resolvent modes to $E_{uu} (y, \blambda, c)$ is determined by substituting $\hat{u}$ from~(\ref{eq.u}) in~(\ref{eq.Euu-u}), see~\cite{moajovtroshamckPOF14},
	\be
	\ba{rcl}
	E_{uu} (y, \blambda, c)
	&\!\! = \!\!&
	\mbox{Re} 
	\Big\{
	\mbox{tr}
	\big(
	A_{uu} (y, \blambda, c)
	\,
	X (\blambda, c) 
	\big)
	\Big\}.
	\ea
	\label{eq.Euu-trace}
	\ee
Here, $\mbox{Re}$ is the real part of a complex number, $\mbox{tr}(\cdot)$ is the matrix trace, $A_{uu} (y, \blambda, c)$ is the $N \times N$ energy density matrix whose $ij$-th element is determined by the resolvent modes,
	\be
	A_{uu,ij} (y, \blambda, c)
	\; = \;
	(2\pi/\lambda_x)^2 \, (2\pi/\lambda_z) \;
	\sigma_i(\blambda, c) \,
	\sigma_j(\blambda, c) \,
	\hat{u}_i (y, \blambda, c) \, \overline{\hat{u}}_j (y, \blambda, c),
	\label{eq.Euu-tilde-ij}
	\ee
and $X (\blambda, c)$ is the $N \times N$ weight matrix whose $ij$-th element is determined by
	\be
	X_{ij} (\blambda, c)
	\; = \;
	\chi_i (\blambda, c)
	\,
	\overline{\chi}_j (\blambda, c).
	\label{eq.X-ij}
	\ee
It follows from~(\ref{eq.X-ij}) that $X$ is rank-1 and positive semi-definite. The expressions~(\ref{eq.Euu-u})-(\ref{eq.Euu-tilde-ij}) for the wall-normal and spanwise energy spectra $E_{vv}$ and $E_{ww}$ and the Reynolds stress co-spectra $E_{uv}$, $E_{uw}$, and $E_{vw}$ are obtained similarly. As discussed earlier, we emphasize the following use of terminologies in this paper: The `energy spectrum', e.g. $E_{uu}$, is obtained by the inner product of the `energy density matrix', e.g. $A_{uu}$, and the `weight matrix' $X$.

    \begin{figure}
    \begin{center}
    \includegraphics[width=0.8\columnwidth]
    {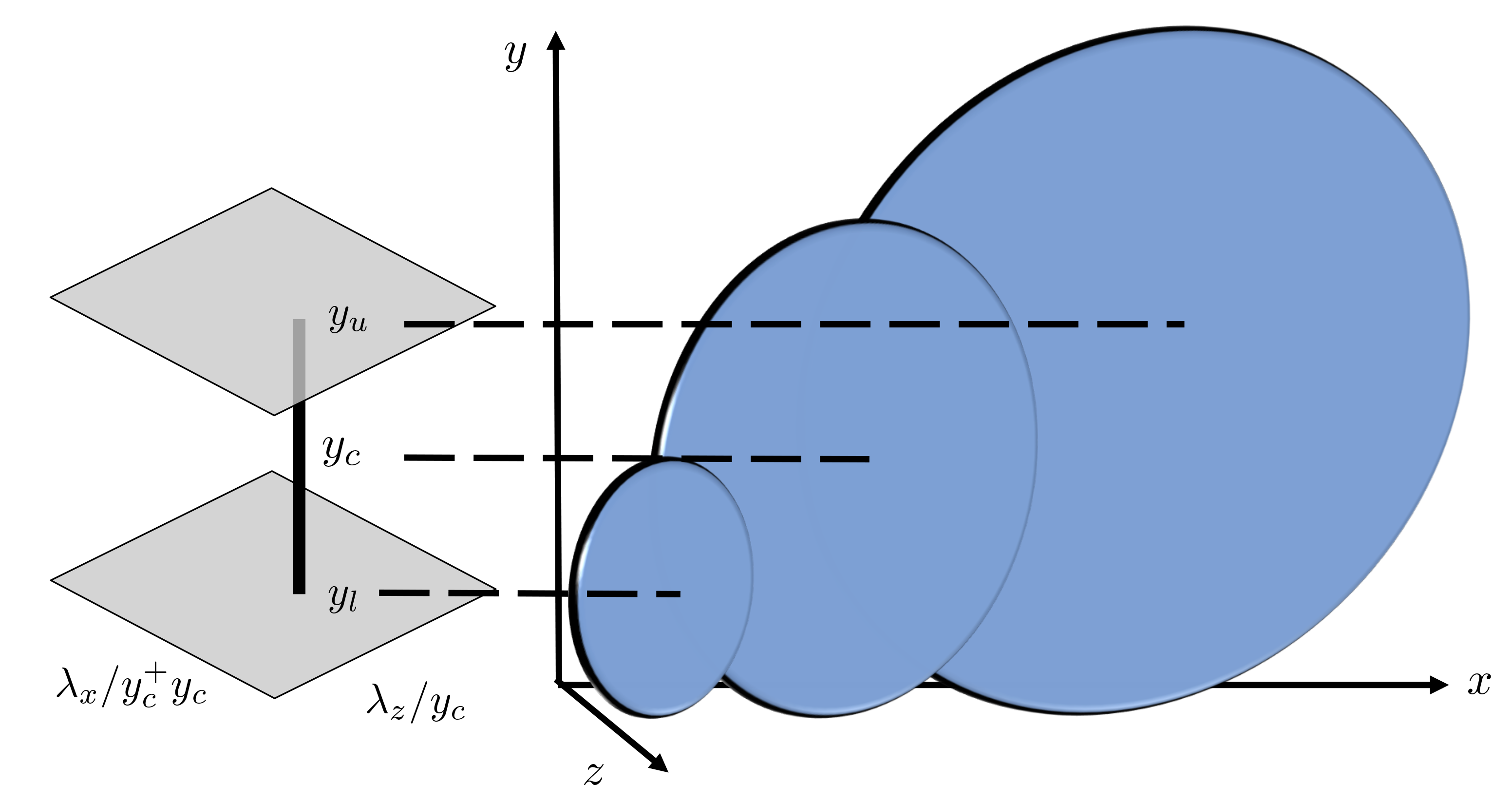}
    \end{center}
    \caption{Schematic showing that any mode in a given hierarchy (shown by the vertical line in the left figure) is self-similar with the largest mode in that hierarchy, and thus, can be expressed in terms of the largest mode. The largest mode is centered around $y_u$ and corresponds to speed $c_u = U(y_u)$, i.e. the top of the logarithmic region.}
    \label{fig.self-similar}
    \end{figure}

We show that the energy density matrix for all the modes in a given hierarchy at all Reynolds numbers can be obtained from the largest mode in that hierarchy. The self-similar scalings can be used to express the resolvent modes in a given hierarchy, i.e. $(\blambda, c) \in \cS(\blambda_u)$, in terms of the largest mode in $\cS(\blambda_u)$. In other words, the mode shapes and their amplification can be determined from the modes whose speed corresponds to the top of the logarithmic region. This is schematically shown in figure~\ref{fig.self-similar}. Specifically, we have
	\be
	\ba{rcl}
	g_1 (y, \blambda, c)
	&\!\! = \!\!&
	\sqrt{y_u/y_c}
	\;
	g_1 \big(
	(y_u/y_c) y, \blambda_u, c_u
	\big),
	\\[0.2cm]
	g_2 (y, \blambda, c)
	&\!\! = \!\!&
	(y_u^+/y_c^+) \sqrt{y_u/y_c}
	\;
	g_2 \big(
	(y_u/y_c) y, \blambda_u, c_u
	\big),
	\ea
	\label{eq.u-map-cu} 
	\ee
where $g_1$ represents $\hat{u}_j$, $\hat{f}_{2j}$, or $\hat{f}_{3j}$ and $g_2$ represents $\hat{v}_j$, $\hat{w}_{j}$, or $\hat{f}_{1j}$. The corresponding singular values or gains are obtained from
	\be
	\sigma_j (\blambda, c)
	\; = \;
	(y_c^+/y_u^+)^2 (y_c/y_u)
	\;
	\sigma_j (
	\blambda_u, c_u
	),
	\label{eq.sigma-map-cu} 
	\ee
where we recall the distinction between $y_c^+$ and $y_u^+$, cf.~(\ref{eq.ycp-yup}). 

Even though the resolvent modes can be described in terms of any mode in the hierarchy, it is advantageous to use the largest mode as in~(\ref{eq.u-map-cu}) and~(\ref{eq.sigma-map-cu}). To see this, notice that the aspect ratio $\lambda_x/\lambda_z$ decreases as $c$ becomes smaller in the log region. If a mode $m_0$ with speed $c_0$ belongs to a hierarchy, any mode with $c > c_0$ along the hierarchy also satisfies the aspect-ratio constraint and can be used to describe $m_0$. On the other hand, the modes with $c < c_0$ along the hierarchy may violate the aspect-ratio constraint, are excluded from the hierarchy, and cannot be used to retrieve $m_0$. Another advantage of representing the modes in terms of the largest mode is that many flow quantities such as the streamwise intensity are the same at $y = y_u = 0.1$ for all Reynolds numbers~\citep{smimckmar11}.  Therefore, the largest modes in the hierarchies, corresponding to $y_c = y_u$, can be used as a fixed point in studying the scaling of turbulent flows.

For $(\blambda, c) \in \cS(\blambda_u)$, substituting~(\ref{eq.u-map-cu}) and~(\ref{eq.sigma-map-cu}) in~(\ref{eq.Euu-tilde-ij}) yields a geometric scaling for the energy density
	\be
	A_{uu,ij} (y, \blambda, c)
	\; = \;
	(Re_{\tau}/Re_{\tau,u})^2
	A_{uu,ij} \big(
	(y_u/y_c) y, \blambda_u, c_u
	\big).
	\label{eq.Euu-tilde-ij-map-cu}
	\ee
In other words, the energy density matrix of the largest mode (i.e. with speed $c_u$) in a given hierarchy at any reference Reynolds number $Re_{\tau,u}$ is sufficient to determine the energy density matrix of all the modes in that hierarchy at all $Re_\tau$. If the nonlinear forcing was broadband, i.e. all the entries in $X$ were $1$, the energy density would be equivalent to the energy spectrum. In real turbulent flows, scaling of the energy spectrum $E_{uu}$ also depends on scaling of the weight matrix. This requires studying the nonlinear interaction of the resolvent modes which is discussed in~\S~\ref{sec.scaling-weights}. 

For $A_{vv,ij}$, we have
	\be
	A_{vv,ij} (y, \blambda, c)
	\; = \;
	(y_u/y_c)^2
	A_{vv,ij} \big(
	(y_u/y_c) y, \blambda_u, c_u
	\big),
	\label{eq.Evv-tilde-ij-map-cu}
	\ee
and similarly for $A_{ww,ij}$ and $A_{vw,ij}$. For $A_{uv,ij}$, we have
	\be
	A_{uv,ij} (y, \blambda, c)
	\; = \;
	(y_u/y_c) (Re_{\tau}/Re_{\tau,u})
	A_{uv,ij} \big(
	(y_u/y_c) y, \blambda_u, c_u
	\big),
	\label{eq.Euv-tilde-ij-map-cu}
	\ee
and similarly for $A_{uw,ij}$. 
		    	    
    \begin{figure}
    \begin{center}
    \includegraphics[width=0.6\columnwidth]
    {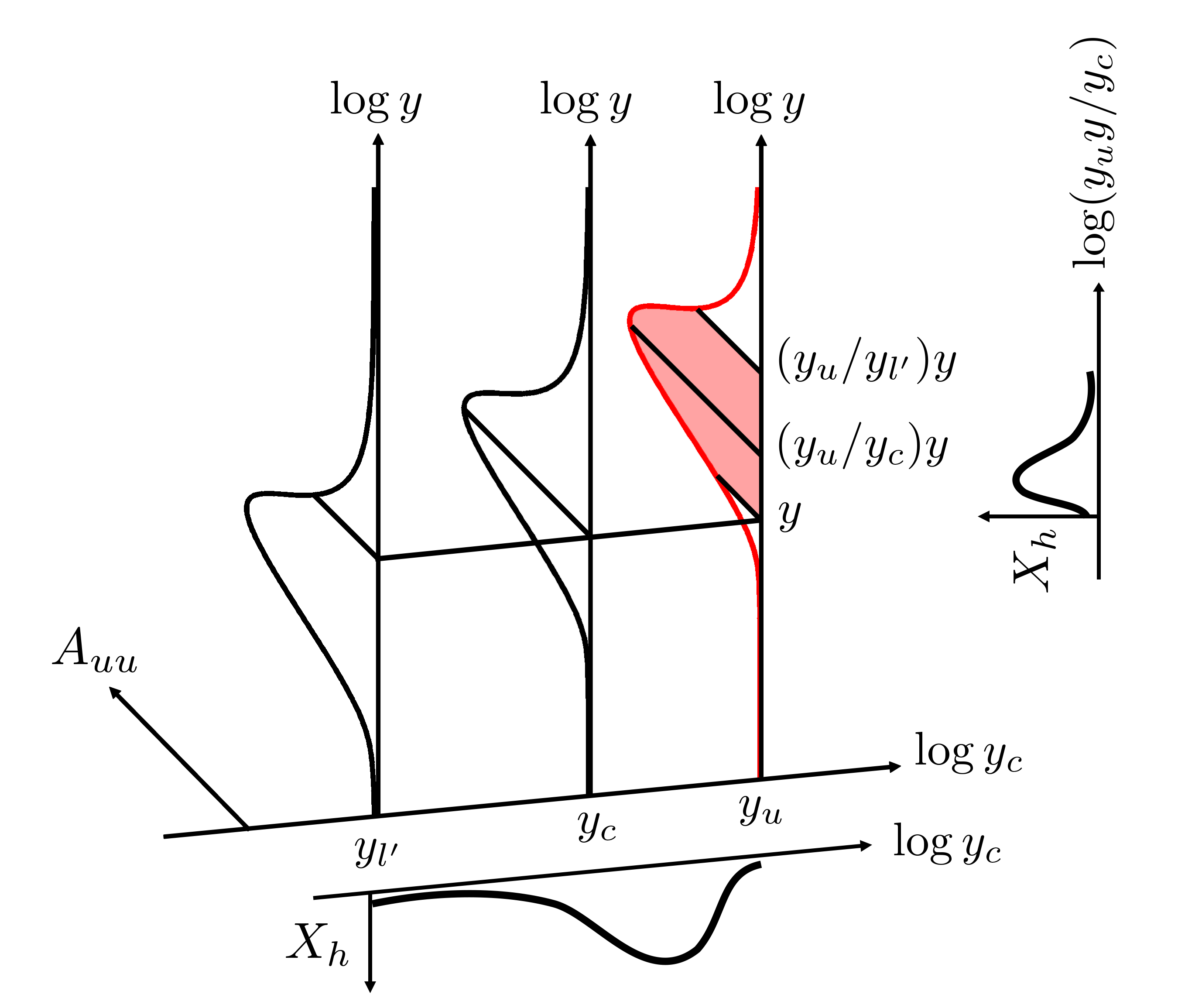}
    \end{center}
    \caption{The contribution of any hierarchy of self-similar modes to the streamwise energy intensity $E_{uu} (y)$ can be obtained by integrating the streamwise energy density $A_{uu}$ of the largest mode in that hierarchy weighted by the weight matrix $X_h$. The energy density matrix of the largest mode (corresponding to $y_u$) is schematically shown in red. The energy density matrices of the smallest mode (corresponding to $y_{l'}$) and a generic mode in the hierarchy (corresponding to $y_c$) are schematically shown in black. The unknown weight matrix is also schematically shown.}
    \label{fig.Euu-hierarchy-analytical}
    \end{figure}
    
\subsection{Normal energy intensities and Reynolds stresses}
\label{sec.analytical-intensity}

The contribution of any hierarchy to the energy intensities and Reynolds stresses at the wall-normal location $y$ can be analytically determined from the largest mode in that hierarchy. To this end, the contribution of hierarchy $\cS (\blambda_u)$ to the streamwise energy intensity is defined as
	\be
	E_{uu,h} (y,\blambda_u)
	\; = \;
	\ds{
	\int_{\cS (\blambda_u)}
	}
	\,
	E_{uu} (y, \blambda, c)
	\,
	\mrd \cS (\blambda_u),
	\label{eq.Euu-hierarchy-1-x}
	\ee	
where the integration is performed over all the modes in the subset $\cS (\blambda_u)$. Substituting $E_{uu}$ from~(\ref{eq.Euu-trace}) in~(\ref{eq.Euu-hierarchy-1-x}) yields
	\be
	E_{uu,h} (y,\blambda_u)
	\; = \;
	\ds{
	\int_{\cS (\blambda_u)}
	}
	\,
	\mbox{Re} 
	\Big\{
	\mbox{tr}
	\big(
	A_{uu} (y, \blambda, c)
	\,
	X (\blambda, c) 
	\big)
	\Big\}
	\,
	\mrd \cS (\blambda_u).
	\label{eq.Euu-hierarchy-2-x}
	\ee	
Since the modes in a given hierarchy are parameterized with $y_c$, it follows from~(\ref{eq.S_h}), (\ref{eq.Euu-tilde-ij-map-cu}), (\ref{eq.Euu-hierarchy-2-x}), and the identity $\mrd c = (1/\kappa) \,\mrd \ln(y_c)$ that
	\be
	\ba{rcl}
	\hskip-0.2cm
	E_{uu,h} (y,\blambda_u)
	&\!\! = \!\!&
	\dfrac{1}{\kappa}
	\dfrac{Re_\tau^2}{Re_{\tau,u}^2}
	\,
	\ds{
	\int_{\ln y_{l'}}^{\ln y_u}
	}
	\,
	\mbox{Re}
	\Big\{
	\mbox{tr}
	\Big(
	A_{uu} \big( \dfrac{y_u}{y_c}y, \blambda_u, c_u \big)
	X_h (\blambda_u, y_c)
	\Big)
	\Big\}
	\,
	\mrd \ln(y_c),
	\ea
	\label{eq.hierarchy-analytical}
	\ee
where $X_h (\blambda_u, y_c)$ denotes the weight matrix along the hierarchy $\cS (\blambda_u)$ at speed $c = U(y_c)$
	\be
	\ba{rcl}
	X_h (\blambda_u, y_c)
	&\!\! = \!\!&
	X \big( 
	(y_c^+ y_c/y_u^+ y_u) \lambda_{x,u}, 
	(y_c/y_u)\lambda_{z,u}, 
	U(y_c) 
	\big).
	\ea
	\label{eq.Xh}
	\ee
Equation~(\ref{eq.hierarchy-analytical}) states that the contribution of any hierarchy to the streamwise energy intensity at the wall-normal location $y$ and the Reynolds number $Re_\tau$ can be analytically determined from the streamwise energy density of the largest mode in that hierarchy computed at the Reynolds number $Re_{\tau,u}$. This is schematically illustrated in figure~\ref{fig.Euu-hierarchy-analytical} where the streamwise energy densities $A_{uu}$ of the largest mode (corresponding to $y_u$), the smallest mode (corresponding to $y_{l'}$), and a generic mode (corresponding to $y_{c}$) in a given hierarchy are shown. The contribution of each mode to the streamwise energy density at the wall-normal location $y$ are marked with solid lines. Owing to the self-similarity of the modes, these contributions are respectively equal to the contribution of the largest mode at the wall-normal locations $y$, $(y_u/y_{l'}) y$, and $(y_u/y_c) y$. The weight matrix is schematically shown for $y_{l'} \leq y_c \leq y_u$. The contribution of all the modes to the energy intensity at $y$ can be obtained by integrating the energy density of the largest mode from $y$ to $(y_{u}/y_{l'}) y$ (the red shaded region) weighted by the weight matrix $X_h$. 

The equations for $E_{vv,h}$, $E_{ww,h}$, and $E_{vw,h}$ are similarly obtained using~(\ref{eq.Evv-tilde-ij-map-cu}), e.g.
	\be
	\ba{rcl}
	E_{vv,h} (y,\blambda_u)
	&\!\! = \!\!&
	\dfrac{1}{\kappa}
	\,
	\ds{
	\int_{\ln y_{l'}}^{\ln y_u}
	}
	\,
	\dfrac{y_u^2}{y_c^2}
	\,
	\mbox{Re}
	\Big\{
	\mbox{tr}
	\Big(
	A_{vv} \big( \dfrac{y_u}{y_c}y, \blambda_u, c_u \big)
	X_h (\blambda_u, y_c)
	\Big)
	\Big\}
	\,
	\mrd \ln(y_c),
	\ea
	\label{eq.Evv-hierarchy-analytical-x}
	\ee
and the equations for $E_{uv,h}$ and $E_{uw,h}$ are obtained using~(\ref{eq.Euv-tilde-ij-map-cu}), e.g.
	\be
	\ba{rcl}
	\hskip-0.4cm
	E_{uv,h} (y,\blambda_u)
	&\!\!\! = \!\!\!&
	\dfrac{1}{\kappa}
	\dfrac{Re_\tau}{Re_{\tau,u}}
	\,
	\ds{
	\int_{\ln y_{l'}}^{\ln y_u}
	}
	\,
	\dfrac{y_u}{y_c}
	\,
	\mbox{Re}
	\Big\{
	\mbox{tr}
	\Big(
	A_{uv} \big( \dfrac{y_u}{y_c}y, \blambda_u, c_u \big)
	X_h (\blambda_u, y_c)
	\Big)
	\Big\}
	\,
	\mrd \ln(y_c).
	\hskip-0.3cm
	\ea
	\label{eq.Euv-hierarchy-analytical-x}
	\ee

    \begin{figure}
    \begin{center}
    \includegraphics[width=0.45\columnwidth]
    {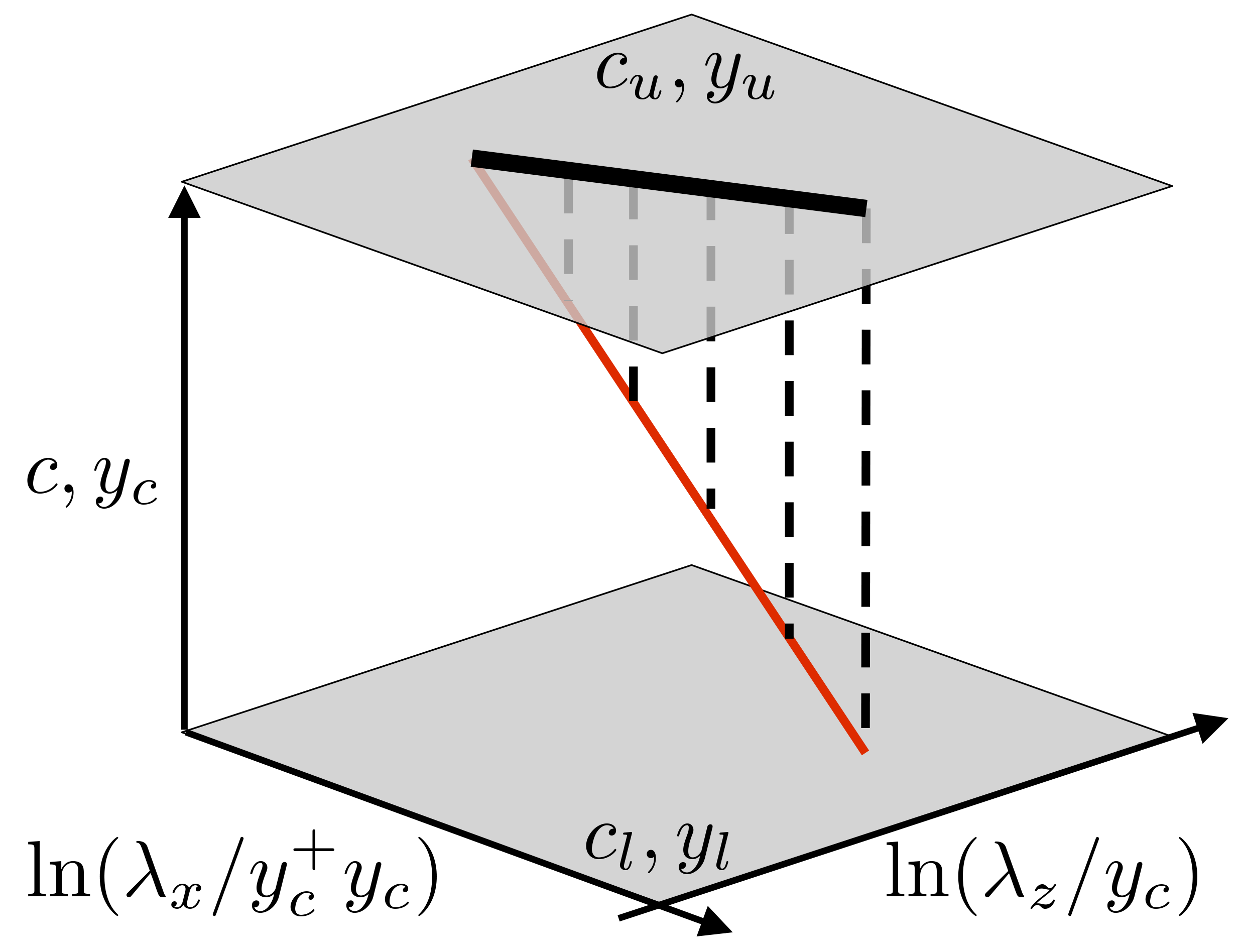}
    \end{center}
    \caption{Schematic showing that the contribution of the self-similar modes to the two-dimensional spectrum at a given $\blambda$ can be obtained from the resolvent modes with speed $c_u$. The thin red line corresponds to fixed $\blambda$ as $c$ increases from $c_l$ to $c_u$. The dashed vertical lines show the hierarchies that correspond to the modes on the thin red line, and the thick black line corresponds to the largest modes in these hierarchies. The two-dimensional spectrum is obtained by averaging over the thin red line or, alternatively, over the thick black line.}
    \label{fig.hierarchy-fixed-lambda}
    \end{figure}	 
    	
\subsection{Two-dimensional energy spectra and co-spectra}
\label{sec.time-averaged-spectra}

We show that the largest modes in the self-similar hierarchies are sufficient to determine the contribution of the self-similar modes to the two-dimensional energy spectrum at a given $\blambda$. For example, the contribution of the self-similar modes to the pre-multiplied mean streamwise energy spectrum is obtained by integrating the three-dimensional spectra over $c$ from $c_{l'}$ to $c_u$,
	\be
	E_{uu,s} (y, \blambda)
	\; = \;
	\ds{\int_{c_{l'}}^{c_u}}
	\,
	E_{uu} (y, \blambda, c)
	\,
	\mrd c,
	\label{eq.Euu-t-avg}
	\ee
where the subscript $s$ denotes the self-similar contribution to $E_{uu}$. From~(\ref{eq.Euu-trace}) and~(\ref{eq.Euu-tilde-ij-map-cu}), we have	
	\be
	\ba{rcl}
	E_{uu,s} (y, \blambda)
	&\!\! = \!\!&
	\dfrac{1}{\kappa}
         \dfrac{Re_\tau^2}{Re_{\tau,u}^2}
	\,
	\ds{
	\int_{y_{l'}}^{y_u}
	}
	\,
	\mbox{Re}
	\Big\{
	\mbox{tr}
	\Big(
	A_{uu} \big(\dfrac{y_u}{y_c} y, \blambda_u(y_c), c_u\big)
	X (\blambda, U(y_c))
	\Big)
	\Big\}
	\,
	\mrd \ln (y_c),
	\ea
	\label{eq.Euu-t-avg-analytical}
	\ee
where the dependence of $\blambda_u$ on $y_c$ is emphasized. In other words, the contribution of the self-similar modes to the two-dimensional energy spectrum at a given $\blambda$ can be obtained by integrating over a subset of the largest modes in the self-similar hierarchies. This is schematically illustrated in figure~\ref{fig.hierarchy-fixed-lambda} where the thin red line corresponds to fixed $\blambda$ as $c$ increases from $c_{l'}$ to $c_u$. The dashed vertical lines show the hierarchies that correspond to the modes on the thin red line, and the thick black line corresponds to the largest modes in these hierarchies. Following~(\ref{eq.Euu-t-avg-analytical}), the two-dimensional spectrum is obtained by integration over the thin red line or, alternatively, over the thick black line.    

The equations for $E_{vv,s}$, $E_{ww,s}$, and $E_{vw,s}$ are similarly obtained using~(\ref{eq.Evv-tilde-ij-map-cu}), e.g.
	\be
	\ba{rcl}
	E_{vv,s} (y, \blambda)
	&\!\! = \!\!&
	\dfrac{1}{\kappa}
	\,
	\ds{
	\int_{y_{l'}}^{y_u}
	}
	\,
	\dfrac{y_u^2}{y_c^2}
	\,
	\mbox{Re}
	\Big\{
	\mbox{tr}
	\Big(
	A_{vv} \big(\dfrac{y_u}{y_c} y, \blambda_u(y_c), c_u\big)
	X (\blambda, U(y_c))
	\Big)
	\Big\}
	\,
	\mrd \ln (y_c),
	\ea
	\label{eq.Evv-t-avg-analytical}
	\ee	
and the equations for $E_{uv,s}$ and $E_{uw,s}$ are obtained using~(\ref{eq.Euv-tilde-ij-map-cu}), e.g.
	\be
	\ba{rcl}
	\hskip-0.4cm
	E_{uv,s} (y, \blambda)
	&\!\! = \!\!&
	\dfrac{1}{\kappa}
      \dfrac{Re_\tau}{Re_{\tau,u}}
	\,
	\ds{
	\int_{y_{l'}}^{y_u}
	}
	\,
	\dfrac{y_u}{y_c}
	\,
	\mbox{Re}
	\Big\{
	\mbox{tr}
	\Big(
	A_{uv} \big(\dfrac{y_u}{y_c} y, \blambda_u(y_c), c_u\big)
	X (\blambda, U(y_c))
	\Big)
	\Big\}
	\,
	\mrd \ln (y_c).
	\hskip-0.3cm
	\ea
	\label{eq.Euv-t-avg-analytical}
	\ee

\subsection{Energy budget}
\label{sec.energy-budget}

The energy budget for the modes in the self-similar hierarchies can be determined from the energy budget of the largest mode in that hierarchy. The energy budget is obtained from the wall-normal integral of the inner product of $\hat{\bu}$ with both sides of equation~(\ref{eq.NS-lin}), see Appendix~\ref{sec.energy-budget-der}
	\be
	E_P (\blambda, c) 
	\, - \,
	E_D (\blambda, c) 
	\; = \;
	E_T (\blambda, c). 
	\label{eq.budget-brief}
	\ee
Here, $E_P$ and $E_D$ are the pre-multiplied energy production and dissipation at the mode with $\blambda$ and $c$, and $E_T$ is the pre-multiplied transported energy from all the modes to the mode with $\blambda$ and $c$
	\be
	\ba{rcl}
	E_P (\blambda,c)
	&\!\! = \!\!&
	-(\dfrac{2\pi}{\lambda_x})^2 \, (\dfrac{2\pi}{|\lambda_z|}) \;
	\ds{\int_{0}^{2}}
	\mbox{Re} 
	\big\{
	U'(y) \, 
	\overline{\hat{u}} (y, \blambda,c) \,
	\hat{v} (y, \blambda,c)
	\big\}
	\mrd y,
	\\[0.25cm]
	E_D (\blambda,c)
	&\!\! = \!\!&
	\dfrac{1}{Re_\tau} \, 
	(\dfrac{2\pi}{\lambda_x})^2 \, (\dfrac{2\pi}{|\lambda_z|}) \;
	\ds{\int_{0}^{2}}
	\big(
	\hat{\bu}' (y, \blambda,c)^* \,
	\hat{\bu}' (y, \blambda,c) 
	\, + \,
	\kappa^2 \,
	\hat{\bu} (y, \blambda,c)^* \,
	\hat{\bu} (y, \blambda,c) 
	\big)
	\,
	\mrd y,
	\\[0.25cm]
	E_T (\blambda,c)
	&\!\! = \!\!&
	-(\dfrac{2\pi}{\lambda_x})^2 \, (\dfrac{2\pi}{|\lambda_z|}) \;
	\ds{\int_{0}^{2}}
	\mbox{Re} 
	\big\{
	\hat{\bu} (y, \blambda,c)^* \,
	\hat{\fvec} (y, \blambda,c)
	\big\}
	\mrd y.
	\ea
	\label{eq.budget-def}
	\ee
The contribution of the first $N$ resolvent modes to $E_P$, $E_D$, and $E_T$ is determined by substituting $\hat{\fvec}$ and $\hat{\bf u}$ from~(\ref{eq.f}) and~(\ref{eq.u}) in~(\ref{eq.budget-def}),
	\be
	\ba{rcl}
	E_P (\blambda, c)
	&\!\! = \!\!&
	\mbox{Re} 
	\Big\{
	\mbox{tr}
	\big(
	A_P (\blambda, c)
	\,
	X (\blambda, c) 
	\big)
	\Big\},
	\\[0.15cm]
	E_D (\blambda, c)
	&\!\! = \!\!&
	\mbox{tr}
	\big(
	A_D (\blambda, c)
	\,
	X (\blambda, c) 
	\big),
	\\[0.15cm]
	E_T (\blambda, c)
	&\!\! = \!\!&
	\mbox{Re} 
	\Big\{
	\mbox{tr}
	\big(
	A_T (\blambda, c)
	\,
	X (\blambda, c) 
	\big)
	\Big\}.
	\ea
	\label{eq.budget-trace}
	\ee
Here, $X$ is the weight matrix defined in~(\ref{eq.X-ij}) and $A_P$, $A_D$, and $A_T$ are the $N \times N$ energy-production, energy-dissipation, and energy-transport matrices whose $ij$-th elements are determined by the resolvent modes,
	\be
	\ba{rcl}
	A_{P,ij} (\blambda, c)
	&\!\! = \!\!&
	-(\dfrac{2\pi}{\lambda_x})^2 \, (\dfrac{2\pi}{|\lambda_z|}) \;
	\sigma_i(\blambda, c) \,
	\sigma_j(\blambda, c) \,
	\ds{\int_{0}^{2}}
	U'(y) \,
	\overline{\hat{u}}_i (y, \blambda,c) \,
	\hat{v}_j (y, \blambda,c) \,
	\mrd y,
	\\[0.25cm]
	A_{D,ij} (\blambda, c)
	&\!\! = \!\!&
	\dfrac{1}{Re_\tau} \, 
	(\dfrac{2\pi}{\lambda_x})^2 \, (\dfrac{2\pi}{|\lambda_z|}) \;
	\sigma_i(\blambda, c) \sigma_j(\blambda, c)
	\, \times
	\\[0.25cm]
	&&
	\ds{\int_{0}^{2}}
	\big(
	\hat{\bpsi}'_i (y, \blambda,c)^*
	\hat{\bpsi}'_j (y, \blambda,c) 
	+
	\kappa^2
	\hat{\bpsi}_i (y, \blambda,c)^*
	\hat{\bpsi}_j (y, \blambda,c)
	\big)
	\mrd y,
	\\[0.25cm]
	A_{T,ij} (\blambda, c)
	&\!\! = \!\!&
	-(\dfrac{2\pi}{\lambda_x})^2 \, (\dfrac{2\pi}{|\lambda_z|}) \;
	\sigma_i(\blambda, c) \,
	\ds{\int_{0}^{2}}
	\hat{\bpsi}_i (y, \blambda,c)^* \,
	\hat{\bphi}_j (y, \blambda,c) \,
	\mrd y.
	\ea
	\label{eq.budget-A}
	\ee	
Notice that the energy dissipation $E_D$ is always positive since it is given by the trace of the product of two positive semi-definite matrices $A_D$ and $X$. 

Following the scalings in~(\ref{eq.u-map-cu}) and~(\ref{eq.sigma-map-cu}), $A_P$, $A_D$, and $A_T$ for any mode in a hierarchy $\cS(\blambda_u)$ can be obtained from the largest mode in that hierarchy
	\be
	\ba{rcl}
	A_P (\blambda,c)
	&\!\! = \!\!&
	(Re_\tau/Re_{\tau,u}) \, (y_u/y_c)
	\,
	A_P (\blambda_u,c_u),
	\\[0.2cm]
	A_D (\blambda,c)
	&\!\! = \!\!&
	(Re_\tau/Re_{\tau,u}) \, (y_u/y_c)
	\,
	A_D (\blambda_u,c_u),
	\\[0.2cm]
	A_T (\blambda,c)
	&\!\! = \!\!&
	(y_u^+/y_c^+) \, (y_u/y_c)^2
	\,
	A_T (\blambda_u,c_u),
	\ea
	\label{eq.budget-scaling}
	\ee
where we recall the distinction between $y_c^+$ and $y_u^+$, cf.~(\ref{eq.ycp-yup}). Notice that the entries in the production and dissipation matrices scale similarly, increase with $Re_\tau$, and decrease with $y_c$. In addition, the entries in the transport matrix decrease with $y_c^+ y_c^2$. Therefore, as the modes become larger in a hierarchy, the energy production, dissipation, and transport are reduced. However, more energy is produced and dissipated by the mode than the energy that is transported to or from the other modes. The ratio of the entries in the production matrix and the entries in the transport matrix grows with $(y_c^+)^2$.

The contribution of hierarchy $\cS (\blambda_u)$ to the energy production, dissipation, and transport is obtained from
	\be
	\ba{rcl}
	E_{P,h} (\blambda_u)
	&\!\! = \!\!&
	\dfrac{1}{\kappa}
	\,
	\dfrac{Re_\tau}{Re_{\tau,u}}
	\,
	\ds{
	\int_{y_{l'}}^{y_u}
	}
	\,
	\dfrac{y_u}{y_c}
	\,
	\mbox{Re}
	\Big\{
	\mbox{tr}
	\big(
	A_P (\blambda_u, c_u)
	\,
	X_h (\blambda_u, y_c)
	\big)
	\Big\}
	\mrd \ln(y_c),
	\\[0.3cm]
	E_{D,h} (\blambda_u)
	&\!\! = \!\!&
	\dfrac{1}{\kappa}
	\,
	\dfrac{Re_\tau}{Re_{\tau,u}}
	\,
	\ds{
	\int_{y_{l'}}^{y_u}
	}
	\,
	\dfrac{y_u}{y_c}
	\,
	\mbox{tr}
	\big(
	A_D (\blambda_u, c_u)
	\,
	X_h (\blambda_u, y_c)
	\big)
	\mrd \ln(y_c),
	\\[0.3cm]
	E_{T,h} (\blambda_u)
	&\!\! = \!\!&
	\dfrac{1}{\kappa}
	\,
	\ds{
	\int_{y_{l'}}^{y_u}
	}
	\,
	\dfrac{y_u^+ y_u^2}{y_c^+ y_c^2}
	\,
	\mbox{Re}
	\Big\{
	\mbox{tr}
	\big(
	A_T (\blambda_u, c_u)
	\,
	X_h (\blambda_u, y_c)
	\big)
	\Big\}
	\mrd \ln(y_c).
	\ea
	\label{eq.budget-hierarchy}
	\ee
	
\section{Implications for the rank-1 model}
\label{sec.rank-1}

\cite{moashatromckJFM13} showed that the resolvent operator for the modes that correspond to the most energetic wavelengths can be approximated using the principal resolvent modes. It was also shown that the resolvent modes come in pairs owing to the channel symmetry relative to the center plane. When $\sigma_1$ and $\sigma_2$ are approximately equal, the corresponding singular functions are approximately symmetric and anti-symmetric counterparts of each other. 

\begin{figure}
    \begin{center}
    \begin{tabular}{cc}
    \subfigure{\includegraphics[width=0.42\columnwidth]
    {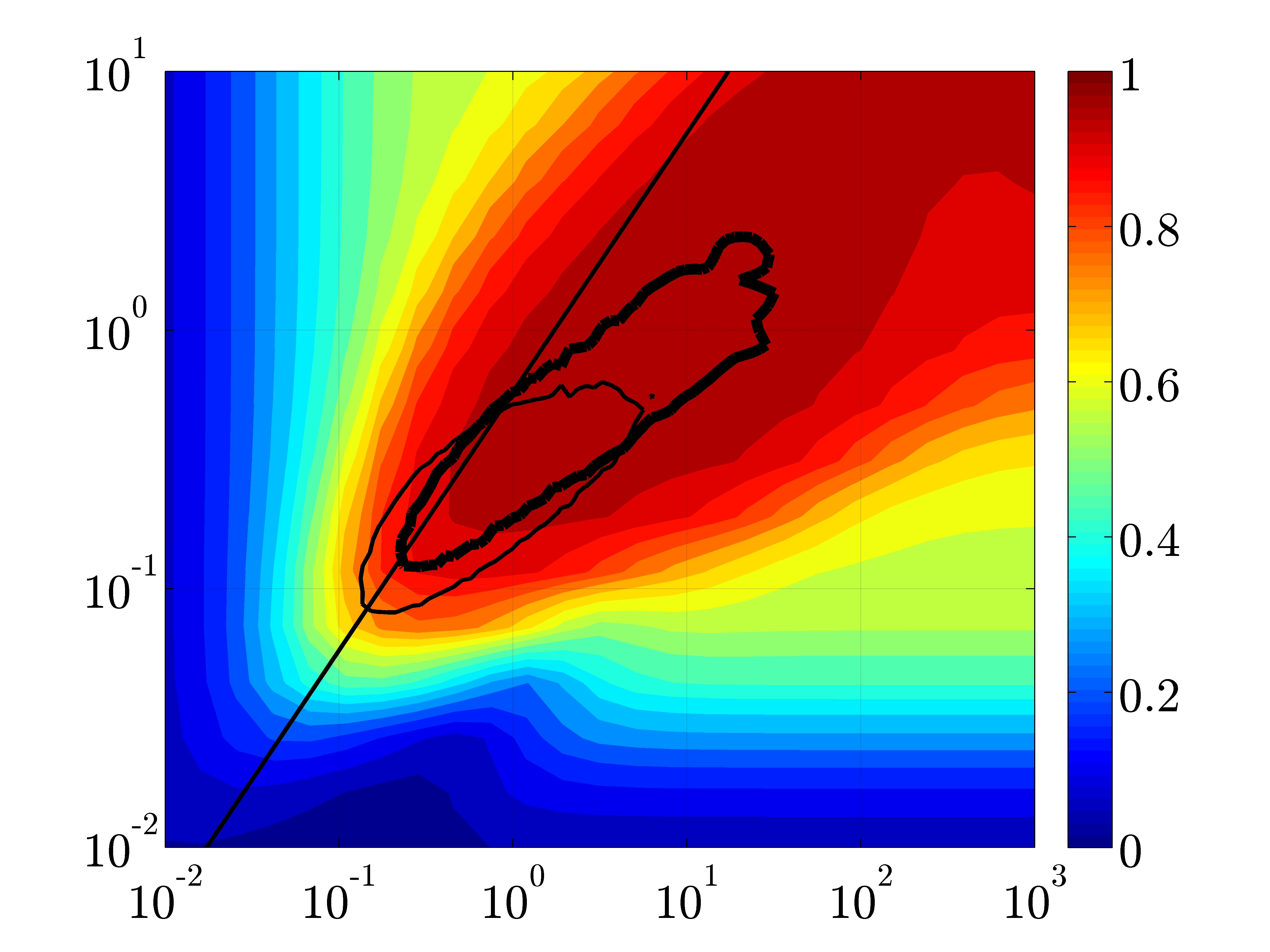}
    \label{fig.EuuthickEuvthinDNS_vs_lx_lz_y0p075_50percent_R2003_sigma1_ratio}}
    &
    \subfigure{\includegraphics[width=0.42\columnwidth]
    {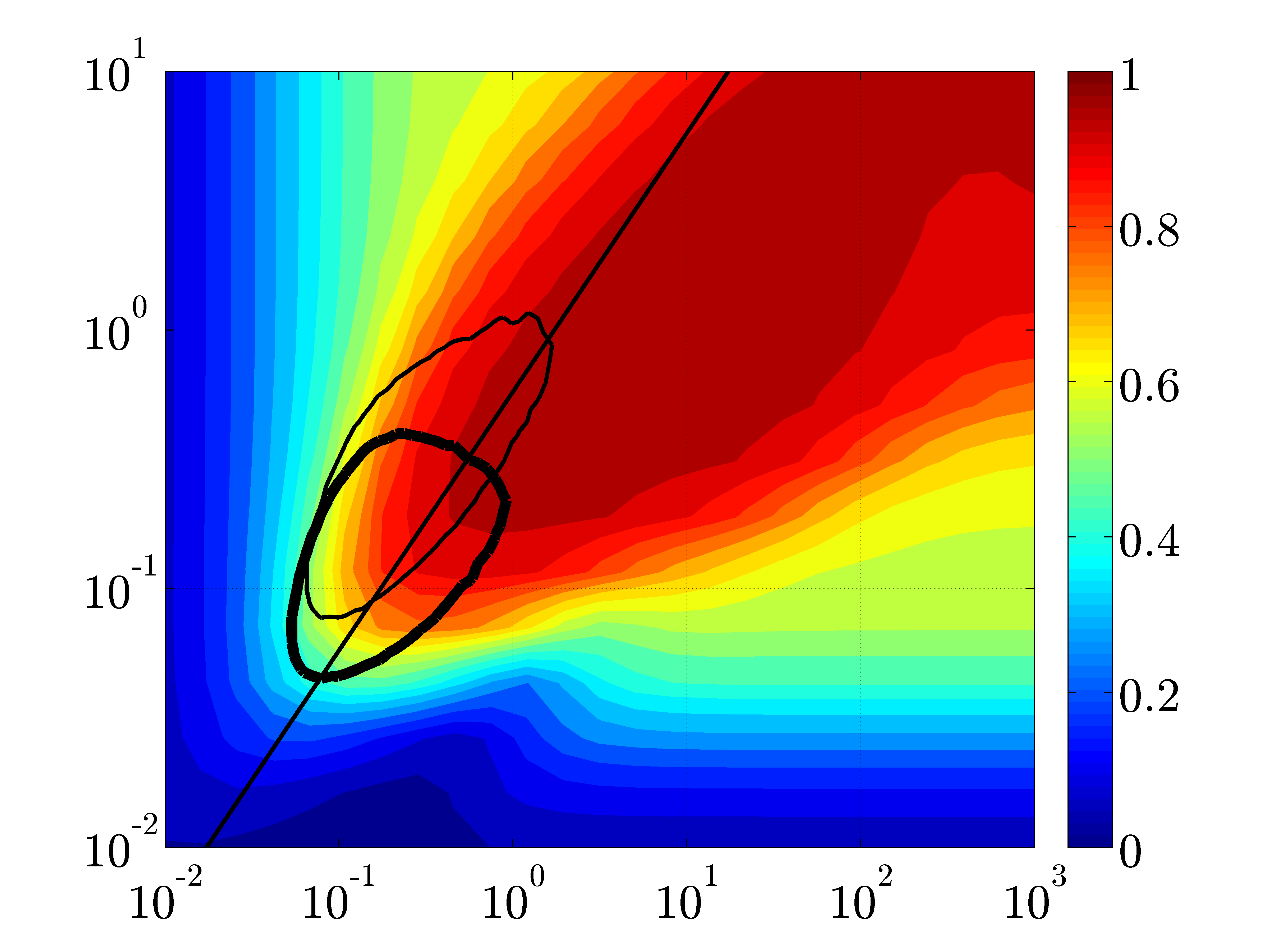}
    \label{fig.EvvthickEwwthinDNS_vs_lx_lz_y0p075_50percent_R2003_sigma1_ratio}}
    \\[0.2cm]
    $(a)$
    &
    $(b)$
    \end{tabular}
    \begin{tabular}{c}
    \\[-3.8cm]
    \begin{tabular}{c}
    \hskip-5.7cm
    \begin{turn}{90}
    \tc{black}{$\lambda_z$}
    \end{turn}
    \hskip5.7cm
    \begin{turn}{90}
    \tc{black}{$\lambda_z$}
    \end{turn}
    \end{tabular}
    \\[2cm]
    \begin{tabular}{c}
    \hskip-0.05cm
    \tc{black}{$\lambda_x$}
    \hskip5.55cm
    \tc{black}{$\lambda_x$}
    \end{tabular}
    \end{tabular}
    \end{center}
    \caption{The black contours show the pre-multiplied (a) $E_{uu}$ (thick), $-E_{uv}$ (thin), (b) $E_{vv}$ (thick), $E_{ww}$ (thin) at a wall-normal location $y = 0.075$ in the overlap region for $Re_\tau = 2003$ from DNS~\citep{hoyjim06}. The contour levels correspond to $50\%$ of the maximum value in each spectrum. The color contours show the energy amplification by the principal forcing modes relative to the total energy amplification by all the forcing modes, i.e. $(\sigma_1^2+\sigma_2^2)/\sum_{j} \sigma_j^2$, for $c = U(y_c = 0.075)$ and $Re_\tau = 2003$. The straight lines correspond to $\lambda_x = \sqrt{3} \lambda_z$.}
    \label{fig.EDNS_vs_lx_lz_y0p075_R2003_sigma1_ratio}
    \end{figure}

The contours in figure~\ref{fig.EDNS_vs_lx_lz_y0p075_R2003_sigma1_ratio} correspond to $50\%$ of the maximum pre-multiplied energy spectra from DNS~\citep{hoyjim06} at $Re_\tau = 2003$ and $y = 0.075$ in the overlap region. The color contours show the energy amplification by the principal forcing mode relative to the total energy amplification, i.e. $(\sigma_1^2+\sigma_2^2)/\sum_{j} \sigma_j^2$, for $c = U(y_c = 0.075)$ at the same $Re_\tau$. Notice that the wavelengths that correspond to more than $50\%$ of the maximum $E_{uu}$ and $E_{uv}$ coincide with the wavelengths for which the principal resolvent pair captures more than $80\%$ of the energy amplification. For the $E_{vv}$ and $E_{ww}$ spectra, the latter number is $60\%$. In addition, the most energetic wavelengths in the $E_{uu}$ and $E_{uv}$ spectra satisfy the necessary aspect-ratio constraint for self-similarity, see the straight lines in figure~\ref{fig.EDNS_vs_lx_lz_y0p075_R2003_sigma1_ratio} corresponding to $\lambda_x = \sqrt{3} \lambda_z$. This motivates studying the implication of the self-similar scalings on the rank-1 model where $\chi_j (\blambda, c) = 0$ for $j \neq 1$. 

For the rank-1 model, we have
	\be
	\hat{\bu}(y, \blambda, c) 
	\; = \;
	\chi_1(\blambda, c)  \,
	\sigma_1(\blambda, c)  \,
	\hat{\bpsi}_1(y, \blambda, c), 
	\non
	\ee		
and the equation for the pre-multiplied streamwise energy spectrum simplifies to, cf.~(\ref{eq.Euu-trace}), 
	\be
	E_{uu} (y, \blambda, c) 
	\; = \;
	|\chi_1 (\blambda, c)|^2  \,
	A_{uu} (y, \blambda, c).
	\label{eq.Euu-rank1}
	\ee
Here, the energy density $A_{uu} (y, \blambda, c)$ is a scalar number. Notice that the energy density equals the energy spectrum if a broadband forcing is considered, i.e. $\chi_1(\blambda,c) = 1$. Even though the nonlinear forcing in a real turbulent flow is non-broadband, it is instructive to compare the flow under the broadband forcing assumption with the real flow. 

\begin{figure}
    \begin{center}
    \begin{tabular}{cc}
    \subfigure{\includegraphics[width=0.42\columnwidth]
    {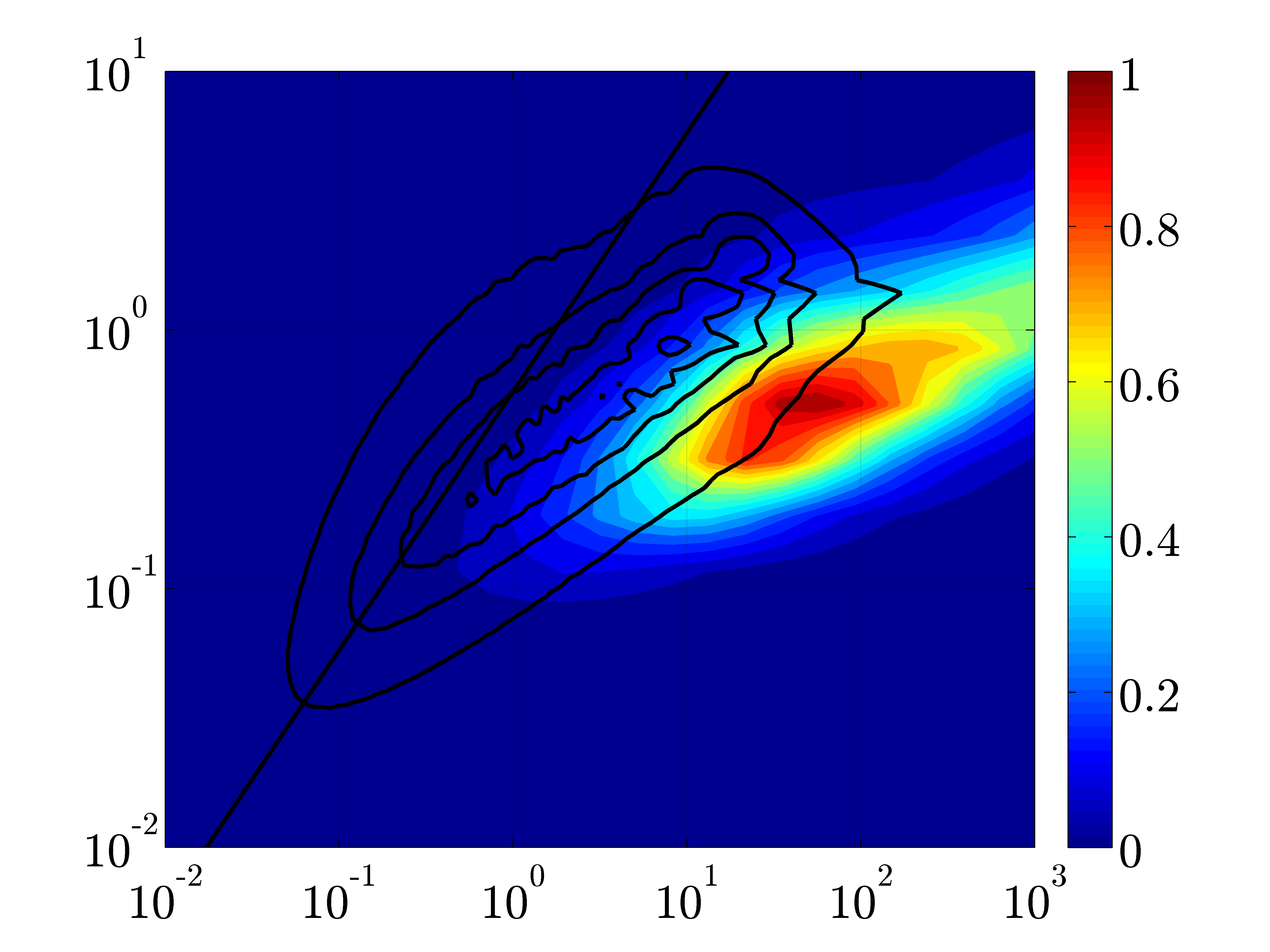}
    \label{fig.EuuDNs_vs_lx_lz_y0p075_R2003_kxkzsigma1u1u1}}
    &
    \subfigure{\includegraphics[width=0.42\columnwidth]
    {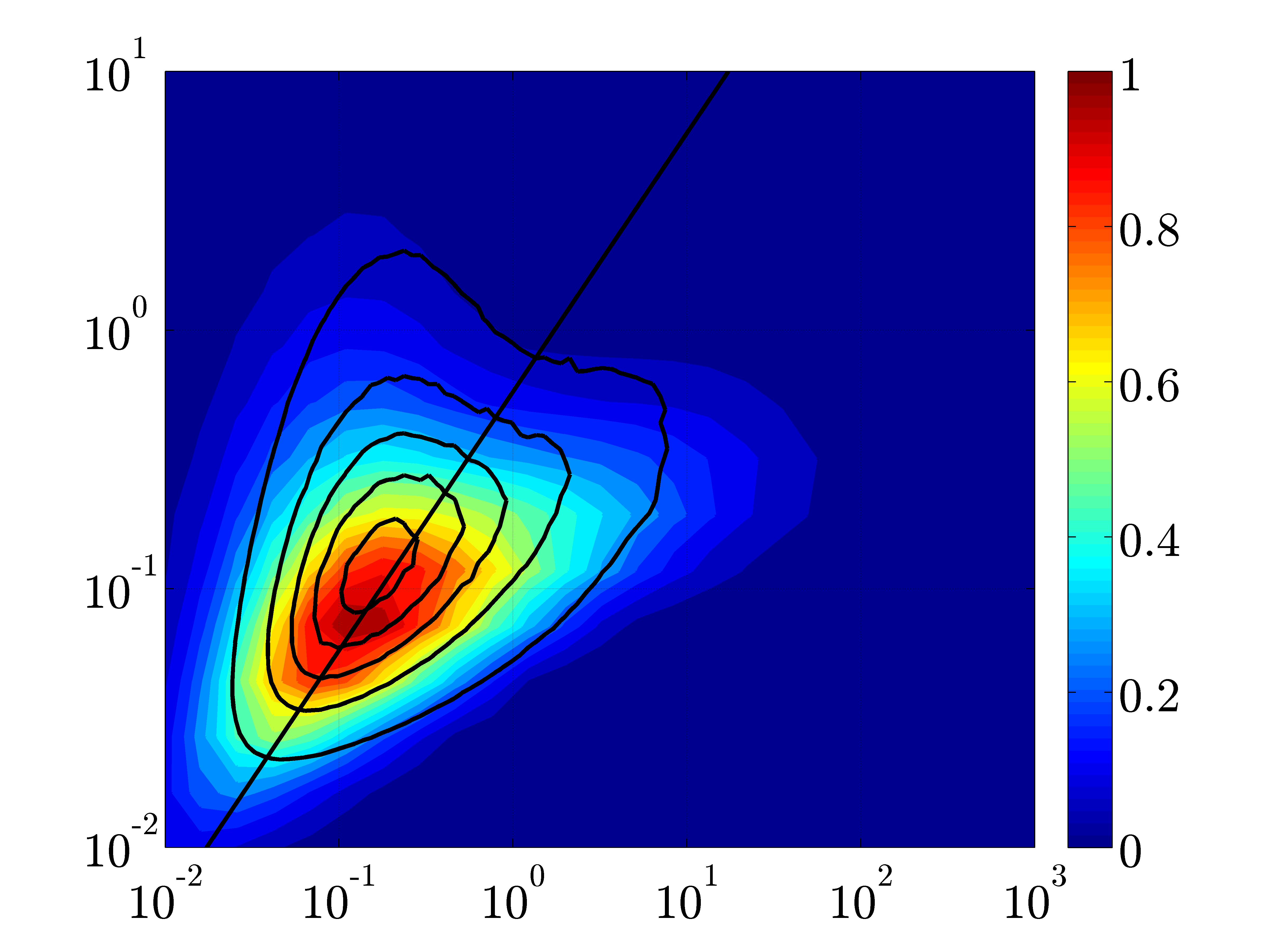}
    \label{fig.EvvDNs_vs_lx_lz_y0p075_R2003_kxkzsigma1v1v1}}
    \\[0.2cm]
    $(a)$
    &
    $(b)$
    \end{tabular}
    \begin{tabular}{c}
    \\[-3.8cm]
    \begin{tabular}{c}
    \hskip-5.7cm
    \begin{turn}{90}
    \tc{black}{$\lambda_z$}
    \end{turn}
    \hskip5.7cm
    \begin{turn}{90}
    \tc{black}{$\lambda_z$}
    \end{turn}
    \end{tabular}
    \\[2cm]
    \begin{tabular}{c}
    \hskip-0.05cm
    \tc{black}{$\lambda_x$}
    \hskip5.55cm
    \tc{black}{$\lambda_x$}
    \end{tabular}
    \end{tabular}
    \\[-0.2cm]
    \begin{tabular}{cc}
    \subfigure{\includegraphics[width=0.42\columnwidth]
    {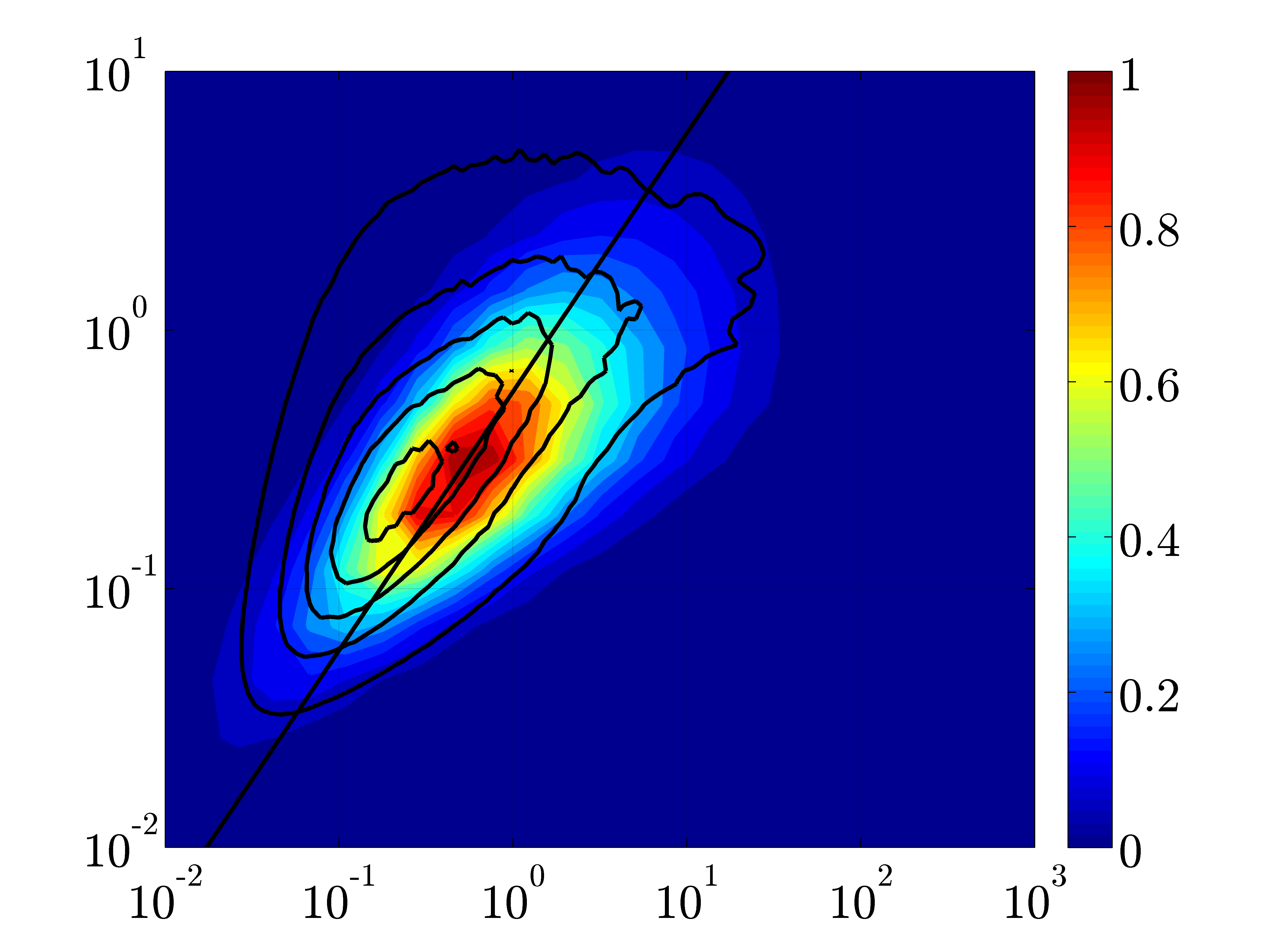}
    \label{fig.EwwDNs_vs_lx_lz_y0p075_R2003_kxkzsigma1w1w1}}
    &
    \subfigure{\includegraphics[width=0.42\columnwidth]
    {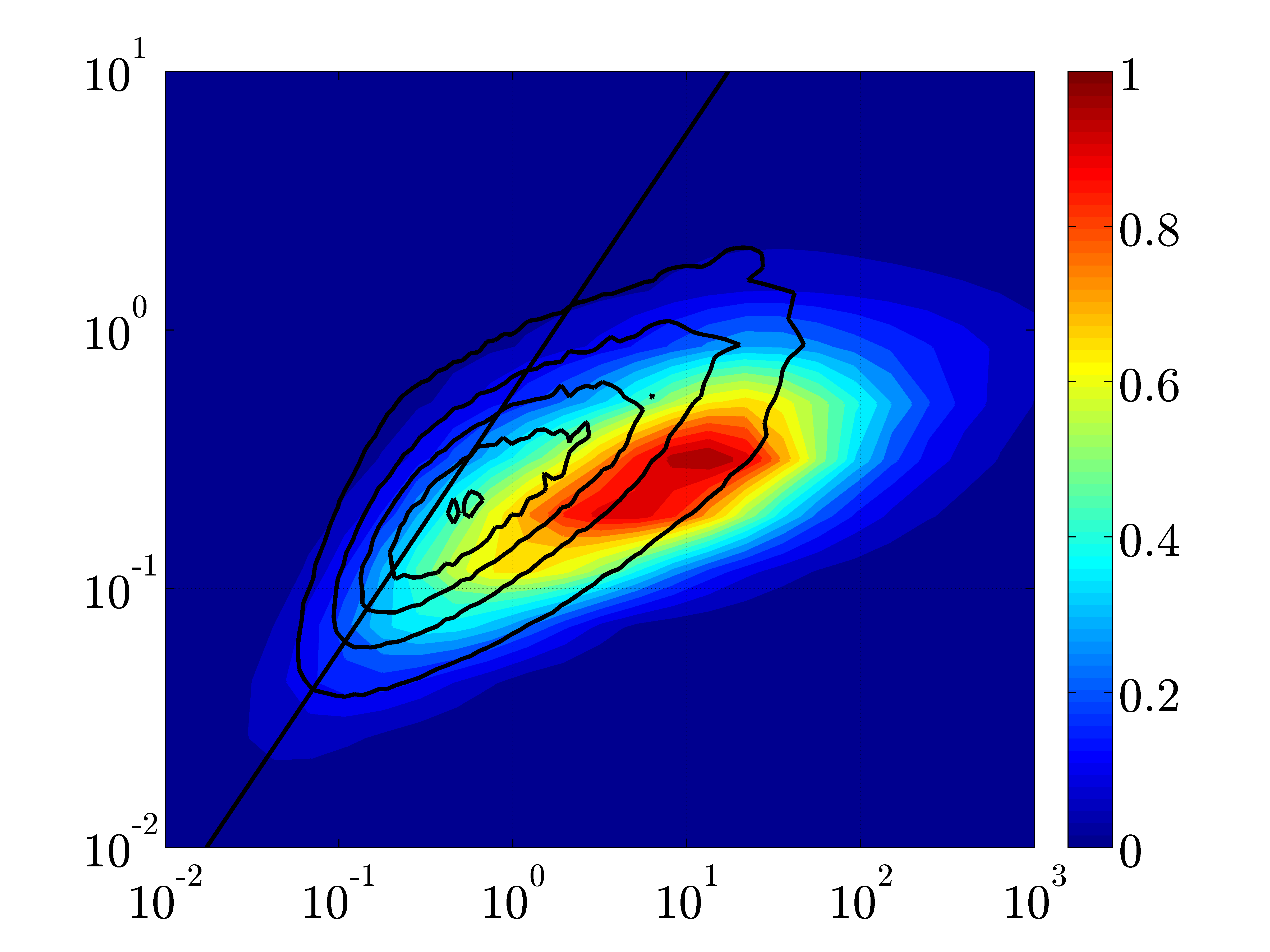}
    \label{fig.EuvDNs_vs_lx_lz_y0p075_R2003_kxkzsigma1u1v1}}
    \\[0.2cm]
    $(c)$
    &
    $(d)$
    \end{tabular}
    \begin{tabular}{c}
    \\[-3.8cm]
    \begin{tabular}{c}
    \hskip-5.7cm
    \begin{turn}{90}
    \tc{black}{$\lambda_z$}
    \end{turn}
    \hskip5.7cm
    \begin{turn}{90}
    \tc{black}{$\lambda_z$}
    \end{turn}
    \end{tabular}
    \\[2cm]
    \begin{tabular}{c}
    \hskip-0.05cm
    \tc{black}{$\lambda_x$}
    \hskip5.55cm
    \tc{black}{$\lambda_x$}
    \end{tabular}
    \end{tabular}
    \end{center}
    \caption{The black contours show the pre-multiplied (a) $E_{uu}$, (b) $E_{vv}$, (c) $E_{ww}$, (d) $-E_{uv}$ at a wall-normal location $y = 0.075$ in the overlap region for $Re_\tau = 2003$ from DNS~\citep{hoyjim06}. The contour levels correspond to $10\%$ to $90\%$ of the maximum value with increments of $20\%$. The color contours show the footprint at $y = 0.075$ of the contribution of the principal resolvent modes with $c = U(y_c = 0.075)$ to the pre-multiplied spectra for $Re_\tau = 2003$, (a) $\kappa_x \kappa_z (\sigma_1 |\hat{u}_1|)^2$, (b) $\kappa_x \kappa_z (\sigma_1 |\hat{v}_1|)^2$, (c) $\kappa_x \kappa_z (\sigma_1 |\hat{w}_1|)^2$, (d) $-\kappa_x \kappa_z \sigma_1^2 \mbox{Re} (\hat{u}_1 \overline{\hat{v}}_1)$. The straight lines correspond to $\lambda_x = \sqrt{3} \lambda_z$.}
    \label{fig.EDNs_vs_lx_lz_y0p075_R2003_kxkzsigma1u1u1}
    \end{figure}

The color contours in figure~\ref{fig.EDNs_vs_lx_lz_y0p075_R2003_kxkzsigma1u1u1} show the contribution of the principal resolvent modes with $c = U(y_c = 0.075)$ to the pre-multiplied energy densities at $y = 0.075$ for $Re_\tau = 2003$. The black contours show the pre-multiplied spectra at $y = 0.075$ for the same $Re_\tau$ from DNS~\citep{hoyjim06}. The contour levels correspond to $10\%$ to $90\%$ of the maximum value with increments of $20\%$ and the straight lines correspond to $\lambda_x = \sqrt{3} \lambda_z$. The approximate agreement between the energy densities and the DNS-based spectra highlights the effectiveness of the rank-1 representation of the flow. In addition, figures~\ref{fig.EuuDNs_vs_lx_lz_y0p075_R2003_kxkzsigma1u1u1} and~\ref{fig.EuvDNs_vs_lx_lz_y0p075_R2003_kxkzsigma1u1v1} show that the resolvent modes that violate the aspect-ratio constraint $\lambda_x/\lambda_z > \sqrt{3}$ have a negligible footprint on the streamwise and the streamwise/wall-normal energy densities. We also note that the modes with the largest contribution to $E_{uu}$ and $E_{uv}$ have a larger aspect-ratio than $\sqrt{3}$. This observation further justifies the relevance of self-similar modes for representation of the $E_{uu}$ and $E_{uv}$ spectra.
   
    \begin{figure}
    \begin{center}
    \begin{tabular}{cc}
    \subfigure{\includegraphics[width=0.47\columnwidth]
    {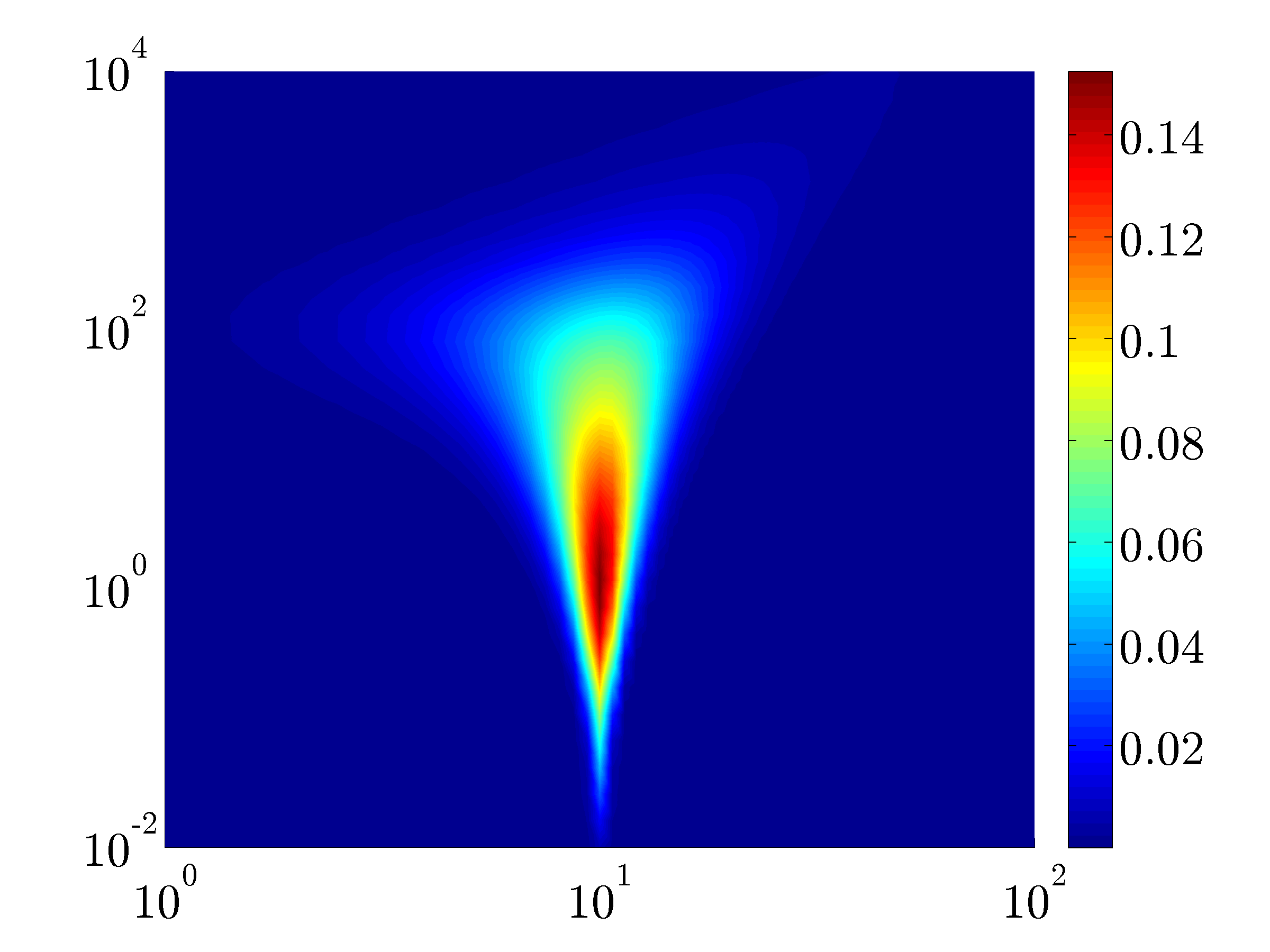}
    \label{fig.Euu-vs-yh-lx-R10000-exact}}
    &
    \subfigure{\includegraphics[width=0.47\columnwidth]
    {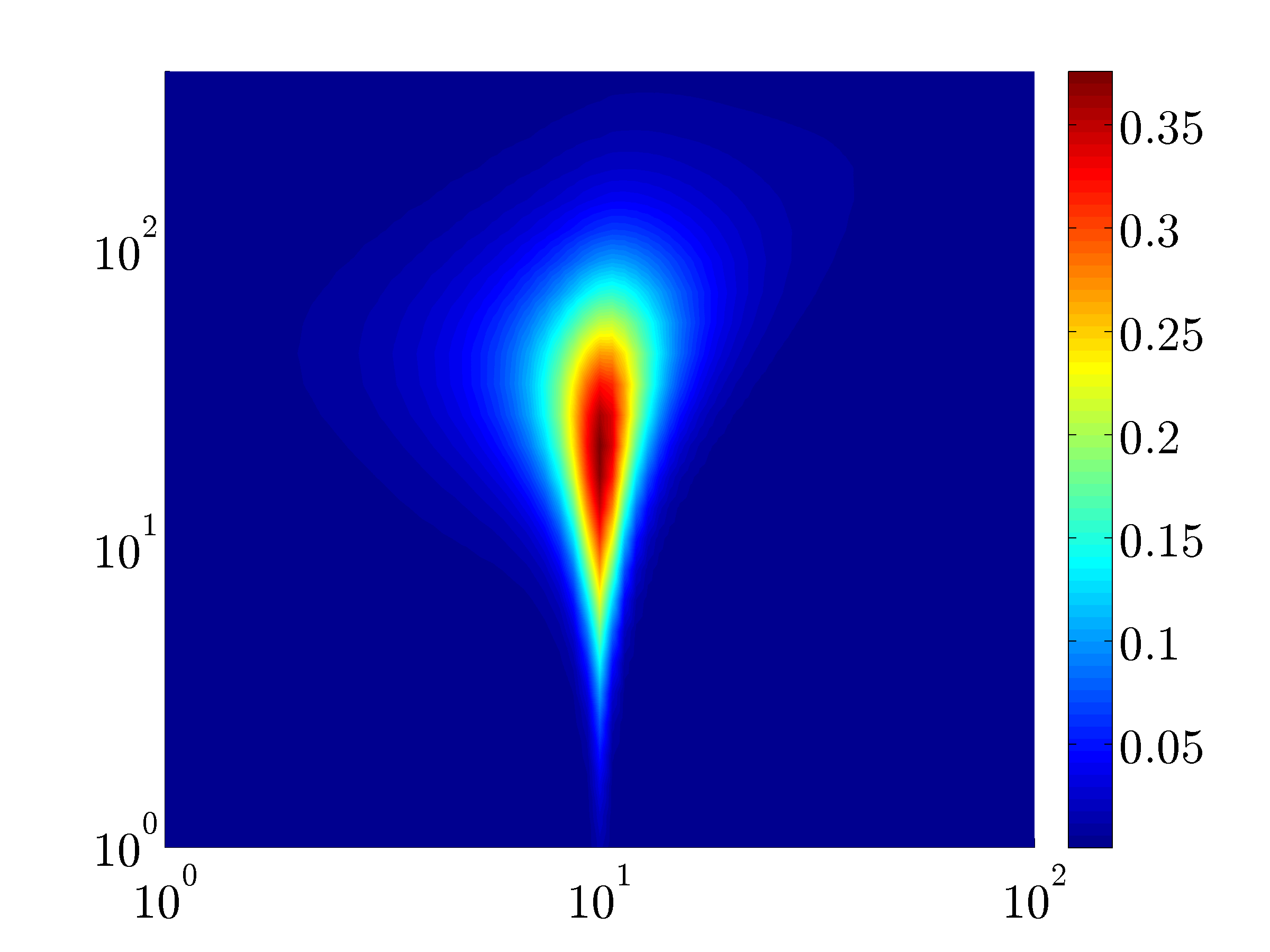}
    \label{fig.Euu-vs-yh-lzh-R10000-exact}}
    \\[0.2cm]
    $(a)$
    &
    $(b)$
    \end{tabular}
    \begin{tabular}{c}
    \\[-4.4cm]
    \begin{tabular}{c}
    \hskip-6.5cm
    \begin{turn}{90}
    \tc{black}{$~~~\lambda_x$}
    \end{turn}
    \hskip6.2cm
    \begin{turn}{90}
    \tc{black}{$\sqrt{\lambda_z^+ \lambda_z}$}
    \end{turn}
    \end{tabular}
    \\[2.2cm]
    \begin{tabular}{c}
    \hskip-0.2cm
    \tc{black}{$\sqrt{y^+ y}$}
    \hskip5.6cm
    \tc{black}{$\sqrt{y^+ y}$}
    \end{tabular}
    \end{tabular}
    \\[-0.2cm]
    \begin{tabular}{cc}
    \subfigure{\includegraphics[width=0.47\columnwidth]
    {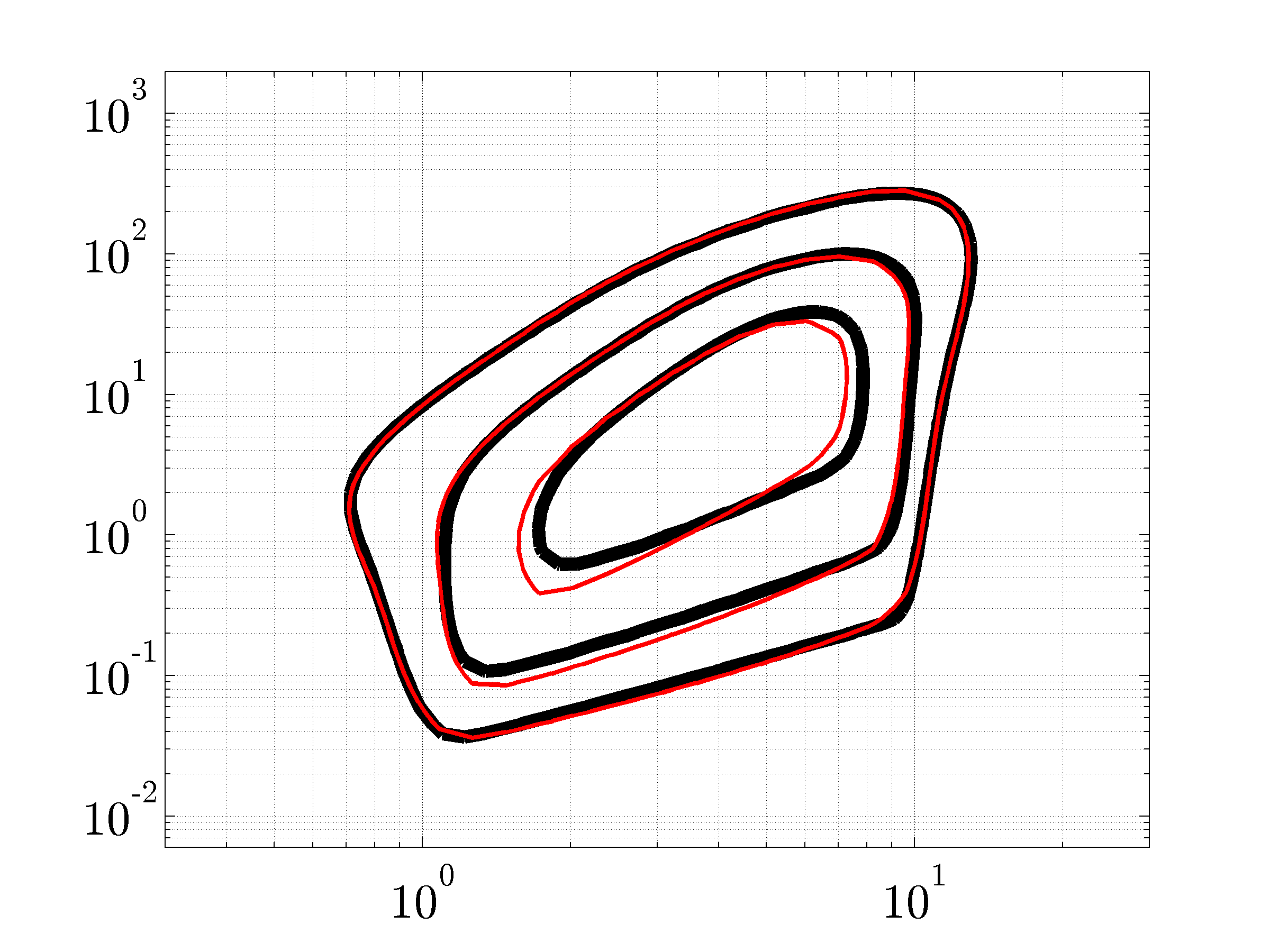}
    \label{fig.Euu-vs-yh-lx-R10000-analytical}}
    &
    \subfigure{\includegraphics[width=0.47\columnwidth]
    {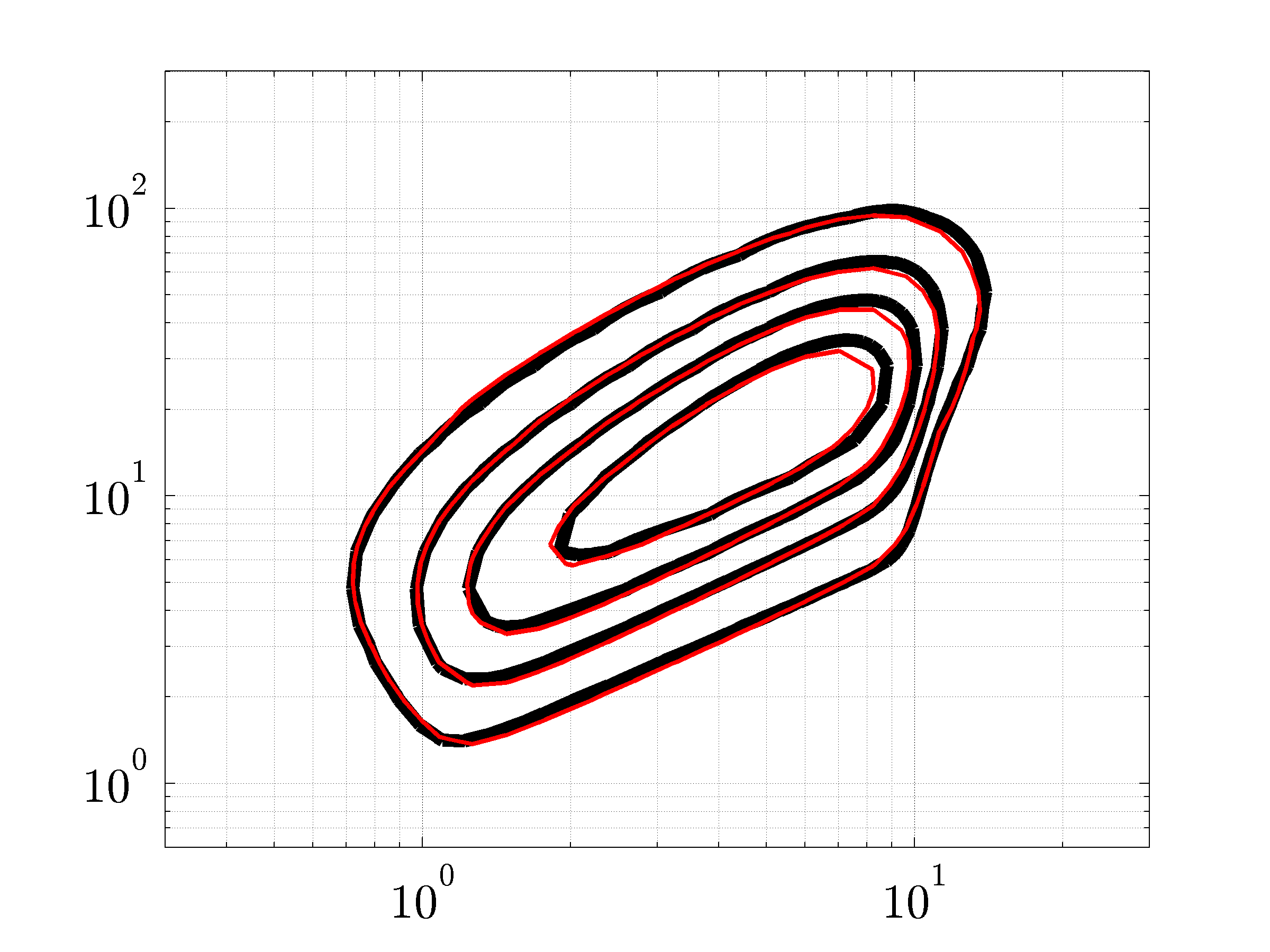}
    \label{fig.Euu-vs-yh-lzh-R10000-analytical}}
    \\[0.2cm]
    $(c)$
    &
    $(d)$
    \end{tabular}
    \begin{tabular}{c}
    \\[-4.4cm]
    \begin{tabular}{c}
    \hskip-6.5cm
    \begin{turn}{90}
    \tc{black}{$~~~\lambda_x$}
    \end{turn}
    \hskip6.2cm
    \begin{turn}{90}
    \tc{black}{$\sqrt{\lambda_z^+ \lambda_z}$}
    \end{turn}
    \end{tabular}
    \\[2.2cm]
    \begin{tabular}{c}
    \hskip-0.2cm
    \tc{black}{$\sqrt{y^+ y}$}
    \hskip5.6cm
    \tc{black}{$\sqrt{y^+ y}$}
    \end{tabular}
    \end{tabular}
    \\[-0.2cm]
    \begin{tabular}{cc}
    \subfigure{\includegraphics[width=0.47\columnwidth]
    {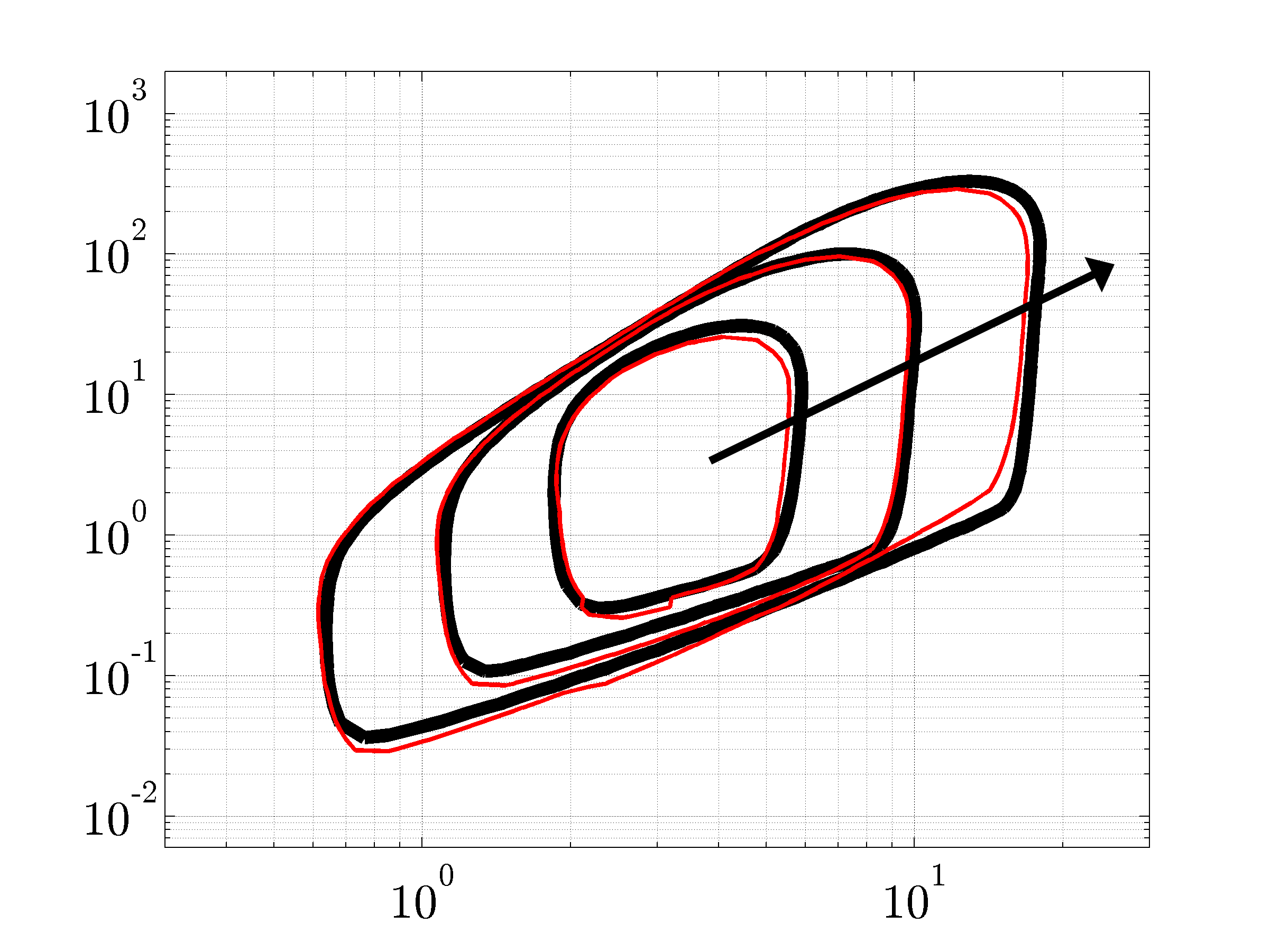}
    \label{fig.Euu-vs-yh-lx-R3333-1e4-3e4-analytical}}
    &
    \subfigure{\includegraphics[width=0.47\columnwidth]
    {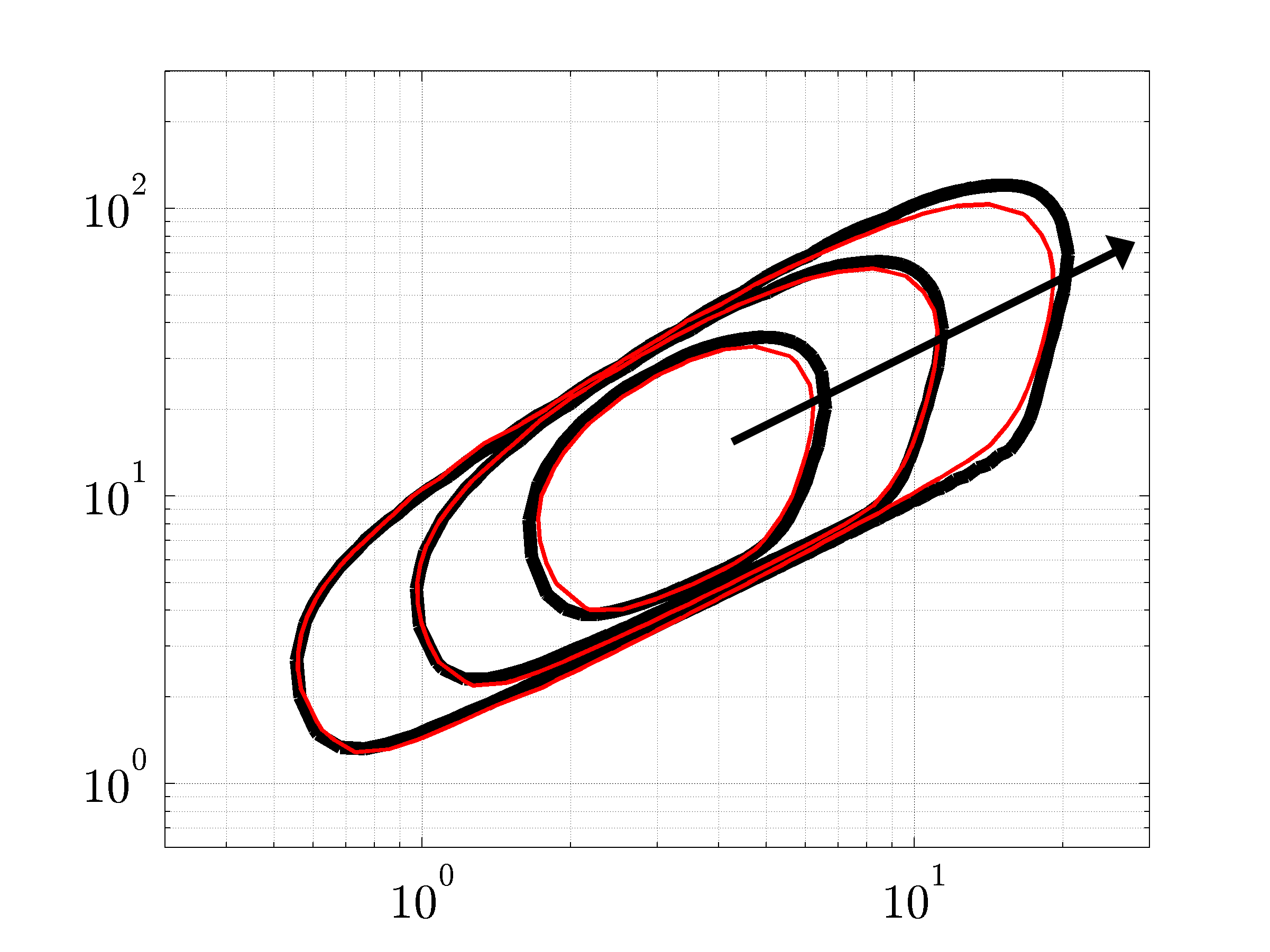}
    \label{fig.Euu-vs-yh-lzh-R3333-1e4-3e4-analytical}}
    \\[0.2cm]
    $(e)$
    &
    $(f)$
    \end{tabular}
    \begin{tabular}{c}
    \\[-4.4cm]
    \begin{tabular}{c}
    \hskip-6.5cm
    \begin{turn}{90}
    \tc{black}{$~~~\lambda_x$}
    \end{turn}
    \hskip6.2cm
    \begin{turn}{90}
    \tc{black}{$\sqrt{\lambda_z^+ \lambda_z}$}
    \end{turn}
    \end{tabular}
    \\[2.2cm]
    \begin{tabular}{c}
    \hskip-0.2cm
    \tc{black}{$\sqrt{y^+ y}$}
    \hskip5.6cm
    \tc{black}{$\sqrt{y^+ y}$}
    \end{tabular}
    \end{tabular}
    \end{center}
    \caption{Illustration showing that the largest modes in the self-similar hierarchies can be used to analytically compute the energy density of the modes in the logarithmic region at any $Re_\tau$. 
    (a, b) The normalized energy densities $Re_\tau^{-2} A_{uu} (y,\lambda_{x,u},c_u)$, (a), and $Re_\tau^{-2} A_{uu} (y,\lambda_{z,u},c_u)$, (b) for the rank-1 model and $Re_{\tau,u} = 10^4$. (c-f) The thin red contours are the analytically-computed normalized energy densities using the largest self-similar modes for $Re_{\tau,u} = 10^4$. The thick black contours are computed using the resolvent modes with $c_l \leq c \leq c_u$ for $Re_\tau = 10^4$ in (c,d) and $Re_\tau = 3333$, $10^4$, and $3\times10^4$ in the direction of the arrows in (e,f). The contour levels are $\{0.05,0.1,0.15\}$ in (c), $\{0.1, 0.2, 0.3, 0.4\}$ in (d), $0.1$ in (e), and $0.2$ in (f).
    \tc{white}{The thick black contours are computed using the resolvent modes with $c_l \leq c \leq c_u$ for $Re_\tau = 10^4$ in (c,d) and $Re_\tau = 3333$, $10^4$, and $3\times10^4$ in the direction of the arrows in (e,f). The contour levels are $\{0.05,0.1,0.15\}$ in (c), $\{0.1, 0.2, 0.3, 0.4\}$ in (d), $0.1$ in (e), and $0.2$ in (f)}
    }
    \label{fig.Euu-vs-yh-lx-lzh-analytical}
    \end{figure} 

Figures~\ref{fig.Euu-vs-yh-lx-R10000-exact} and~\ref{fig.Euu-vs-yh-lzh-R10000-exact} show the normalized energy densities for the largest modes in the self-similar hierarchies and $Re_{\tau,u} = 10^4$ as a function of the scaled distance from the wall. To eliminate the effect of the non-logarithmic parts of the mean velocity on the energy density of the largest modes, we have considered a logarithmic mean velocity throughout the channel. The energy densities are localized around $\lambda_x \approx \cO(1)$, $\lambda_z \approx \cO(0.1)$, and $y = y_u = 0.1$. The $\lambda_x\/$-axis is scaled with outer units and the $\lambda_z\/$- and $y\/$-axes are scaled with the geometric mean of the inner and outer bounds of the logarithmic region. They correspond to the observed scalings of the VLSM peak in the turbulent energy spectrum~\citep{mathutmar09}.

It follows from~(\ref{eq.Euu-tilde-ij-map-cu}) that the energy densities of the self-similar modes with speeds $c_{l'} \leq c \leq c_u$ can be analytically obtained by shifting the energy densities in figures~\ref{fig.Euu-vs-yh-lx-R10000-exact} and~\ref{fig.Euu-vs-yh-lzh-R10000-exact} by $y_c/y_u$ in the wall-normal and spanwise directions and by $y_c^+ y_c/ y_u^+ y_u$ in the streamwise direction. Notice that this eliminates the need for direct computation of the resolvent modes with speeds $c_{l'} \leq c \leq c_u$ even for flows with different Reynolds numbers. The thin red contours in figures~\ref{fig.Euu-vs-yh-lx-R10000-analytical} and~\ref{fig.Euu-vs-yh-lzh-R10000-analytical} show the analytically computed and normalized energy densities $A_{uu} (y,\lambda_x)$ and $A_{uu} (y,\lambda_z)$ for the modes with $c_{l'} \leq c \leq c_u$ and $Re_\tau = 10^4$. These results are in close agreement with the energy densities computed using the resolvent modes with $c_l \leq c \leq c_u$ at the same $Re_\tau$ (thick black contours). Therefore, the modes that violate the aspect-ratio constraint have negligible contributions to the energy density relative to the modes that satisfy the constraint. This is in agreement with figure~\ref{fig.EuuDNs_vs_lx_lz_y0p075_R2003_kxkzsigma1u1u1}.

To show the favorable scaling of the present computational method with the Reynolds number, figures~\ref{fig.Euu-vs-yh-lx-R3333-1e4-3e4-analytical} and~\ref{fig.Euu-vs-yh-lzh-R3333-1e4-3e4-analytical} compare the analytically computed energy densities with the energy densities obtained using the resolvent modes for $Re_\tau = 3333$, $10^4$, and $3\times10^4$ in the direction of the arrows. Notice that the analytical results for these Reynolds numbers are obtained using the largest self-similar modes for $Re_{\tau,u} = 10^4$. The close agreement between these contours illustrates that the largest modes in the self-similar hierarchies can be used to analytically compute the energy density of the modes in the logarithmic region at any $Re_\tau$. The small discrepancies between the analytical and numerical energy densities for $Re_\tau = 3333$ and $3\times10^4$ are due to edge effects discussed in Appendix~\ref{sec.edge}. 

Since the logarithmic region expands in physical space as $Re_\tau$ increases, the energy density of the self-similar modes covers a larger range of wavelengths. In addition, notice that the peak of the rank-1 streamwise energy density takes place close to the location of the VLSM peak for all $Re_\tau$. We highlight that the analytical computations are significantly cheaper since they do not require the resolvent modes for all $\blambda$ and $c$ in the logarithmic region. For $Re_\tau = 3\times10^4$, the computational cost is reduced by four orders of magnitude, one minute vs. $1000$ hours, on a $2.4$ GHz laptop. This enhances the saving that is achieved by a low-order representation of the flow relative to the numerical resolution used in DNS, see~\cite{moajovtroshamckPOF14}. Further cost reduction can be expected for larger $Re_\tau$ since the range of admissible speeds for the self-similar modes becomes larger.

\begin{figure}
    \begin{center}
    \begin{tabular}{cc}
    \subfigure{\includegraphics[width=0.47\columnwidth]
    {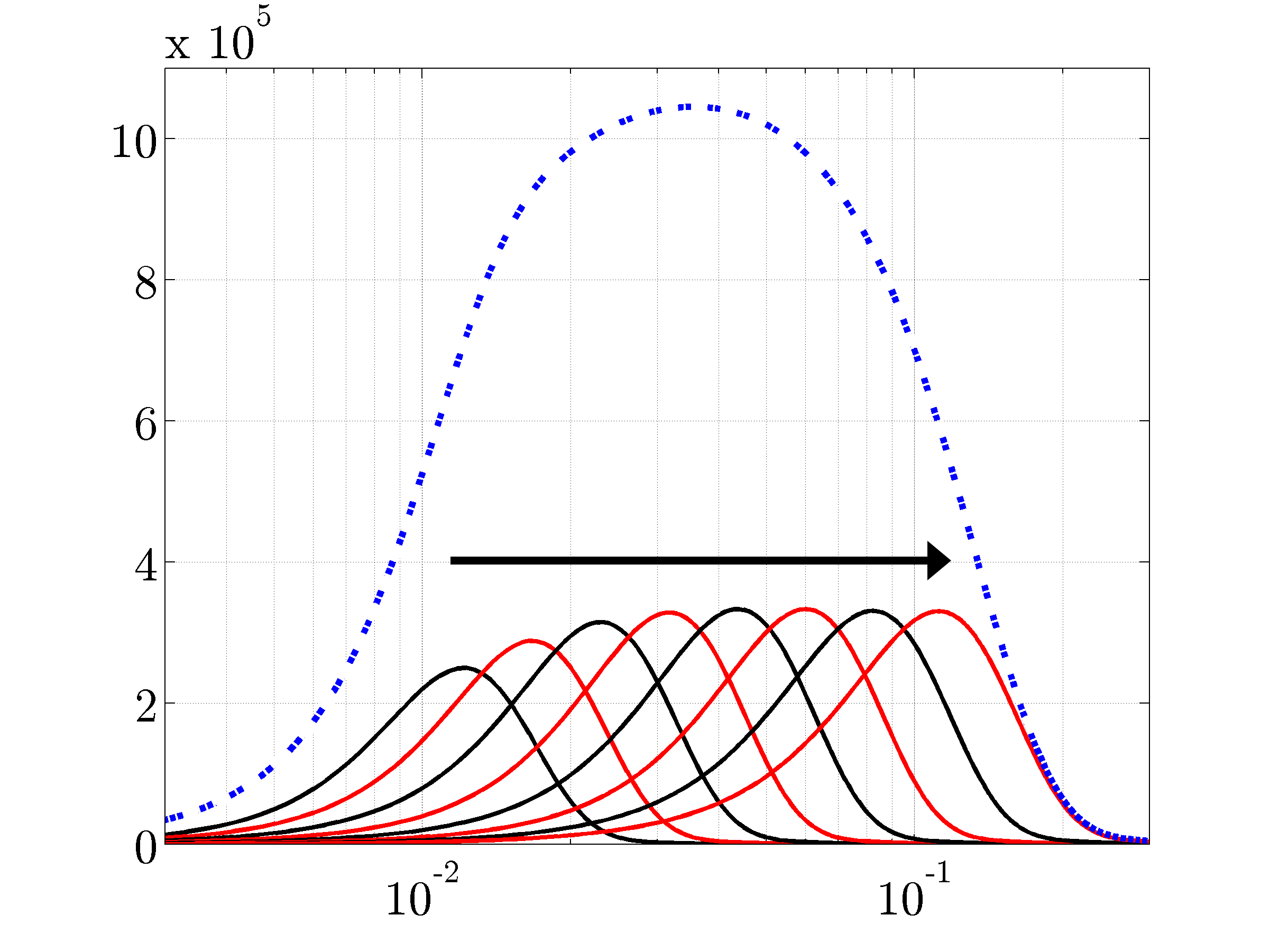}
    \label{fig.Euu-vs-y-hierarchy}}
    &
    \subfigure{\includegraphics[width=0.47\columnwidth]
    {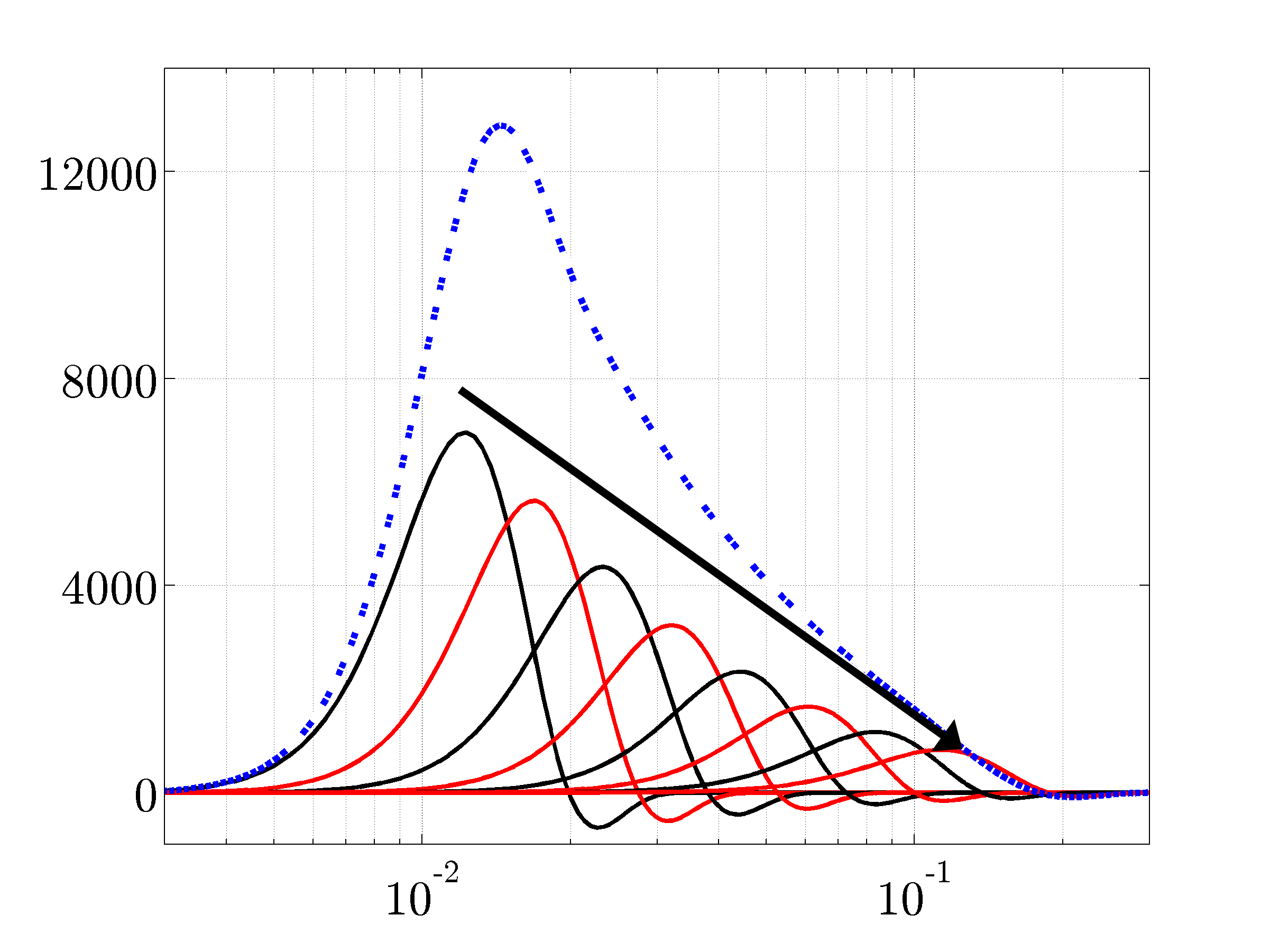}
    \label{fig.Euv-vs-y-hierarchy}}
    \\[0.2cm]
    $(a)$
    &
    $(b)$
    \end{tabular}
    \begin{tabular}{c}
    \\[-4.6cm]
    \begin{tabular}{c}
    \hskip-6.4cm
    \begin{turn}{90}
    \tc{black}{$~~~~A_{uu}$}
    \end{turn}
    \hskip6.15cm
    \begin{turn}{90}
    \tc{black}{$-\mbox{Re} (A_{uv})$}
    \end{turn}
    \end{tabular}
    \\[2.1cm]
    \begin{tabular}{c}
    \tc{black}{$y$}
    \hskip6.35cm
    \tc{black}{$y$}
    \end{tabular}
    \end{tabular}
    \end{center}
    \caption{The solid curves show the energy densities $A_{uu} (y, \blambda, c)$, (a), and $\mbox{Re} \big(A_{uv} (y, \blambda, c) \big)$, (b), for the rank-1 model and the modes in the hierarchy that passes through the representative VLSM mode with $\lambda_x = 6$, $\lambda_z = 0.6$, and $c = (2/3) U_c$ at $Re_\tau = 10^4$. The arrows show the direction of increasing $c$ in the hierarchy. The dotted curves show the sum of the energy densities.
    }
    \label{fig.Euu-Euv-vs-y-hierarchy}
    \end{figure}

Following~(\ref{eq.Euu-hierarchy-1-x}) and~(\ref{eq.Euu-rank1}), the contribution of a given hierarchy of modes $\cS (\blambda_u)$ to the streamwise energy intensity and the Reynolds stress for the rank-1 model is determined by integrating the contribution of all the modes in the hierarchy
	\be
	\ba{rcl}
	E_{uu,h} (y,\blambda_u)
	&\!\! = \!\!&
	\ds{
	\int_{\cS (\blambda_u)}
	}
	\,
	|\chi_1 (\blambda, c)|^2 
	\,
	A_{uu} (y, \blambda, c)
	\,
	\mrd \cS (\blambda_u),
	\\[0.15cm]
	E_{uv,h} (y,\blambda_u)
	&\!\! = \!\!&
	\ds{
	\int_{\cS (\blambda_u)}
	}
	\,
	|\chi_1 (\blambda, c)|^2 
	\,
	\mbox{Re} 
	\big(
	A_{uv} (y, \blambda, c)
	\big)
	\,
	\mrd \cS (\blambda_u).
	\ea
	\label{eq.Euu-Euv-rank1}
	\ee	
The solid curves in figure~\ref{fig.Euu-Euv-vs-y-hierarchy} show the energy densities $A_{uu} (y, \blambda, c)$ and $\mbox{Re} \big(A_{uv} (y, \blambda, c) \big)$ for the rank-1 model and the modes in the hierarchy that passes through the representative VLSM mode with $\lambda_x = 6$, $\lambda_z = 0.6$, and $c = (2/3) U_c$ at $Re_\tau = 10^4$. The arrows show the direction of increasing $c$ in the hierarchy. The contribution of each mode to the streamwise energy intensity and the Reynolds stress is localized around the critical layer $y_c$ and decays to zero away from it. This agrees with the notion of active and inactive motions introduced by~\cite{tow76}. In addition, we see that the streamwise energy density of the modes remains constant while the Reynolds stress density decreases with $y_c$ as the modes become larger, see also~(\ref{eq.hierarchy-analytical}) and~(\ref{eq.Euv-hierarchy-analytical-x}). 

Several theoretical arguments~\citep{tow76,percho82} and experimental studies~\citep{nicmarhafhutcho07,marmonhulsmi13} have shown that $E_{uu} (y)$ is a logarithmically decaying function of $y$ and $E_{uv} (y)$ is approximately constant in the logarithmic region. Following~(\ref{eq.Euu-Euv-rank1}), the contribution of a given hierarchy to the streamwise energy intensity and the Reynolds stress is obtained by a $|\chi_1|^2\/$-weighted integral of the energy densities. The dotted curves in figure~\ref{fig.Euu-Euv-vs-y-hierarchy} show that integrating the energy densities with unit weights (i.e. broadband forcing) yields an approximately constant $E_{uu} (y)$ and a decaying $E_{uv} (y)$ in the logarithmic region. In order to obtain a constant $E_{uv}$, per experimental observations, the weights should increase with $c$ in the hierarchy. However, this yields an increasing function $E_{uu}$. Therefore, the rank-1 model cannot simultaneously capture the experimentally observed $E_{uu} (y)$ and $E_{uv} (y)$ in the logarithmic region even if a non-broadband forcing is used. 

The above discussion highlights the importance of higher-order resolvent modes for representing the turbulent spectra.~\cite{moajovtroshamckPOF14} used convex optimization to show that a $12\/$th-order resolvent model per wall-parallel wavenumber pair and mode speed can be used to approximate the turbulent spectra obtained from DNS at $Re_\tau = 2003$. The spectra were captured with $22 \%$ and $62 \%$ deviation error, respectively, for the inner-scaled peak ($\lambda_x^+ = 700$, $\lambda_z^+ = 100$) and the outer-scaled peak ($\kappa_x = 0.6$, $\kappa_z = 6$) of the streamwise spectrum. It was also shown that including higher-order modes yields diminishing improvements. 

The developments of~\S~\ref{sec.log-region} can be used to study a resolvent model with arbitrary order.  The compatibility of higher-order models with experimental observations is discussed in~\S~\ref{sec.exp-scaling}. One of the main challenges that remains to be addressed concerns the weight matrix that, as discussed in~\S~\ref{sec.resolvent}, represents the scaling influence of the nonlinear interaction of the resolvent modes. A foundation for studying the scaling of the weight matrix is provided in~\S~\ref{sec.scaling-weights}.
	
\section{On scaling of the weight matrix}
\label{sec.scaling-weights}

The nonlinear interactions in wall turbulence redistribute the turbulent kinetic energy across different scales and different wall-normal locations. The modes that directly interact through the quadratic nonlinearity in the NSE are triadically consistent meaning that their streamwise wavenumbers, their spanwise wavenumbers, and their temporal frequencies sum to zero. For homogeneous isotropic turbulence, a tractable method for manipulating the triadic interactions was proposed by~\cite{chezak14}. This was done by formulating the nonlinear terms in a canonical form and representing the nonlinear operators in terms of the energy density instead of a convolution operator in the Fourier space. For wall-bounded turbulent flows,~\cite{mckshajac13} showed that the resolvent formulation allows for direct study of the triadic interactions. In this section, we study the scaling of the weight matrix using the nonlinear interaction of the resolvent modes.  

It follows from~(\ref{eq.f-phi}) that the weight $\chi_j (\blambda, c)$ is obtained by projecting the forcing $\hat{\fvec} (y, \blambda, c)$ onto the forcing mode $\hat{\bphi}_j (y, \blambda, c)$. Since $\fvec = -\bu \cdot \nabla \bu = -\nabla \cdot (\bu \bu^T)$, the Fourier-transformed forcing is given by the gradient of the convolution of the triadically-consistent modes,
	\be
	\ba{l}
	\overline{\hat{\fvec}} (y, \blambda, c)
	\; = \;
	-
	\nabla \cdot 
	\ds{
	\iint
	}
	\,
	(\dfrac{2\pi}{\lambda_x'})^2 \dfrac{2\pi}{|\lambda_z'|}
	\,
	\hat{\bu} (y, \blambda', c')
	\,
	\hat{\bu}^* (y, \blambda'', c'')
	\,
	\mrd \ln \blambda' \,
	\mrd c'
	,
	\ea
	\label{eq.f-convolution}
	\ee
where 
	\be
	\lambda_x'' \, = \, 	\dfrac{\lambda_x \lambda_x'}{\lambda_x' + \lambda_x},~~
	\lambda_z'' \, = \, \dfrac{\lambda_z \lambda_z'}{\lambda_z' + \lambda_z},~~
	c'' \, = \, \dfrac{c \lambda_x' + c' \lambda_x}{\lambda_x' + \lambda_x},
	\label{eq.mode-pp} 
	\ee
are defined for notational simplicity, and the symmetry relationships $\hat{\fvec} (y, -\blambda, c) = \overline{\hat{\fvec}} (y, \blambda, c)$ and $\hat{\bu} (y, -\blambda, c) = \overline{\hat{\bu}} (y, \blambda, c)$ are used. The mode speeds are confined to the interval $0 < c < U_{cl}$.

\subsection{The interaction coefficient}
\label{sec.coupling}

Following~\cite{mckshajac13}, an explicit equation for the weights can be obtained by substituting~(\ref{eq.f-convolution}) in~(\ref{eq.f-phi}) and using the symmetry relationship $\chi_i (-\blambda, c) = \overline{\chi}_i (\blambda, c)$,
	\be
	\ba{l}
	\overline{\chi}_l (\blambda, c)
	\; = \;
	\ds{
	\sum_{i,j = 1}^{N}
	\;
	\iint
	}
	\,
	\cN_{lij} (\blambda,c,\blambda',c') \,
	\chi_i(\blambda', c') \,
	\overline{\chi}_j(\blambda'', c'') \,
	\;
	\mrd \ln \blambda' \,
	\mrd c'.
	\ea
	\label{eq.f-convolution-expand}
	\ee
Here, $\cN_{lij} (\blambda,c,\blambda',c')$ denotes the interaction coefficient between the resolvent modes and does not depend on the resolvent weights, see Appendix~\ref{sec.N-details} for details.

The interaction coefficient identifies the resolvent modes that have the largest interaction with each other. In addition, causality arguments can be made based on the interaction coefficient since $\cN_{lij} (\blambda,c,\blambda',c')$ represents the coupling of the resolvent mode $\hat{\bpsi}_i(\blambda',c')$ with $\hat{\bpsi}_j(-\blambda'',c'')$ to nonlinearly force the resolvent mode $\hat{\bpsi}_l(-\blambda,c)$. We highlight that $\cN_{lij}$ quantifies the interaction coefficient per unit weights $\chi_i$ and $\chi_j$. In addition, notice that the expression for $\cN_{lij}$ is not symmetric with respect to swapping $i$ and $j$. Therefore, the coupling of $\hat{\bpsi}_i(\blambda',c')$ with $\hat{\bpsi}_j(-\blambda'',c'')$ to force $\hat{\bpsi}_l(-\blambda,c)$ is not necessarily equal to the coupling of $\hat{\bpsi}_j(-\blambda'',c'')$ with $\hat{\bpsi}_i(\blambda',c')$ to force $\hat{\bpsi}_l(-\blambda,c)$, i.e. $\cN_{lij} (\blambda,c,\blambda',c') \neq \cN_{lji} (\blambda,c,-\blambda'',c'')$. The total coupling of $\hat{\bpsi}_i(\blambda',c')$ and $\hat{\bpsi}_j(-\blambda'',c'')$ to force $\hat{\bpsi}_l(-\blambda,c)$ is defined as
	\be
	\cN_{lij}^t (\blambda,c,\blambda',c')
	\; = \;
	\cN_{lij} (\blambda,c,\blambda',c')
	\, + \,
	\cN_{lji} (\blambda,c,-\blambda'',c'').
	\non
	\ee
	
    \begin{figure}
    \begin{center}
    \begin{tabular}{cc}
    \subfigure{\includegraphics[width=0.4\columnwidth]
    {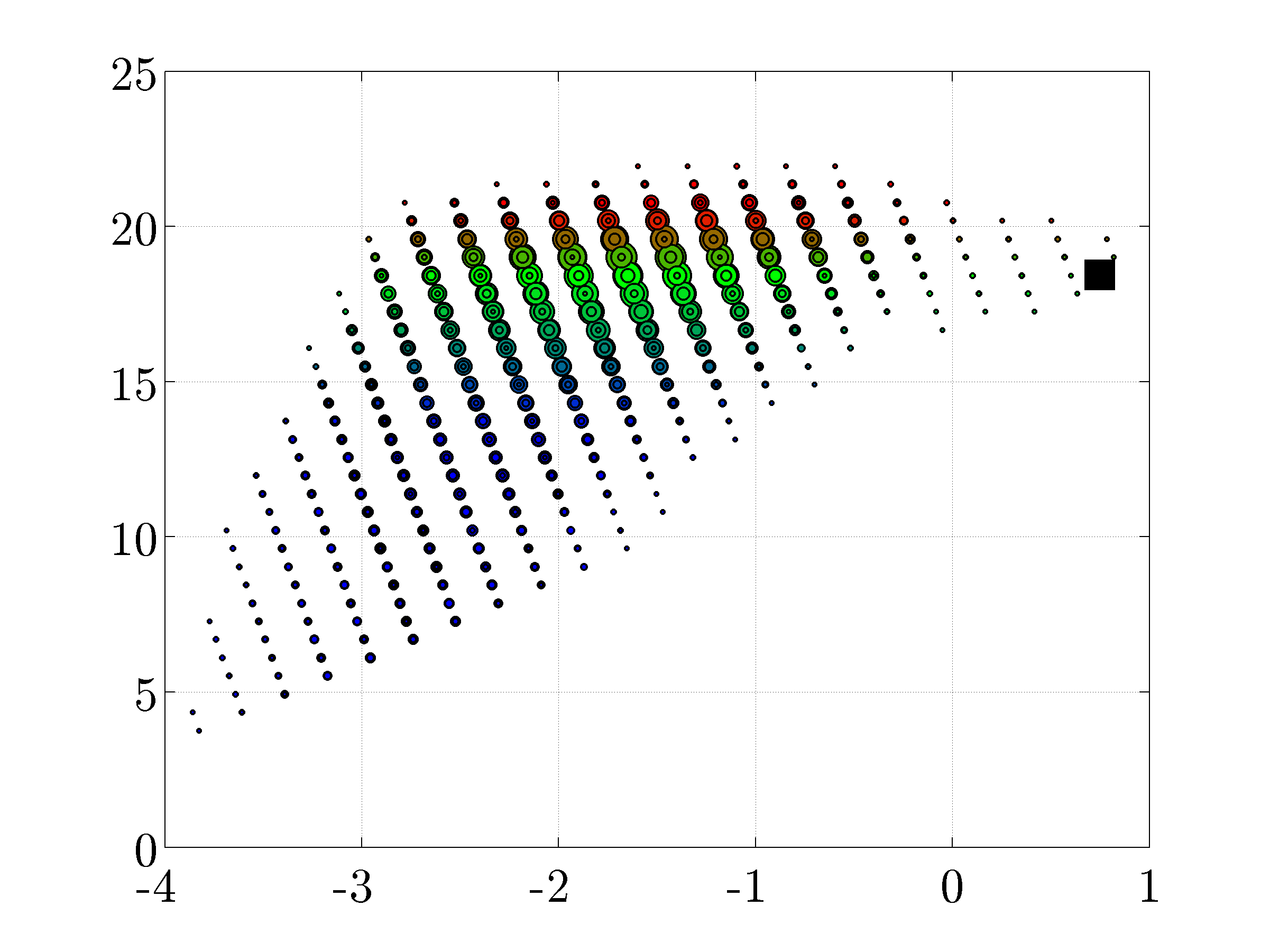}
    \label{fig.absN-vs-cp-vs-lxp-lx5p7-lz0p6-c23rdUc-R1e4-Q1}}
    &
    \subfigure{\includegraphics[width=0.4\columnwidth]
    {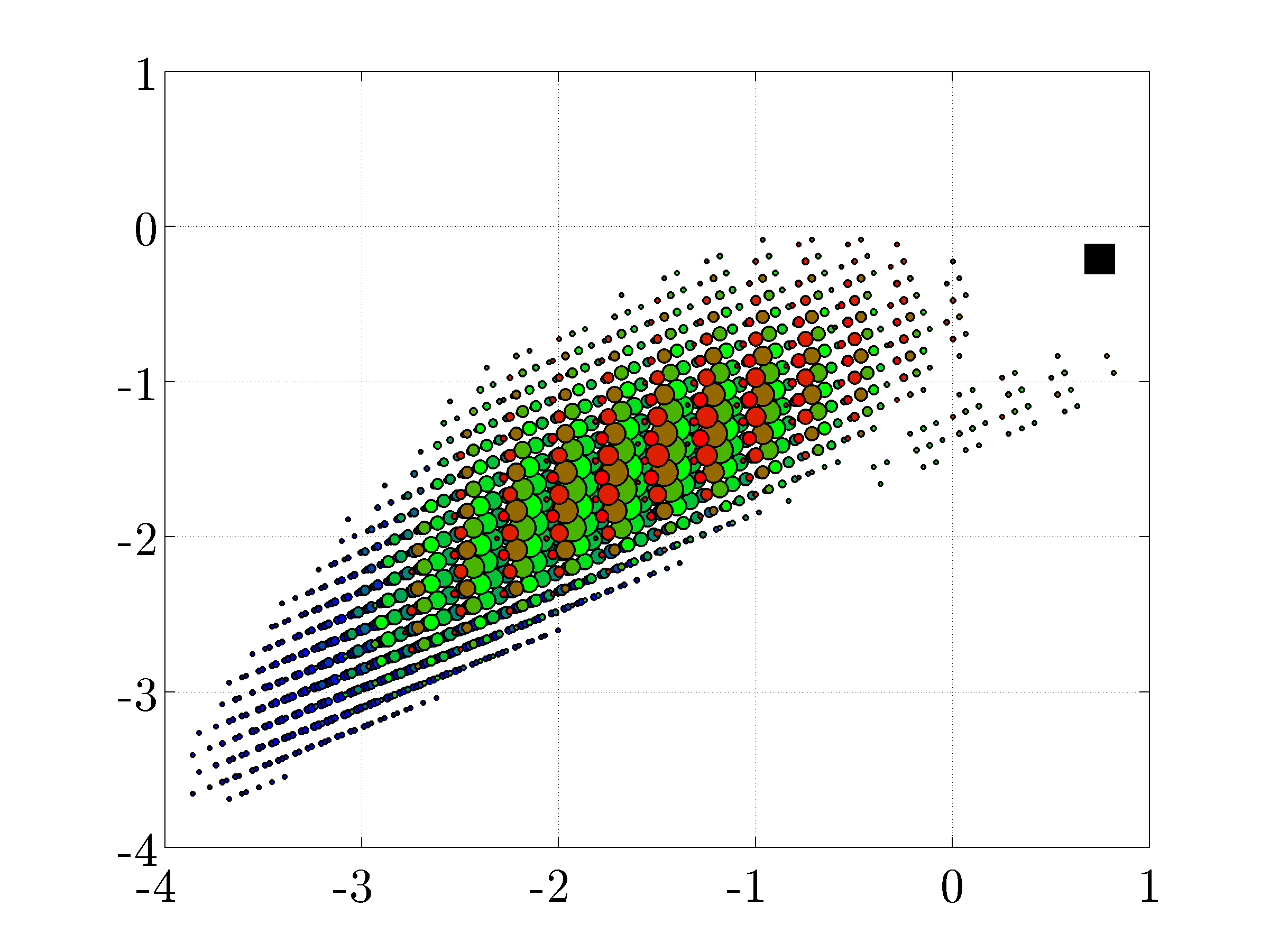}
    \label{fig.absN-vs-lxp-vs-lzp-lx5p7-lz0p6-c23rdUc-R1e4-Q1}}
    \\[0.2cm]
    $(a)$
    &
    $(b)$
    \end{tabular}
    \begin{tabular}{c}
    \\[-3.8cm]
    \begin{tabular}{c}
    \hskip-5.4cm
    \begin{turn}{90}
    \tc{black}{$~~~c'$}
    \end{turn}
    \hskip5.3cm
    \begin{turn}{90}
    \tc{black}{$\log |\lambda_z'|$}
    \end{turn}
    \end{tabular}
    \\[1.7cm]
    \begin{tabular}{c}
    \hskip0cm
    \tc{black}{$\log |\lambda_x'|$}
    \hskip4.7cm
    \tc{black}{$\log |\lambda_x'|$}
    \end{tabular}
    \end{tabular}
    \\[-0.2cm]
    \begin{tabular}{cc}
    \subfigure{\includegraphics[width=0.4\columnwidth]
    {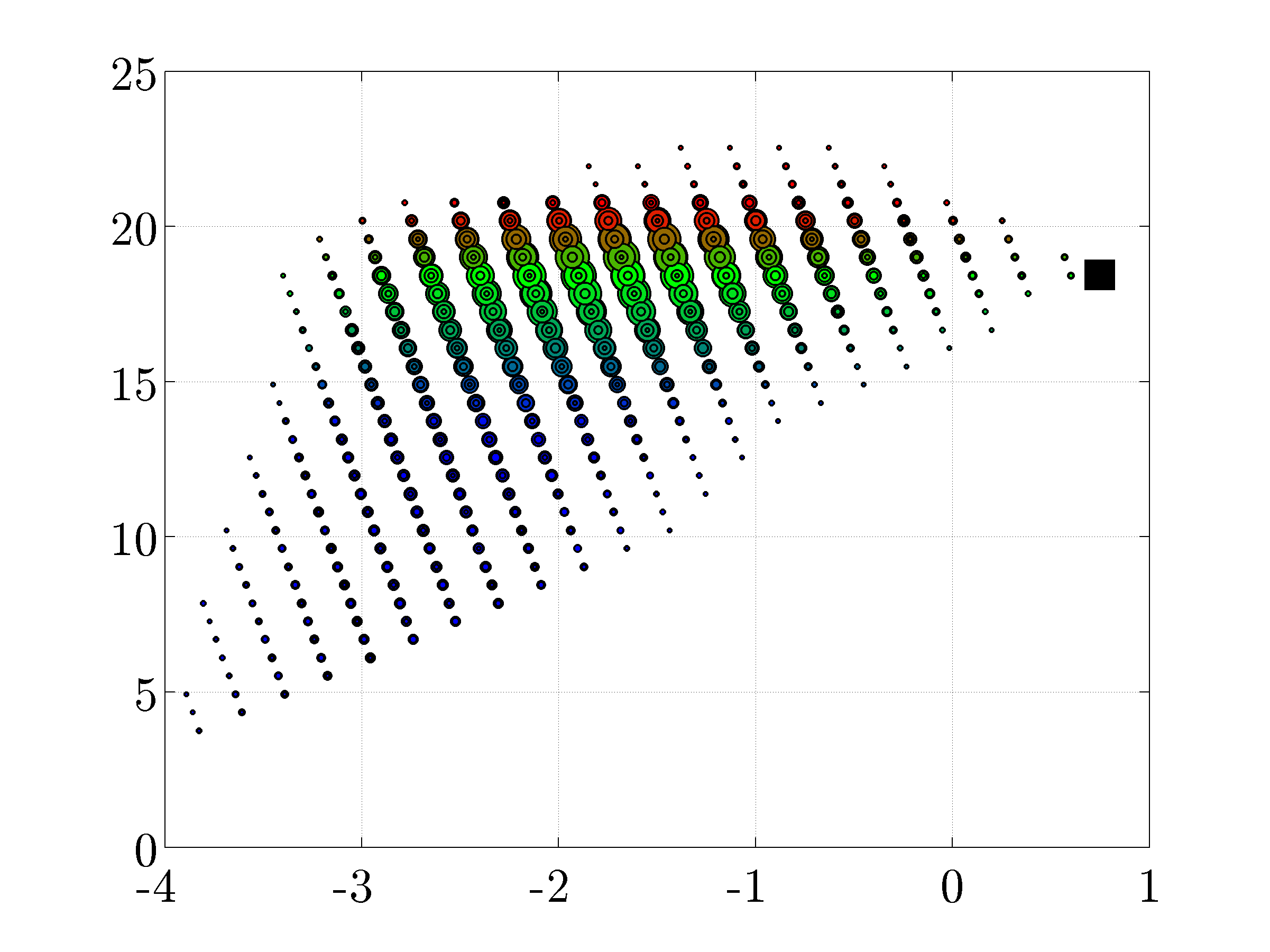}
    \label{fig.absN-vs-cp-vs-lxp-lx5p7-lz0p6-c23rdUc-R1e4-Q2}}
    &
    \subfigure{\includegraphics[width=0.4\columnwidth]
    {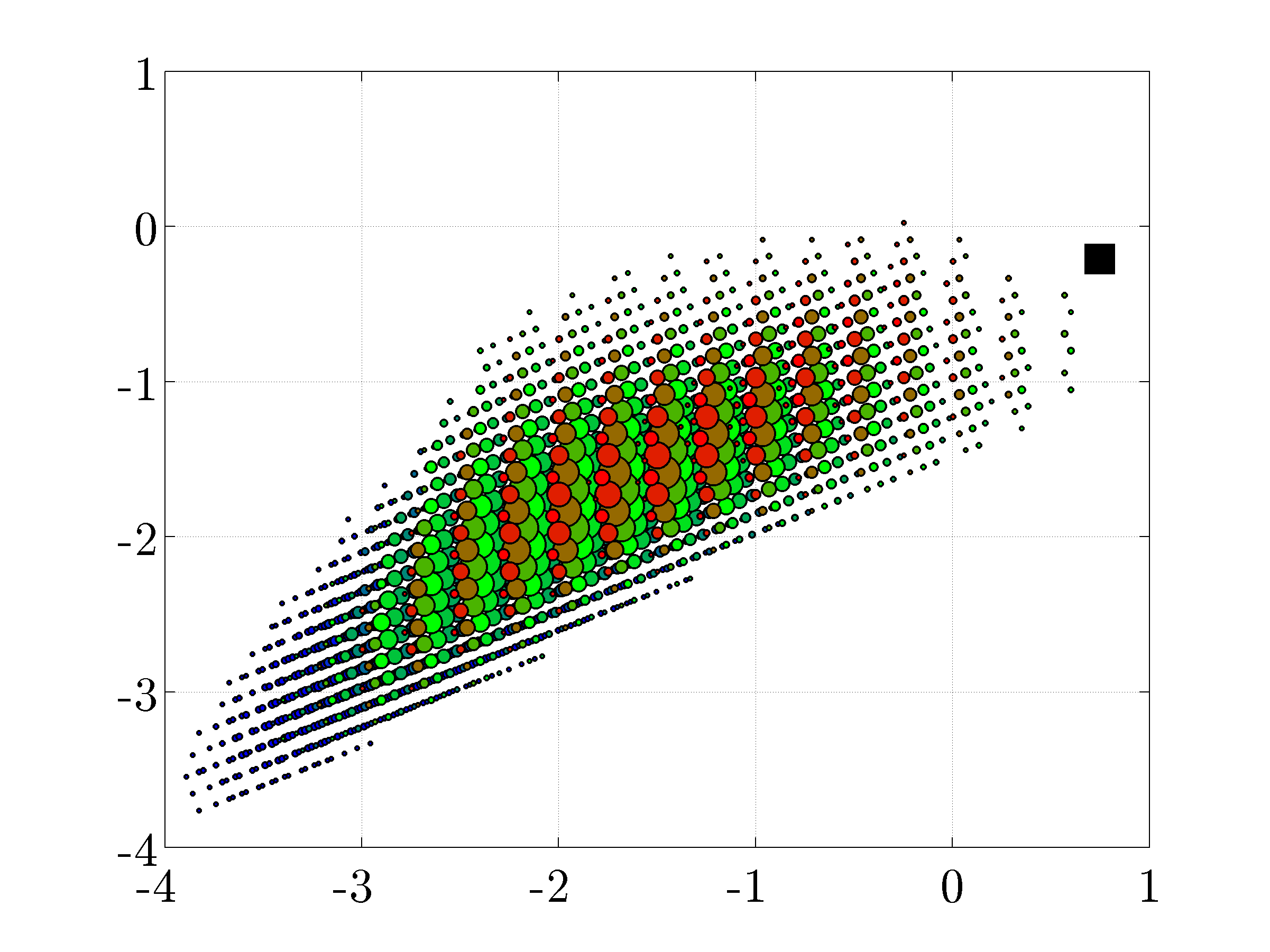}
    \label{fig.absN-vs-lxp-vs-lzp-lx5p7-lz0p6-c23rdUc-R1e4-Q2}}
    \\[0.2cm]
    $(c)$
    &
    $(d)$
    \end{tabular}
    \begin{tabular}{c}
    \\[-3.8cm]
    \begin{tabular}{c}
    \hskip-5.4cm
    \begin{turn}{90}
    \tc{black}{$~~~c'$}
    \end{turn}
    \hskip5.3cm
    \begin{turn}{90}
    \tc{black}{$\log |\lambda_z'|$}
    \end{turn}
    \end{tabular}
    \\[1.7cm]
    \begin{tabular}{c}
    \hskip0cm
    \tc{black}{$\log |\lambda_x'|$}
    \hskip4.7cm
    \tc{black}{$\log |\lambda_x'|$}
    \end{tabular}
    \end{tabular}
    \\[-0.2cm]
    \begin{tabular}{cc}
    \subfigure{\includegraphics[width=0.4\columnwidth]
    {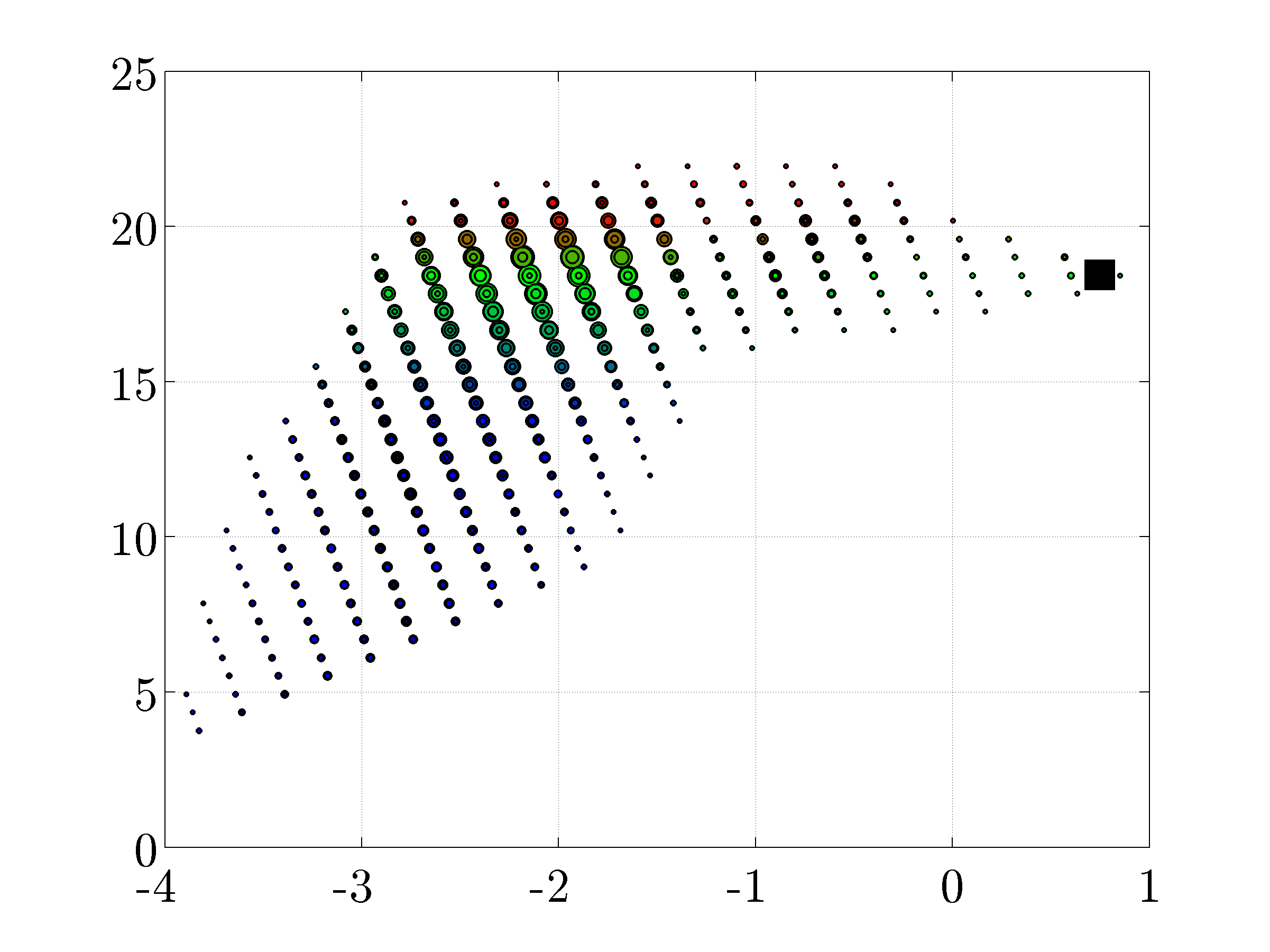}
    \label{fig.absN-vs-cp-vs-lxp-lx5p7-lz0p6-c23rdUc-R1e4-Q3}}
    &
    \subfigure{\includegraphics[width=0.4\columnwidth]
    {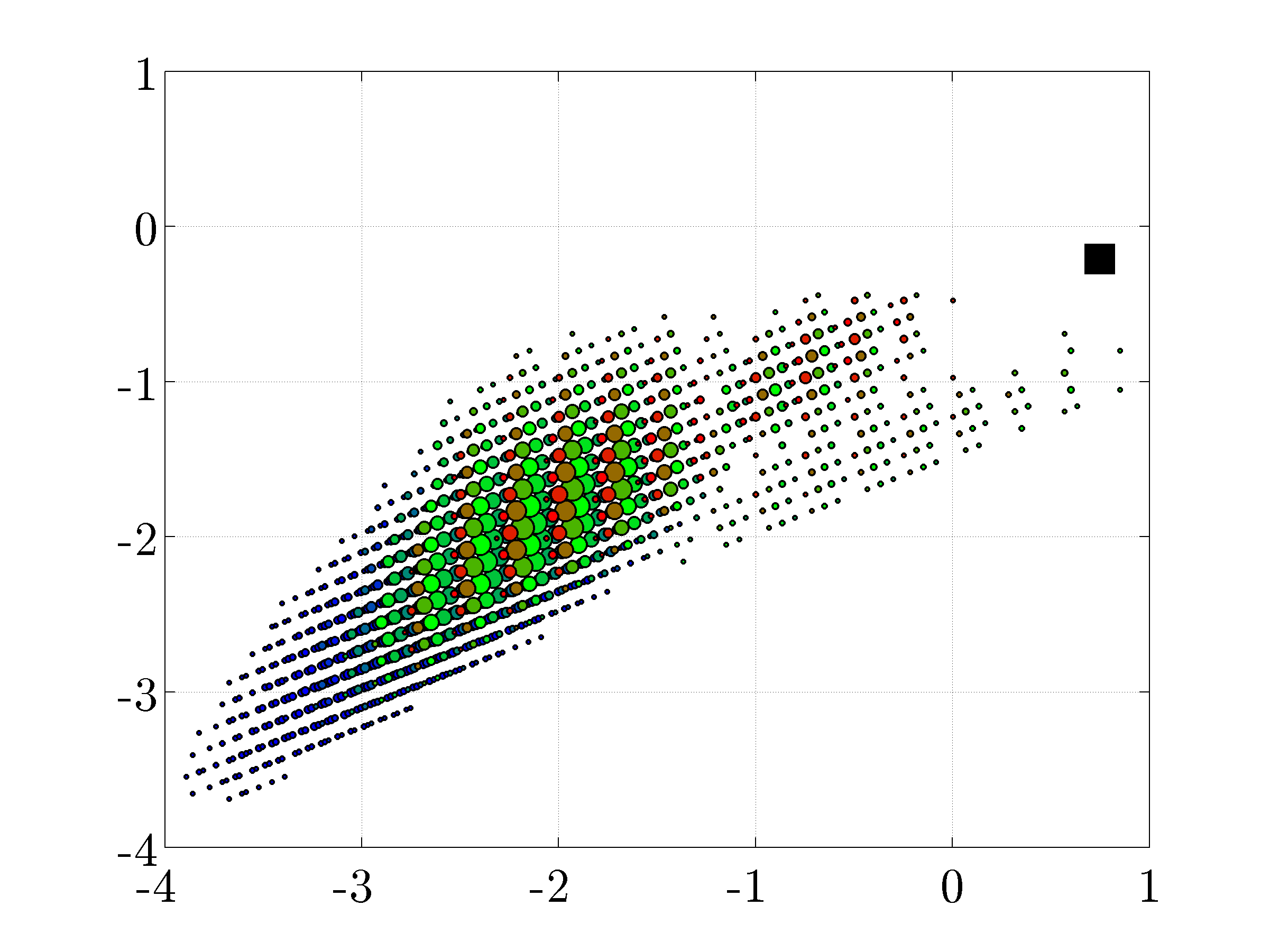}
    \label{fig.absN-vs-lxp-vs-lzp-lx5p7-lz0p6-c23rdUc-R1e4-Q3}}
    \\[0.2cm]
    $(e)$
    &
    $(f)$
    \end{tabular}
    \begin{tabular}{c}
    \\[-3.8cm]
    \begin{tabular}{c}
    \hskip-5.4cm
    \begin{turn}{90}
    \tc{black}{$~~~c'$}
    \end{turn}
    \hskip5.3cm
    \begin{turn}{90}
    \tc{black}{$\log |\lambda_z'|$}
    \end{turn}
    \end{tabular}
    \\[1.7cm]
    \begin{tabular}{c}
    \hskip0cm
    \tc{black}{$\log |\lambda_x'|$}
    \hskip4.7cm
    \tc{black}{$\log |\lambda_x'|$}
    \end{tabular}
    \end{tabular}
    \\[-0.2cm]
    \begin{tabular}{cc}
    \subfigure{\includegraphics[width=0.4\columnwidth]
    {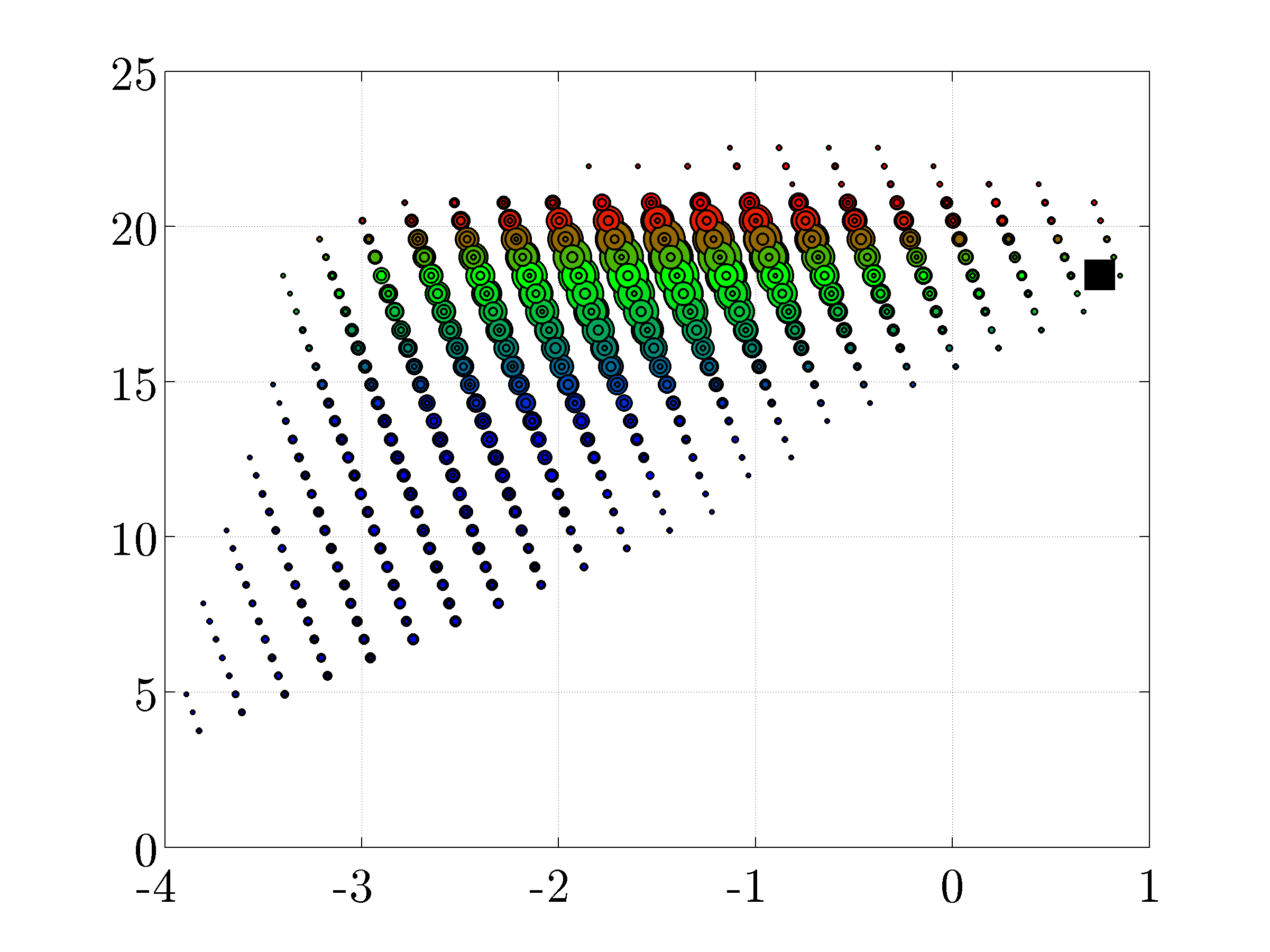}
    \label{fig.absN-vs-cp-vs-lxp-lx5p7-lz0p6-c23rdUc-R1e4-Q4}}
    &
    \subfigure{\includegraphics[width=0.4\columnwidth]
    {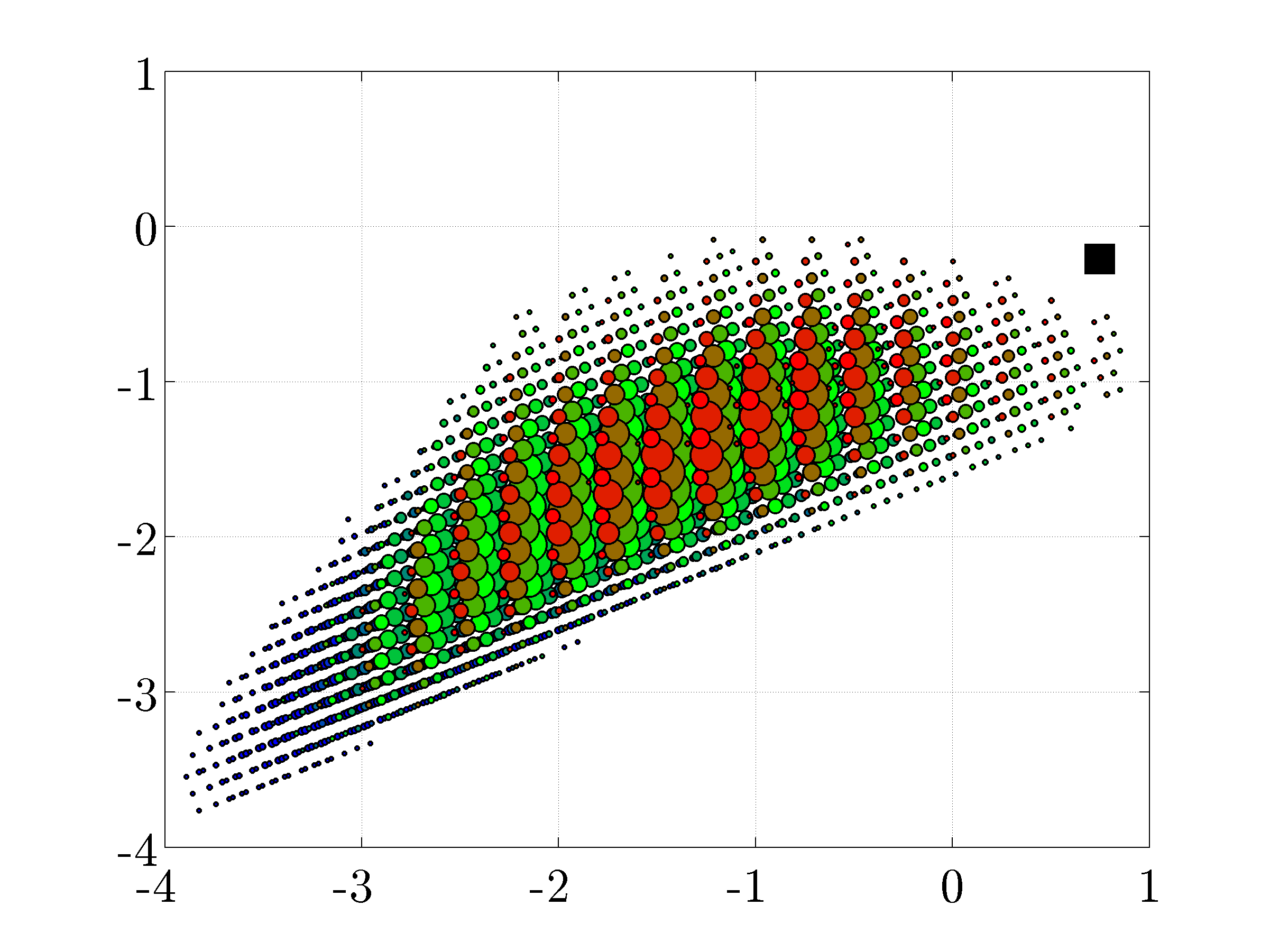}
    \label{fig.absN-vs-lxp-vs-lzp-lx5p7-lz0p6-c23rdUc-R1e4-Q4}}
    \\[0.2cm]
    $(g)$
    &
    $(h)$
    \end{tabular}
    \begin{tabular}{c}
    \\[-3.8cm]
    \begin{tabular}{c}
    \hskip-5.4cm
    \begin{turn}{90}
    \tc{black}{$~~~c'$}
    \end{turn}
    \hskip5.3cm
    \begin{turn}{90}
    \tc{black}{$\log |\lambda_z'|$}
    \end{turn}
    \end{tabular}
    \\[1.7cm]
    \begin{tabular}{c}
    \hskip0cm
    \tc{black}{$\log |\lambda_x'|$}
    \hskip4.7cm
    \tc{black}{$\log |\lambda_x'|$}
    \end{tabular}
    \end{tabular}
    \end{center}
    \caption{The size of the circles is proportional to the absolute value of the interaction coefficient $|\cN_{111} (\blambda,c,\blambda',c')|$ for $\lambda_x = 5.7$, $\lambda_z = 0.6$, and $c = 18.4$ (black square) for $Re_\tau = 10^4$. The largest and smallest circles correspond to $|\cN_{111}| = 8.1\times10^6$ and $8.1\times10^4$, respectively. The colors indicate $c'$: blue for $c' < c$, green for $c' \approx c$, and red for $c' > c$. (a, b), $\lambda_x' > 0$, $\lambda_z' > 0$; (c, d), $\lambda_x' < 0$, $\lambda_z' > 0$; (e, f), $\lambda_x' < 0$, $\lambda_z' < 0$; (g, h), $\lambda_x' > 0$, $\lambda_z' < 0$.}
    \label{fig.absN-vs-cp-vs-lxp-lx5p7-lz0p6-c23rdUc-R1e4}
    \end{figure}  
    
Figure~\ref{fig.absN-vs-cp-vs-lxp-lx5p7-lz0p6-c23rdUc-R1e4} shows the interaction coefficient $\cN_{lij} (\blambda,c,\blambda',c')$ for all the modes that force a representative VLSM mode, i.e. $\lambda_x = 5.7$, $\lambda_z = 0.6$, and $c = 18.4$ (black square) for $Re_\tau = 10^4$. The size of the circles is proportional to the absolute value of $\cN$ for the principal resolvent modes ($l = i = j = 1$). The largest and smallest circles correspond to $|\cN_{111}| = 8.1\times10^6$ and $8.1\times10^4$, respectively. The colors indicate $c'$, i.e. blue for $c' < c$, green for $c' \approx c$, and red for $c' > c$. The different rows in figure~\ref{fig.absN-vs-cp-vs-lxp-lx5p7-lz0p6-c23rdUc-R1e4} correspond to different signs of $\lambda_x'$ and $\lambda_z'$, e.g. $\lambda_x', \lambda_z' > 0$ in figures~\ref{fig.absN-vs-cp-vs-lxp-lx5p7-lz0p6-c23rdUc-R1e4-Q1} and~\ref{fig.absN-vs-lxp-vs-lzp-lx5p7-lz0p6-c23rdUc-R1e4-Q1}. It is evident from the figures on the left that the largest interaction takes place for the modes with similar speeds, i.e. $c' \approx c$. This highlights the role of critical layers and is intuitive since the modes with similar wall-normal localization are expected to interact more with each other. In addition, the modes that do not overlap in the physical space cannot interact. Figure~\ref{fig.absN-vs-cp-vs-lxp-lx5p7-lz0p6-c23rdUc-R1e4} also shows that the interaction is strongest where $|\blambda'|/|\blambda| \approx 10^{-2}-10^{-3}$. For this range of $\blambda'$, it follows from~(\ref{eq.mode-pp}) that $\blambda'' \approx \blambda'$ and $c'' \approx c'$. Therefore, we see that two relatively short modes contribute the most to the long mode. The strongest interaction takes place for positive $\lambda_x'$ and negative $\lambda_z'$, cf.~figure~\ref{fig.absN-vs-lxp-vs-lzp-lx5p7-lz0p6-c23rdUc-R1e4-Q4}, and the weakest interaction takes place for the case where both $\lambda_x'$ and $\lambda_z'$ are negative, cf.~figure~\ref{fig.absN-vs-lxp-vs-lzp-lx5p7-lz0p6-c23rdUc-R1e4-Q3}.

\subsection{Scaling of the interaction coefficient for the self-similar modes}
\label{sec.coupling-scaling}

\begin{figure}
    \begin{center}
    \begin{tabular}{cc}
    \subfigure{\includegraphics[width=0.41\columnwidth]
    {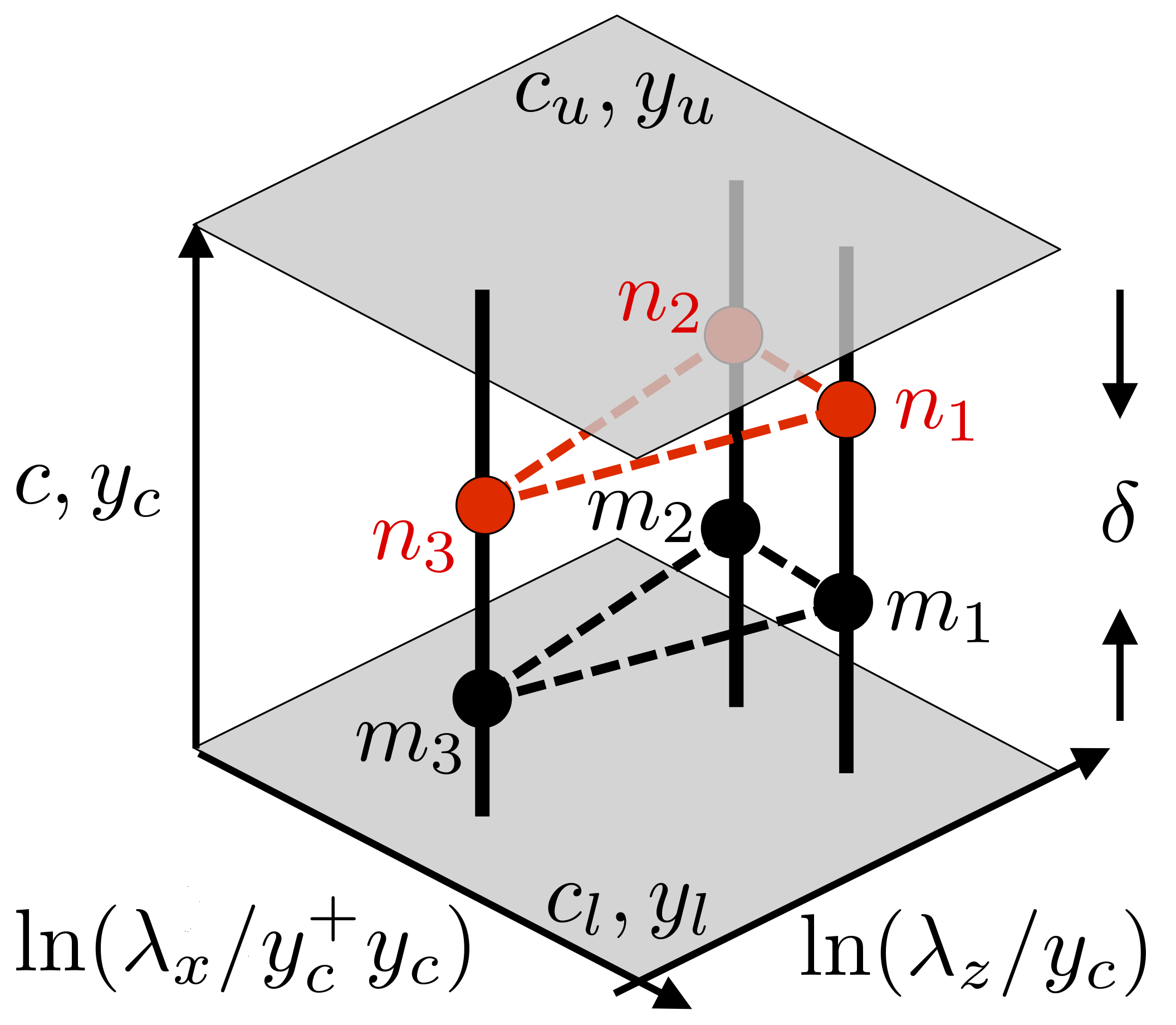}
    \label{fig.hierarchy-triad-1}}
    &
    \subfigure{\includegraphics[width=0.45\columnwidth]
    {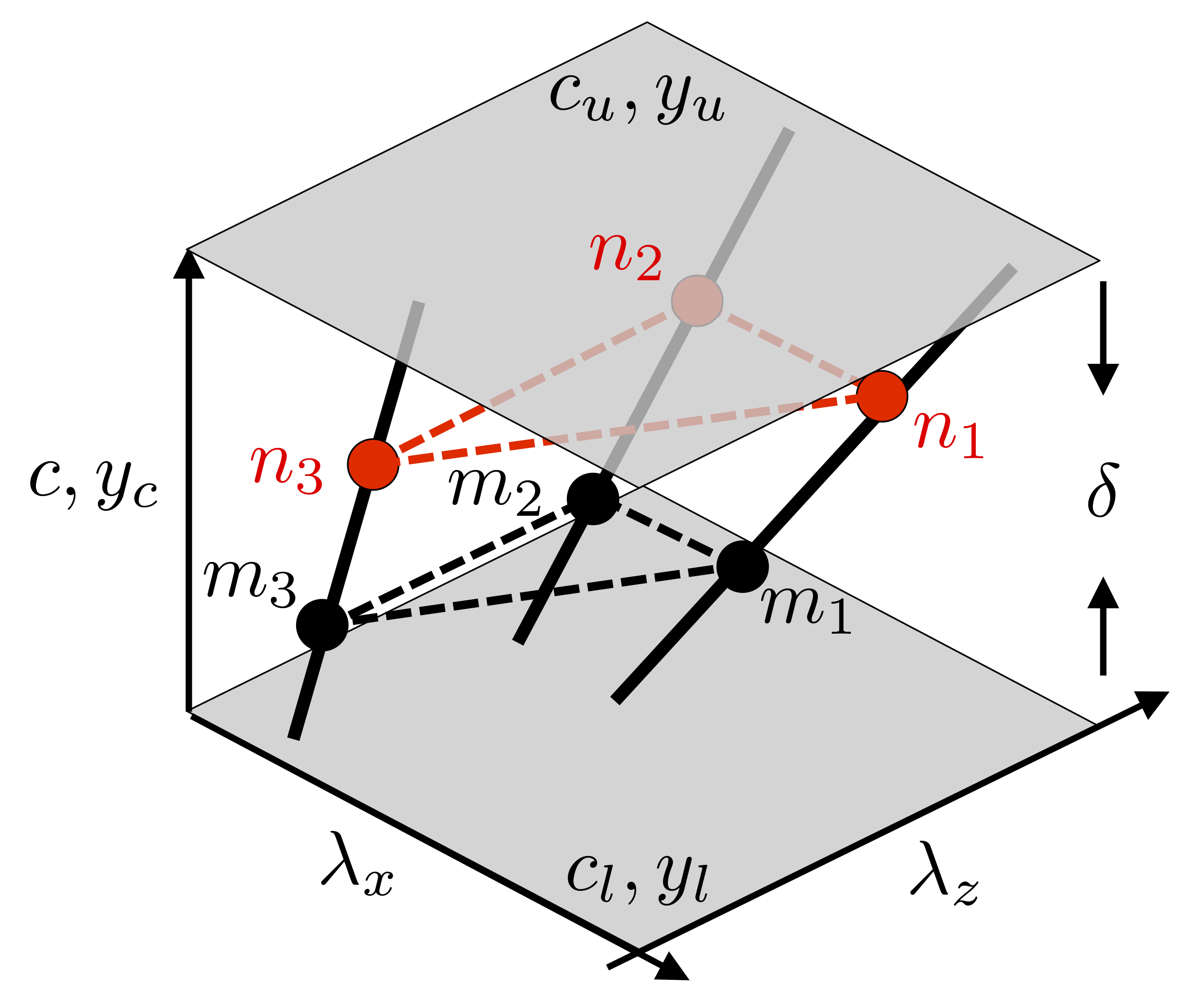}
    \label{fig.hierarchy-triad-alt}}
    \\[0.cm]
    $(a)$
    &
    $(b)$
    \end{tabular}
    \end{center}
    \caption{Schematic showing triadically consistent self-similar hierarchies. The set of modes $m_1$, $m_2$, and $m_3$ are triadically consistent. The set of modes $n_1$, $n_2$, and $n_3$ are obtained by increasing the speeds of modes $m_1$, $m_2$, and $m_3$ along the corresponding hierarchies (vertical lines). As shown in table~\ref{table.triad}, the set of modes $n_1$, $n_2$, and $n_3$ are also triadically consistent. (a) Normalized wavelengths and (b) non-normalized wavelengths.}
    \label{fig.hierarchy-triad}
    \end{figure}

The self-similar hierarchies exhibit an important property that facilitates studying the scaling of the weight matrix in the self-similar region. For any three modes that nonlinearly interact with each other, three new modes with higher/lower speeds on the same hierarchies will also nonlinearly interact. In other words, by starting from any set of triadically-consistent modes and moving an equal distance along the corresponding hierarchies, we arrive at a new set of triadically-consistent modes. This is illustrated in figure~\ref{fig.hierarchy-triad} and further explained in table~\ref{table.triad} where the parameters for three triadically-consistent modes $m_1$, $m_2$, and $m_3$ are outlined. The modes $n_1$, $n_2$ and $n_3$ are respectively obtained by moving along the hierarchies that include the modes $m_1$, $m_2$, and $m_3$. This is done by increasing the mode speed by $\delta = (1/\kappa) \ln (\alpha^+)$ which moves the mode centers away from the wall by $\alpha$ in outer units and $\alpha^+$ in inner units and increases the mode wavelengths accordingly, cf. figure~\ref{fig.hierarchy-triad-alt}. Notice that the modes $n_1$, $n_2$ and $n_3$ are triadically consistent and thus directly interact with each other. Therefore, the notion of triadically-interacting modes is extended to triadically-interacting hierarchies. This will be used to identify the scaling of the interaction coefficients associated with the self-similar modes.

     \begin{table}
      \centering
      \begin{tabular}{ccccc}
	mode
	&
	$\lambda_x$
	&
	$\lambda_z$
	&
	$\omega$
	&
	$c$
	\\[0.3cm]
	$m_1$
	&
	$\lambda_{x}$
	&
	$\lambda_{z}$
	&
	$\dfrac{2\pi c}{\lambda_{x}}$
	&
	$c$
	\\[0.3cm]
	$m_2$ 
	&
	$\lambda_{x}'$ 
	&
	$\lambda_{z}'$ 
	&
	$\dfrac{2\pi c'}{\lambda_{x}'}$ 
	&
	$c'$
	\\[0.3cm]
	$m_3$ 
	&
	\hskip0.3cm
	$-\dfrac{\lambda_{x} \lambda_{x}'}{\lambda_{x} + \lambda_{x}'}$ 
	\hskip0.3cm
	&
	\hskip0.3cm
	$-\dfrac{\lambda_{z} \lambda_{z}'}{\lambda_{z} + \lambda_{z}'}$ 
	\hskip0.3cm
	&
	\hskip0.3cm
	$-\dfrac{2\pi (c' \lambda_{x} + c \lambda_{x}')}{\lambda_{x} \lambda_{x}'}$ 
	\hskip0.3cm
	&
	\hskip0.3cm
	$\dfrac{c' \lambda_{x} + c \lambda_{x}'}{\lambda_{x} + \lambda_{x}'}$
	\\[0.3cm]
	$n_1$
	&
	$\alpha^+\alpha \lambda_{x}$
	&
	$\alpha \lambda_{z}$
	&
	$\dfrac{2\pi (c + \delta)}{\alpha^+\alpha \lambda_{x}}$
	&
	$c + \delta$
	\\[0.3cm]
	$n_2$ 
	&
	$\alpha^+\alpha \lambda_{x}'$ 
	&
	$\alpha \lambda_{z}'$ 
	&
	$\dfrac{2\pi (c' + \delta)}{\alpha^+\alpha \lambda_{x}'}$ 
	&
	$c' + \delta$
	\\[0.3cm]
	$n_3$ 
	&
	\hskip0.3cm
	$-\dfrac{\alpha^+\alpha \lambda_{x} \lambda_{x}'}{\lambda_{x} + \lambda_{x}'}$ 
	\hskip0.3cm
	&
	\hskip0.3cm
	$-\dfrac{\alpha \lambda_{z} \lambda_{z}'}{\lambda_{z} + \lambda_{z}'}$ 
	\hskip0.3cm
	&
	\hskip0.3cm
	$-\dfrac{2\pi ((c' + \delta) \lambda_{x} + (c + \delta) \lambda_{x}')}{\alpha^+\alpha \lambda_{x} \lambda_{x}'}$ 
	\hskip0.3cm
	&
	\hskip0.3cm
	$\dfrac{c' \lambda_{x} + c \lambda_{x}'}{\lambda_{x} + \lambda_{x}'} + \delta$
	\end{tabular}	
	\caption{A set of triadically-consistent modes $m_1$, $m_2$, and $m_3$ and the set of modes $n_1$, $n_2$, and $n_3$ that are obtained by respectively moving along the hierarchies that include $m_1$, $m_2$, and $m_3$ such that the mode speeds increase with $\delta$. Relative to any of the modes $m_1$, $m_2$, and $m_3$, the centers of modes $n_1$, $n_2$, and $n_3$ move away from the wall by $\alpha$ in outer units and $\alpha^+$ in inner units where $\delta = (1/\kappa) \ln (\alpha^+)$. Notice that $n_1$, $n_2$, and $n_3$ are triadically consistent themselves. See also figure~\ref{fig.hierarchy-triad}.
	}
	\label{table.triad}
	\end{table}

We consider the case where the weights of the modes with speeds in the logarithmic region are primarily determined by the modes in the logarithmic region, i.e. all the interacting modes are self-similar. This is justified by the local interaction of the modes with each other as discussed in~\S~\ref{sec.coupling}. The scalings of the resolvent modes, cf.~(\ref{eq.u-map-cu}) and~(\ref{eq.sigma-map-cu}), can be used to express~(\ref{eq.f-convolution-expand}) in terms of the largest modes in the underlying hierarchies, see Appendix~\ref{sec.N-details-scaling} for details,
	\be
	\ba{l}
	\overline{\chi}_l (\blambda, c)
	\; = \;
	\mre^{2.5 \kappa (c_u - c)}
	\;
	\ds{
	\sum_{i,j = 1}^{N}
	\;
	\iint
	}
	\,
	\cM_{lij} (\blambda_u,\blambda_u',c'-c) \,
	\chi_i(\blambda', c') \,
	\overline{\chi}_j(\blambda'', c'') \,
	\;
	\mrd \ln \blambda_u' \,
	\mrd c'.
	\ea
	\label{eq.f-convolution-scale}
	\ee
Here, $\cM_{lij} (\blambda_u,\blambda_u',c'-c)$ is the ``self-similar interaction coefficient'' in the sense that for any modes $(\blambda, c) \in \cS(\blambda_u)$ and $(\blambda', c') \in \cS(\blambda_u')$, we have
	\be
	\cN_{lij} (\blambda,c,\blambda',c')
	\; = \;
	\mre^{2.5 \kappa (c_u - c)}
	\,
	\cM_{lij} (\blambda_u,\blambda_u',c'-c).
	\label{eq.M-N}
	\ee
Notice that $\cM$ only depends on the largest modes in the hierarchies that pass through the coupled modes. Therefore, the interaction coefficient for any set of triadically-consistent modes can be obtained from the interaction coefficient for the largest modes in the corresponding hierarchies. In other words, every interaction in the logarithmic region can be determined by the modes with speed $c_u = U(y_u)$. In addition, it follows from~(\ref{eq.M-N}) that within a set of triadically consistent hierarchies, the interaction coefficient is determined by the speed of the forced mode $c$ and the difference between $c$ and the speed of one of the forcing modes $c'$.

    \begin{figure}
    \begin{center}
    \begin{tabular}{cc}
    \subfigure{\includegraphics[width=0.42\columnwidth]
    {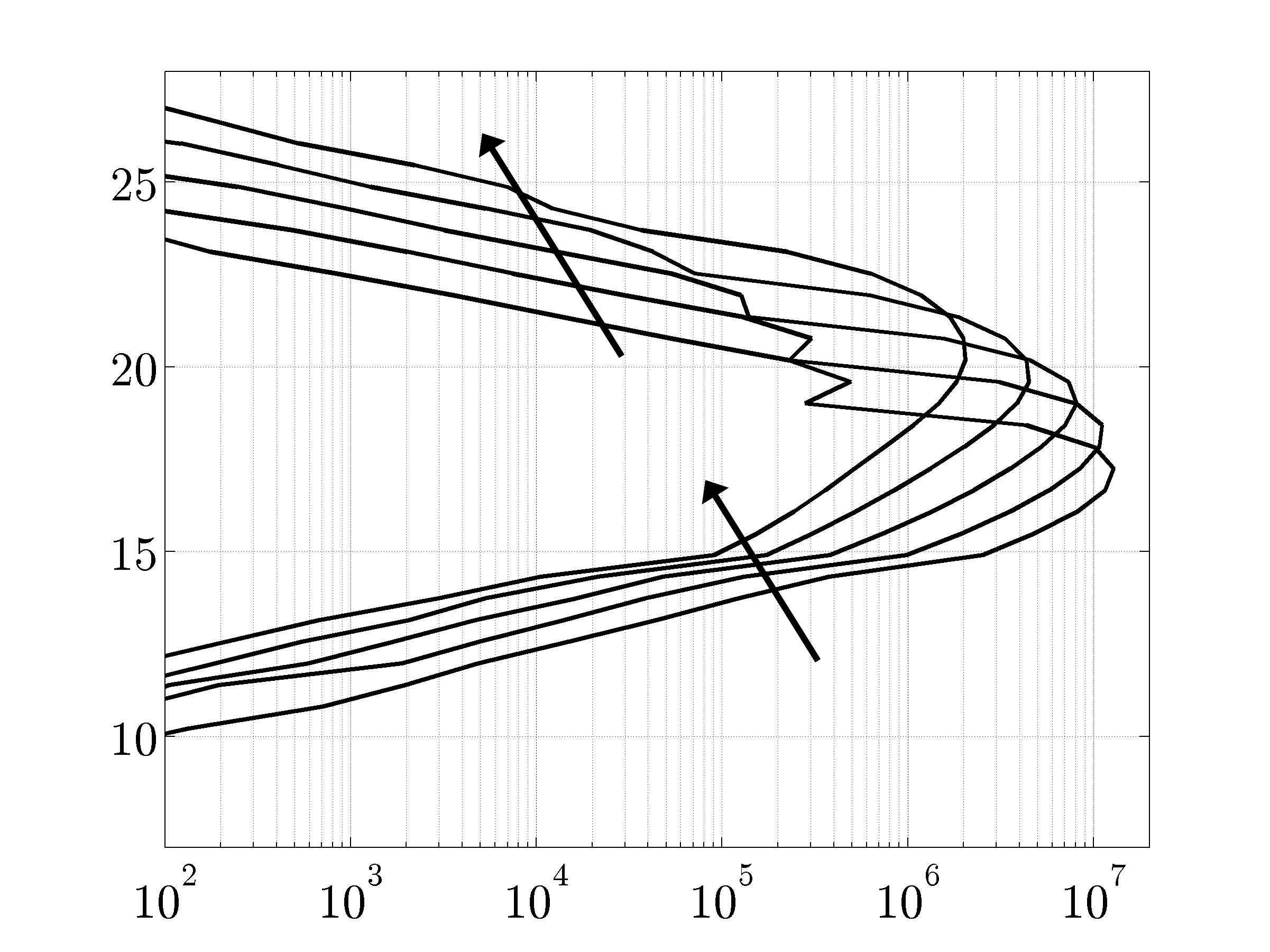}
    \label{fig.absNf-vs-cp2toUc-lx5p7-lz0p6-c23rdUc-lxpu0p35-lzpu0p11-R1e4-Q4-030414}}
    &
    \subfigure{\includegraphics[width=0.42\columnwidth]
    {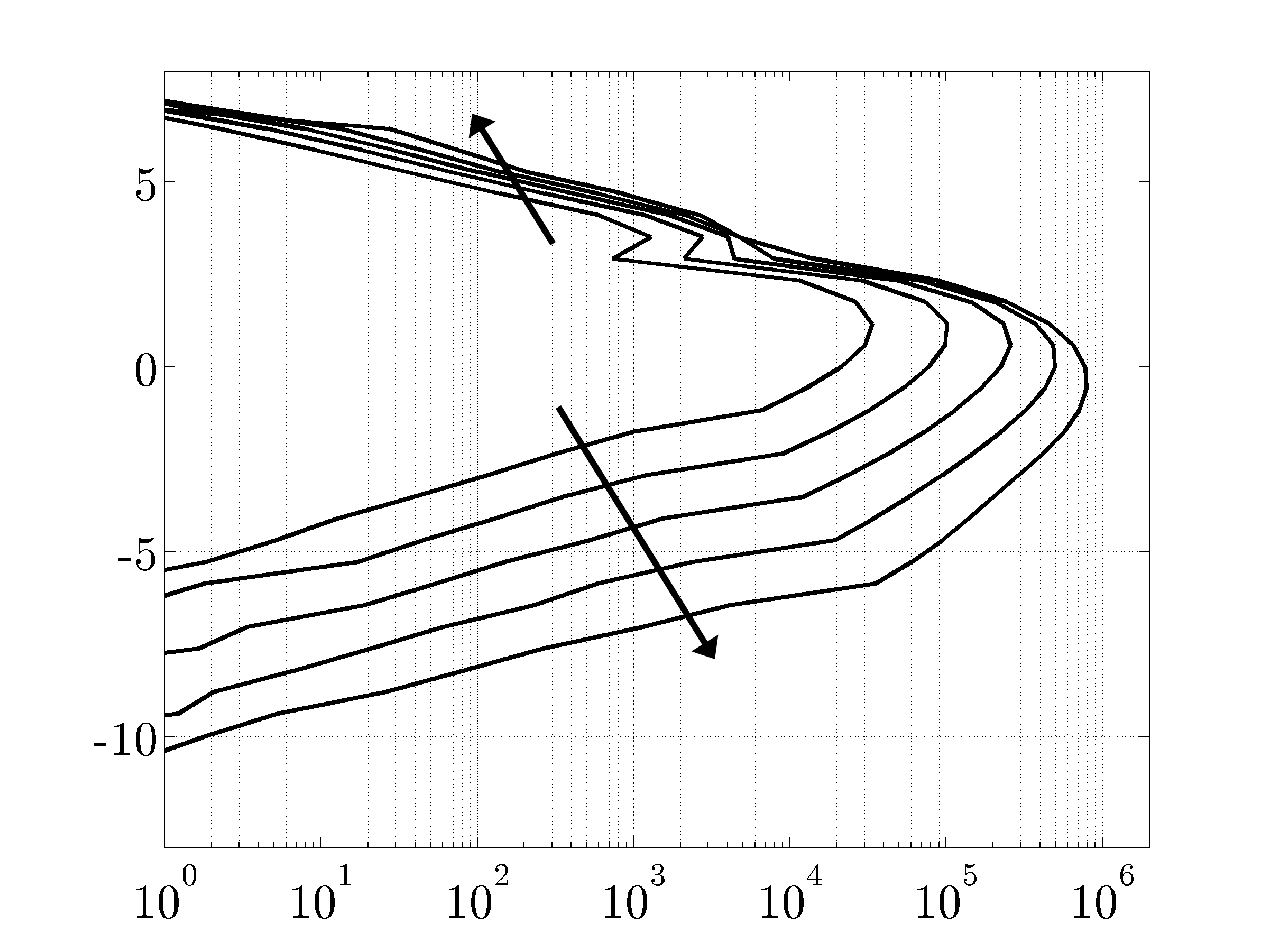}
    \label{fig.absNf-scaled-vs-cpmc1-lx5p7-lz0p6-c23rdUc-lxpu0p35-lzpu0p11-R1e4-Q4-030414}}
    \\[0.2cm]
    $(a)$
    &
    $(b)$
    \end{tabular}
    \begin{tabular}{c}
    \\[-3.8cm]
    \begin{tabular}{c}
    \hskip-5.7cm
    \begin{turn}{90}
    \tc{black}{$~~c'$}
    \end{turn}
    \hskip5.7cm
    \begin{turn}{90}
    \tc{black}{$c' - c$}
    \end{turn}
    \end{tabular}
    \\[1.9cm]
    \begin{tabular}{c}
    \hskip0.1cm
    \tc{black}{$|\cN_{111}|$}
    \hskip5cm
    \tc{black}{$|\cM_{111}|$}
    \end{tabular}
    \end{tabular}
    \\[-0.2cm]
    \begin{tabular}{cc}
    \subfigure{\includegraphics[width=0.42\columnwidth]
    {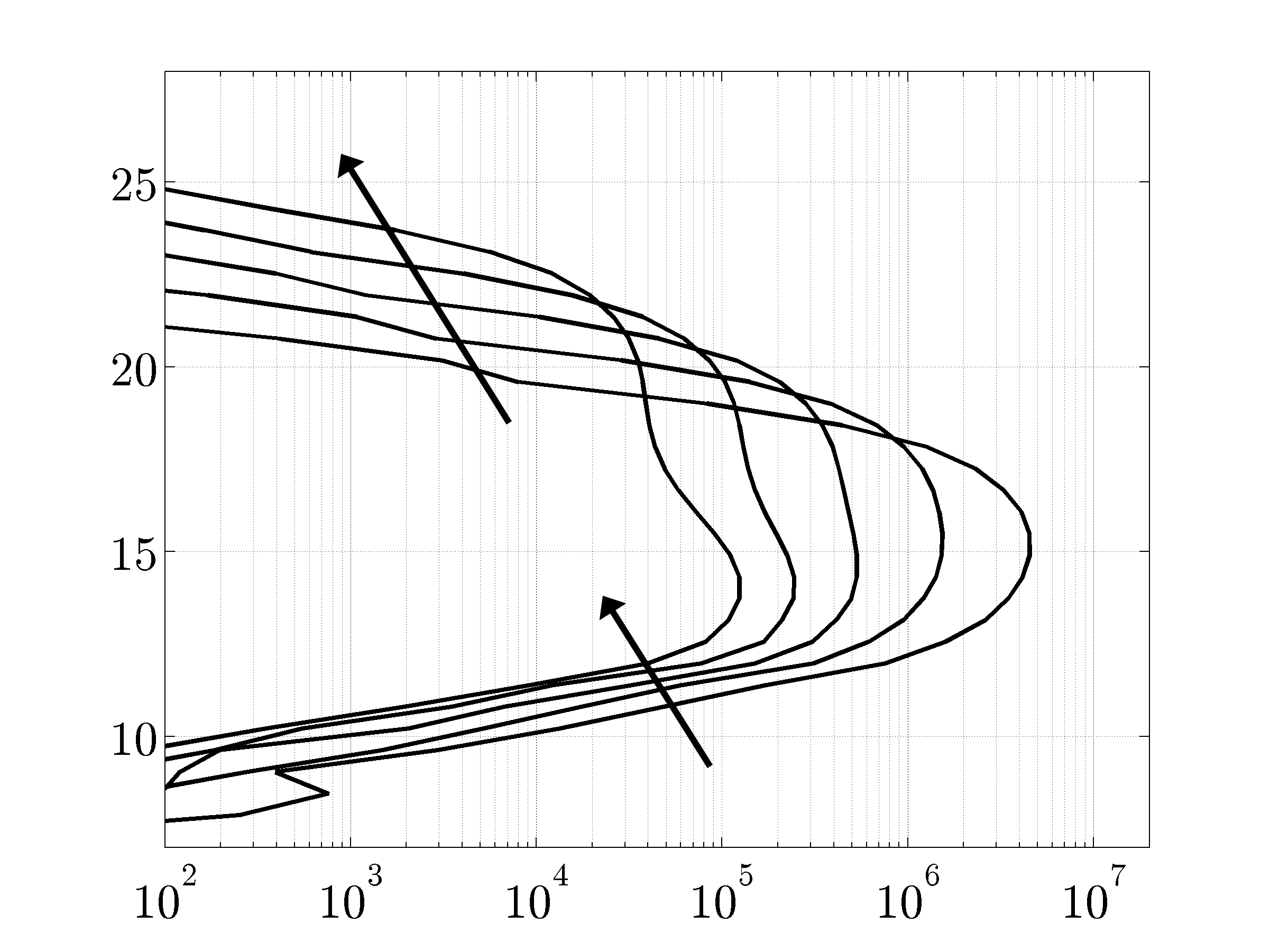}
    \label{fig.absNf-vs-cp2toUc-lx5p7-lz0p6-c23rdUc-lxpu1p11-lzpu0p035-R1e4-Q4-030414}}
    &
    \subfigure{\includegraphics[width=0.42\columnwidth]
    {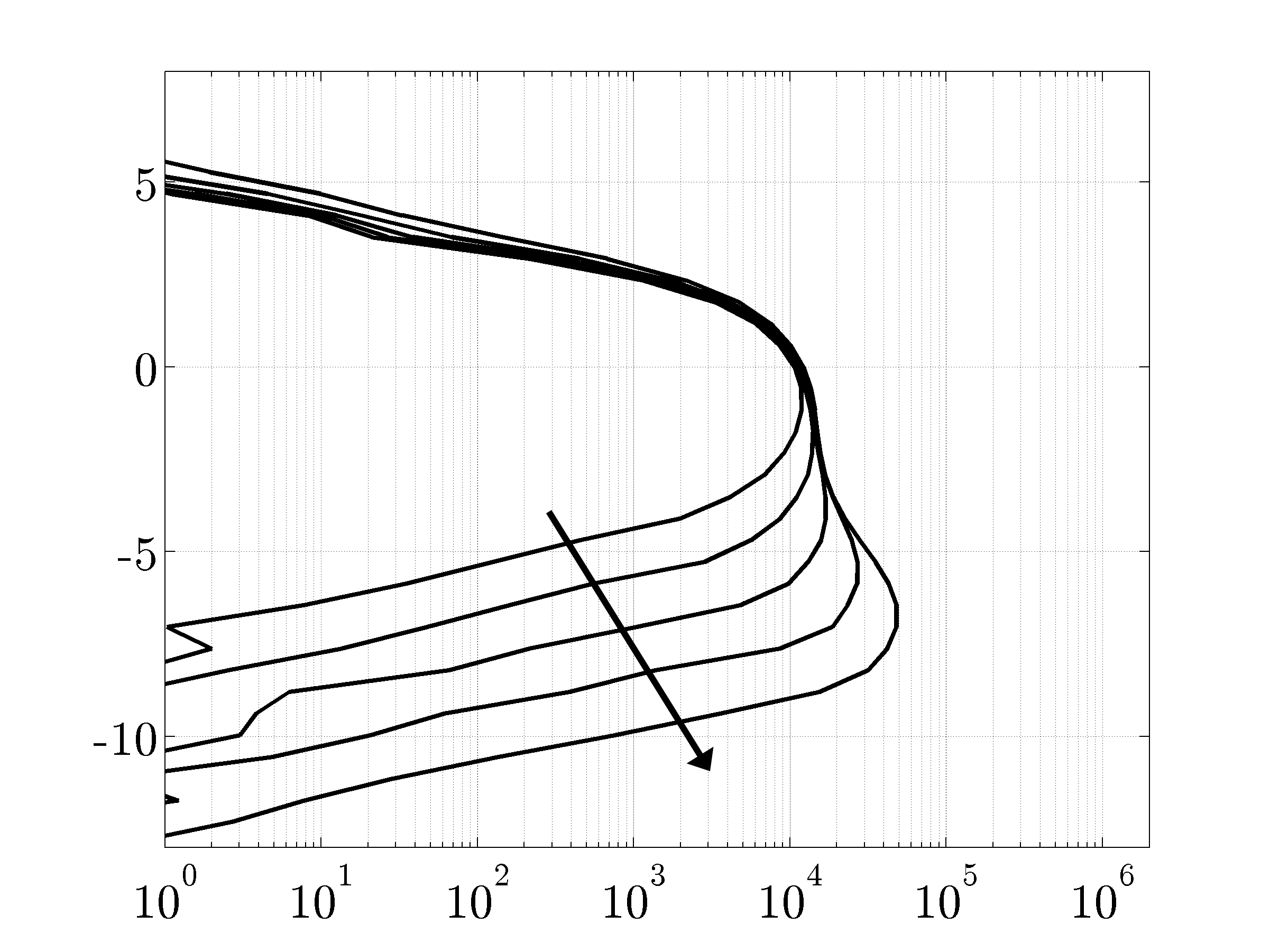}
    \label{fig.absNf-scaled-vs-cpmc1-lx5p7-lz0p6-c23rdUc-lxpu1p11-lzpu0p035-R1e4-Q4-030414}}
    \\[0.2cm]
    $(c)$
    &
    $(d)$
    \end{tabular}
    \begin{tabular}{c}
    \\[-3.8cm]
    \begin{tabular}{c}
    \hskip-5.7cm
    \begin{turn}{90}
    \tc{black}{$~~c'$}
    \end{turn}
    \hskip5.7cm
    \begin{turn}{90}
    \tc{black}{$c' - c$}
    \end{turn}
    \end{tabular}
    \\[1.9cm]
    \begin{tabular}{c}
    \hskip0.1cm
    \tc{black}{$|\cN_{111}|$}
    \hskip5cm
    \tc{black}{$|\cM_{111}|$}
    \end{tabular}
    \end{tabular}
    \\[-0.2cm]
    \begin{tabular}{cc}
    \subfigure{\includegraphics[width=0.42\columnwidth]
    {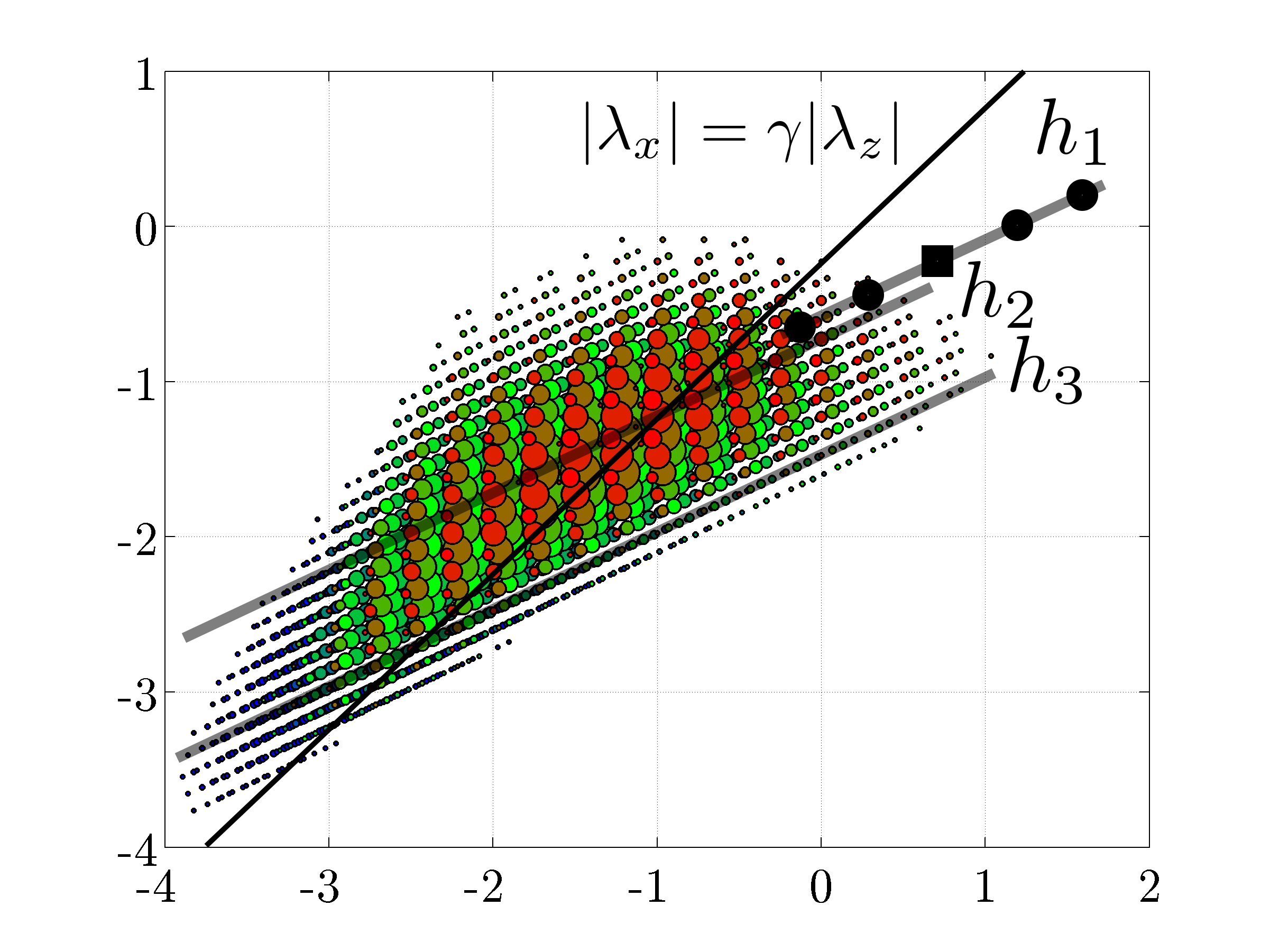}
    \label{fig.absN-vs-lxp-vs-lzp-lx5p7-lz0p6-c23rdUc-R1e4-Q4-h}}
    &
    \subfigure{\includegraphics[width=0.42\columnwidth]
    {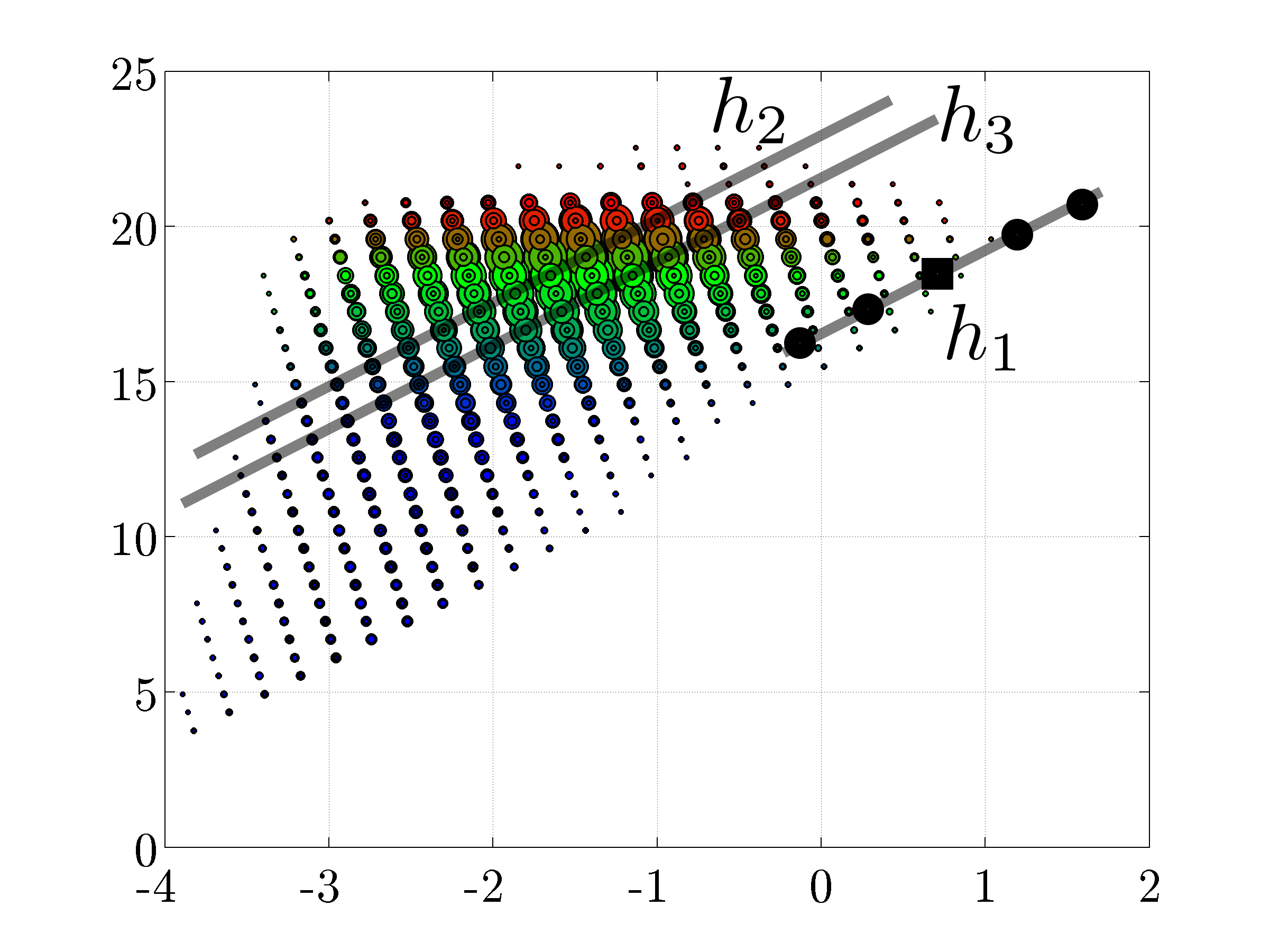}
    \label{fig.absN-vs-cp-vs-lxp-lx5p7-lz0p6-c23rdUc-R1e4-Q4-h}}
    \\[0.2cm]
    $(e)$
    &
    $(f)$
    \end{tabular}
    \begin{tabular}{c}
    \\[-4.6cm]
    \begin{tabular}{c}
    \hskip-5.7cm
    \begin{turn}{90}
    \tc{black}{$\log |\lambda_z|, \log |\lambda_z'|$}
    \end{turn}
    \hskip5.7cm
    \begin{turn}{90}
    \tc{black}{$~~~~~~~~c, c'$}
    \end{turn}
    \end{tabular}
    \\[1.9cm]
    \begin{tabular}{c}
    \hskip0cm
    \tc{black}{$\log |\lambda_x|, \log |\lambda_x'|$}
    \hskip3.7cm
    \tc{black}{$\log |\lambda_x|, \log |\lambda_x'|$}
    \end{tabular}
    \end{tabular}
    \end{center}
    \caption{The absolute value of (a,c) the interaction coefficient $|\cN_{111} (\blambda,c,\blambda',c')|$ and (b,d) the self-similar interaction coefficient $|\cM_{111} (\blambda,\blambda',c'-c)|$ for $Re_\tau = 10^4$. Five forced modes $(\blambda, c)$ that belong to the hierarchy $h_1$, see (e,f), are considered. This hierarchy passes through the representative VLSM mode with $\lambda_x = 5.7$, $\lambda_z = 0.6$, and $c = 18.4$ marked by the square in (e,f). The mode speeds in the direction of the arrows are $c = 16$, $17.2$, $18.4$, $19.6$, and $20.8$, see the circles on the hierarchy $h_1$ in (e,f). The forcing modes $(\blambda', c')$ belong to (a,b) the hierarchy $h_2$ with $\lambda_{x,u}' = 0.35$, $\lambda_{z,u}' = -0.11$ and (c,d) the hierarchy $h_3$ with $\lambda_{x,u}' = 1.11$, $\lambda_{z,u}' = -0.035$. (e,f) The forced hierarchy $h_1$ and the forcing hierarchies $h_2$ and $h_3$ are shown. The size of the colored circles is proportional to $|\cN_{111}|$ for all the modes that force the VLSM mode marked by the square on $h_1$, see also~figures~\ref{fig.absN-vs-lxp-vs-lzp-lx5p7-lz0p6-c23rdUc-R1e4-Q4} and~\ref{fig.absN-vs-cp-vs-lxp-lx5p7-lz0p6-c23rdUc-R1e4-Q4}.}
    \label{fig.absN-scaled-vs-c1-cp-lx5p7-lz0p6-c23rdUc-R1e4}
    \end{figure}    

Figure~\ref{fig.absN-scaled-vs-c1-cp-lx5p7-lz0p6-c23rdUc-R1e4} shows the interaction coefficient for five forced modes in the hierarchy $h_1$ and all the forcing modes in the hierarchies $h_2$ and $h_3$ marked by the shaded lines in figures~\ref{fig.absN-vs-lxp-vs-lzp-lx5p7-lz0p6-c23rdUc-R1e4-Q4-h} and~\ref{fig.absN-vs-cp-vs-lxp-lx5p7-lz0p6-c23rdUc-R1e4-Q4-h}. The hierarchy $h_1$ is selected to pass through the representative VLSM mode with $\lambda_x = 5.7$, $\lambda_z = 0.6$, and $c = 18.4$ marked by the square in figures~\ref{fig.absN-vs-lxp-vs-lzp-lx5p7-lz0p6-c23rdUc-R1e4-Q4-h} and~\ref{fig.absN-vs-cp-vs-lxp-lx5p7-lz0p6-c23rdUc-R1e4-Q4-h}. Figures~\ref{fig.absNf-vs-cp2toUc-lx5p7-lz0p6-c23rdUc-lxpu0p35-lzpu0p11-R1e4-Q4-030414} and~\ref{fig.absNf-scaled-vs-cpmc1-lx5p7-lz0p6-c23rdUc-lxpu0p35-lzpu0p11-R1e4-Q4-030414} show $|\cN_{111} (\blambda,c,\blambda',c')|$ and $|\cM_{111} (\blambda,\blambda',c'-c)|$ for the forcing hierarchy $h_2$ with $\lambda_{x,u}' = 0.35$, $\lambda_{z,u}' = -0.11$. This hierarchy passes through the forcing modes that exhibit the largest interaction coefficient with the representative VLSM mode. As evident from figure~\ref{fig.absNf-vs-cp2toUc-lx5p7-lz0p6-c23rdUc-lxpu0p35-lzpu0p11-R1e4-Q4-030414}, $|\cN_{111}|$ peaks for $c' \approx c$ and decreases as $c$ becomes larger. Figure~\ref{fig.absNf-scaled-vs-cpmc1-lx5p7-lz0p6-c23rdUc-lxpu0p35-lzpu0p11-R1e4-Q4-030414} shows that the interaction coefficients are approximately self-similar for $2 < c' - c < 3$, notice the approximate collapse of $|\cM_{111}|$ in this region. This is because the aspect-ratio constraint for self-similarity of the modes in $h_2$ is satisfied only for large enough values of $c'$, see figure~\ref{fig.absN-vs-lxp-vs-lzp-lx5p7-lz0p6-c23rdUc-R1e4-Q4-h}. 

For comparison, we also consider the forcing hierarchy $h_3$ with $\lambda_{x,u}' = 1.11$, $\lambda_{z,u}' = -0.035$ where the aspect-ratio constraint is satisfied for a larger interval of $c'$ in the logarithmic region, cf. figure~\ref{fig.absN-vs-lxp-vs-lzp-lx5p7-lz0p6-c23rdUc-R1e4-Q4-h}. Figure~\ref{fig.absNf-vs-cp2toUc-lx5p7-lz0p6-c23rdUc-lxpu1p11-lzpu0p035-R1e4-Q4-030414} shows that $|\cN_{111} (\blambda,c,\blambda',c')|$ for $h_3$ locally peaks around $c' \approx c$ while a second peak emerges for $c' \approx 14$ as $c$ increases. Figure~\ref{fig.absNf-scaled-vs-cpmc1-lx5p7-lz0p6-c23rdUc-lxpu1p11-lzpu0p035-R1e4-Q4-030414} shows that the interaction coefficient is self-similar for $-1 < c' - c < 3$ and $c$ in the logarithmic region. Notice that the self-similarity extends to $|c' - c| < 3$ when only larger values of $c$ in the logarithmic region are considered. Even though figure~\ref{fig.absN-scaled-vs-c1-cp-lx5p7-lz0p6-c23rdUc-R1e4} demonstrates that the self-similar interaction coefficients do not necessarily correspond to the largest ones, we emphasize that the forcing is obtained by the product of the interaction coefficient and the weights corresponding to the forcing modes. Studying the combined effect of the interaction coefficient and the weights is the subject of ongoing research.

\begin{figure}
    \begin{center}
    \begin{tabular}{cc}
    \begin{tabular}{c}
    \\[-4.6cm]
    \subfigure{\includegraphics[width=0.52\columnwidth]
    {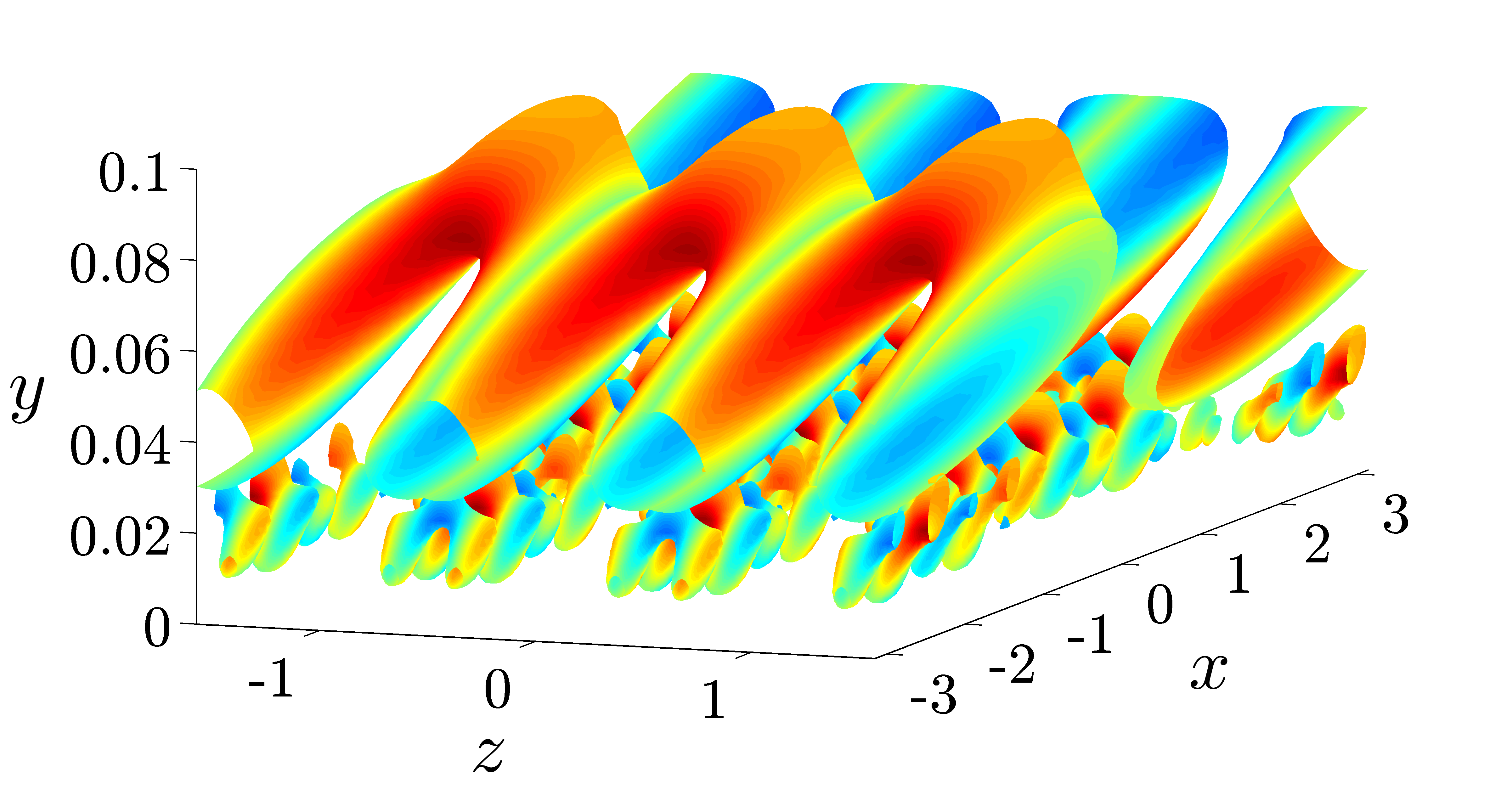}
    \label{fig.triad-swirl-2triads-3modes-3d-081614}}
    \end{tabular}
    &
    \subfigure{\includegraphics[width=0.45\columnwidth]
    {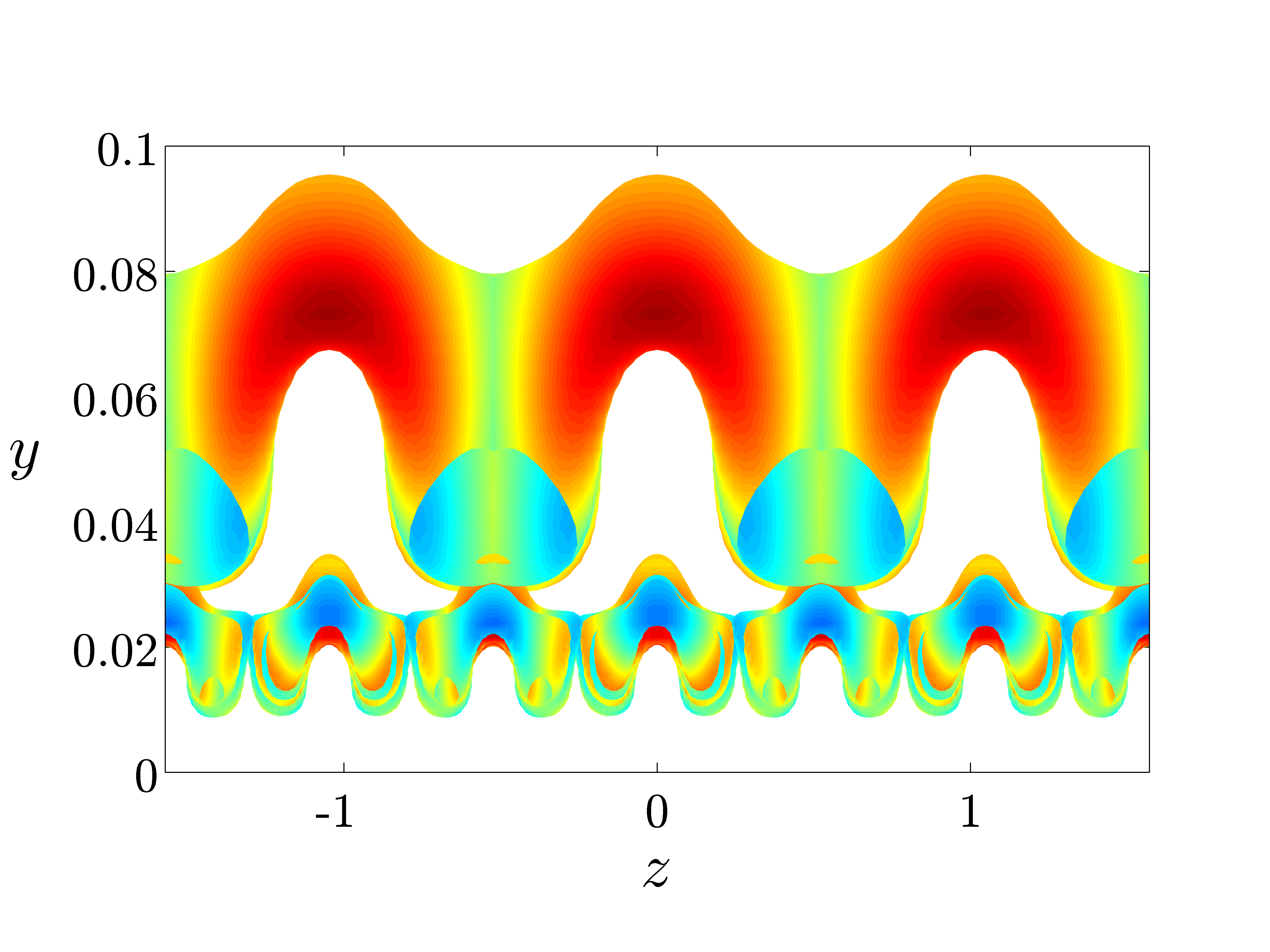}
    \label{fig.triad-swirl-2triads-3modes-yz-081614}}
    \\[-0.2cm]
    $(a)$
    &
    $(b)$
    \end{tabular}
    \end{center}
    \caption{The isosurfuces represent $50\%$ of the maximum swirling strength $\lambda_{ci}$ for two sets of triadically-consistent modes that belong to the same triadically-consistent hierarchies for $Re_\tau = 10^4$, cf.~table~\ref{table.2triads}. The smaller/lower and larger/upper swirl structures respectively correspond to the triad modes $m_i$ and $n_i$ in table~\ref{table.2triads}. The colors show the spanwise vorticity normalized by its maximum value where red (blue) denotes rotation in (opposite) the sense of the mean velocity. (a) Three-dimensional view and (b) cross-stream view.}
    \label{fig.swirl-triad}
    \end{figure}

The identified self-similar triads provide an objective selection process for examining the `turbulence kernel'~\citep{shamcK13} and a quantitative link between its structural discussion and the prediction of the spectra. Specifically, the self-similar scalings provide a systematic way to describe how triadically-consistent mode packets scale geometrically in the log region and with $Re_\tau$. This is evident in figure~\ref{fig.swirl-triad} where the isosurfuces corresponding to $50\%$ of the maximum swirling strength $\lambda_{ci}$~\citep{chabaladr05} are shown for two sets of triadically-consistent modes that belong to the same triadically-consistent hierarchies for $Re_\tau = 10^4$, see~table~\ref{table.2triads} for the mode parameters. Each triad includes two short modes and one long mode in the streamwise direction. The amplitude $\chi_1 \sigma_1$ of the modes is selected arbitrarily for the purpose of this illustration. Notice that the streamwise and spanwise lengths of the larger triad are $3^2$ and $3$ times larger than the smaller triad and the speed of the larger triad is larger by $(1/\kappa) \ln 3$.~\cite{shamcK13} discussed how the modes within each turbulence kernel (triad) may self-excite through nonlinear interactions. Here, we complement their discussion by identifying how the nonlinear interaction coefficient scales for the triad modes in the log region.
 
\begin{table}
\centering
\begin{tabular}{cccccc}
\hspace{0.7cm}
&
\hspace{0.5cm}
$\pm \kappa_x$
\hspace{0.5cm}
&
\hspace{0.5cm}
$\pm \kappa_z$
\hspace{0.5cm}
&
\hspace{0.5cm}
$c = \omega/\kappa_x = U(y_c)$
\hspace{0.5cm}
&
\hspace{0.5cm}
$y_c$
\hspace{0.5cm}
&
\hspace{1.2cm}
$\chi_1 \sigma_1$
\hspace{1.2cm}
\\[0.2cm]
$m_1$
&
$6$
&
$6$
&
$17$
&
$0.016$
&
$(0.05) \mathrm{e}^{-2.6\mathrm{i}}$
\\[0.2cm]
$m_2$
&
$1$
&
$6$
&
$17$
&
$0.016$
&
$1$
\\[0.2cm]
$m_3$
&
$7$
&
$12$
&
$17$
&
$0.016$
&
$(0.045) \mathrm{e}^{-2.1\mathrm{i}}$
\\[0.2cm]
$n_1$
&
$6/3^2$
&
$6/3$
&
$17 + (1/\kappa) \ln 3$
&
$0.048$
&
$(0.05) \mathrm{e}^{-2.6\mathrm{i}}$
\\[0.2cm]
$n_2$
&
$1/3^2$
&
$6/3$
&
$17 + (1/\kappa) \ln 3$
&
$0.048$
&
$1$
\\[0.2cm]
$n_3$
&
$7/3^2$
&
$12/3$
&
$17 + (1/\kappa) \ln 3$
&
$0.048$
&
$(0.045) \mathrm{e}^{-2.1\mathrm{i}}$
     \end{tabular}
	\caption{The parameters for two sets of triadically-consistent modes $\{m_1, m_2, m_3\}$ and $\{n_1, n_2, n_3\}$ that belong to the same triadically-consistent hierarchies for $Re_\tau = 10^4$. Notice that the streamwise and spanwise lengths of the modes $n_i$ are $3^2$ and $3$ times larger than the modes $m_i$ and the speed of the modes $n_i$ is larger than the speed of the modes $m_i$ by $(1/\kappa) \ln 3$. See also figure~\ref{fig.swirl-triad}.
	}
	\label{table.2triads}
	\end{table}

\section{Compatibility with experimental observations}
\label{sec.exp-scaling}

It was shown in~\S~\ref{sec.log-region} that the energy intensities, the Reynolds stresses, and the energy budget can be expressed in terms of the resolvent modes with speed $c_u$ at the top of the logarithmic region. In this formulation, the resolvent weights for the modes with $c < c_u$ appear as weights on the contributions from the resolvent modes with speed $c_u$. Ultimately, we require that the weighted resolvent modes capture the high-Reynolds-number experimental observations in the logarithmic region. This imposes several constraints on the weight matrices and can be used to provide further intuition about their scaling and the role of nonlinearity in the NSE. In this section, we outline two examples for such constraints. 

A constant Reynolds stress in the logarithmic region requires, cf.~(\ref{eq.Euv-hierarchy-analytical-x}),
	\be
	\ba{rcl}
	\dfrac{1}{\kappa}
	\dfrac{Re_\tau}{Re_{\tau,u}}
	\,
	\ds{
	\int_{\ln y_{l'}}^{\ln y_u}
	}
	\,
	\dfrac{y_u}{y_c}
	\,
	\mbox{Re}
	\Big\{
	\mbox{tr}
	\Big(
	A_{uv} \big( \dfrac{y_u}{y_c}y, \blambda_u, c_u \big)
	X_h (\blambda_u, y_c)
	\Big)
	\Big\}
	\,
	\mrd \ln(y_c)
	&\!\! = \!\!&
	b_1,
	\ea
	\label{eq.Euv-log}
	\ee
where $b_1$ is a constant. Since $A_{uv} \big( (y/y_c) y_u, \blambda_u, c_u \big)$ is localized around $\eta = y/y_c = 1$, we consider the case where $A_{uv} \big( \eta y_u, \blambda_u, c_u \big)$ is negligible for $1/d < \eta < d$ and some $d > 1$. Under this assumption,~(\ref{eq.Euv-log}) is rewritten as
	\be
	\ba{rcl}
	\dfrac{1}{\kappa}
	\dfrac{Re_\tau}{Re_{\tau,u}}
	\,
	\ds{
	\int_{-\ln d}^{\ln d}
	}
	\,
	\dfrac{\eta y_u}{y}
	\,
	\mbox{Re}
	\Big\{
	\mbox{tr}
	\Big(
	A_{uv} \big( \eta y_u, \blambda_u, c_u \big)
	X_h (\blambda_u, y/\eta)
	\Big)
	\Big\}
	\,
	\mrd \ln \eta
	&\!\! = \!\!&
	b_1.
	\ea
	\label{eq.Euv-log-reduced}
	\ee
A sufficient condition for the left-hand-side of~(\ref{eq.Euv-log-reduced}) to be constant for all $y$ is that the real part of the inner product of $A_{uv} \big( \eta y_u, \blambda_u, c_u \big)$ and $X_h (\blambda_u, y/\eta)$ is proportional to $y$ for any fixed $\eta$. Since $A_{uv} \big( \eta y_u, \blambda_u, c_u \big)$ is independent of $y$ for fixed $\eta$, the above constraint is solely imposed on the weight matrices $X_h (\blambda_u, y/\eta)$ along the hierarchy $\cS(\blambda_u)$.

Similarly, a logarithmically decreasing streamwise intensity requires, cf.~(\ref{eq.hierarchy-analytical}),
	\be
	\ba{rcl}
	\dfrac{1}{\kappa}
	\dfrac{Re_\tau^2}{Re_{\tau,u}^2}
	\,
	\ds{
	\int_{-\ln d}^{\ln d}
	}
	\,
	\mbox{Re}
	\Big\{
	\mbox{tr}
	\Big(
	A_{uu} \big( \eta y_u, \blambda_u, c_u \big)
	X_h (\blambda_u, y/\eta)
	\Big)
	\Big\}
	\,
	\mrd \ln \eta
	&\!\! = \!\!&
	b_2
	\, - \,
	b_3 \, \ln y,
	\ea
	\label{eq.Euu-log-reduced}
	\ee
where $b_2$ and $b_3 > 0$ are constants. A sufficient condition for~(\ref{eq.Euu-log-reduced}) is that the real part of the inner product of $A_{uu} \big( \eta y_u, \blambda_u, c_u \big)$ and $X_h (\blambda_u, y/\eta)$ is logarithmically decreasing with $y$ for any $\eta$. Since $A_{uu} \big( \eta y_u, \blambda_u, c_u \big)$ is independent of $y$ for fixed $\eta$, this constraint is also solely imposed on the weight matrices along the hierarchy $\cS(\blambda_u)$. 

The above constraints can be used to develop empirical models for the weight matrices without explicitly studying the nonlinear terms in the NSE. Identifying the scaling of the interaction coefficient between the resolvent modes represents an important step towards explicit determination of the weight matrices, as discussed in~\S~\ref{sec.scaling-weights}. Direct computation of the weight matrices from the nonlinear terms is a subject of ongoing research.

\section{Summary and conclusions}
\label{sec.conclusion}

We formulated a framework for analytical developments in the logarithmic region of wall-bounded turbulent flows. The basic elements of this model are the self-similar resolvent modes that are governed by the linear amplification mechanisms in the NSE. The resolvent formulation allows for reduction in complexity of the analysis by deconstructing the flow into propagating modes whose dynamics in response to the nonlinearity can be individually studied. An important property of the resolvent modes is that their Reynolds number scaling and geometric self-similarity are known. The self-similar scalings were used to express the energy intensities, the Reynolds stresses, and the energy budget in terms of the resolvent modes with speeds corresponding to the top of the logarithmic region. In addition, we showed that the interaction coefficient for triadically-consistent resolvent modes in the log region follows an exponential function. This represents a significant step towards the explicit calculation of the resolvent weights which would offer a closed model of the fluctuations in this region. These findings can be combined to better understand the dynamics and interaction of flow structures in the logarithmic region.

\subsection{Implications of the self-similar scalings}
\label{sec.implications}

Self-similarity can be used to yield order reduction of the models in the logarithmic region. This is because any resolvent mode with $c_{l'} < c < c_u$ and $\lambda_x/\lambda_z > \gamma$ can be obtained from the largest mode in the corresponding hierarchy, i.e. the mode with speed $c_u$, cf.~(\ref{eq.u-map-cu}) and~(\ref{eq.sigma-map-cu}). Once the largest resolvent modes are computed, we readily have all the resolvent modes with speeds in the logarithmic region. This property was used to obtain the following results:

\vskip0.3cm

\bi

\item The contribution of any self-similar hierarchy to the time-averaged energy intensity or Reynolds stress at the wall-normal location $y$ can be obtained from a weighted integral of the energy density of the largest mode in that hierarchy from $y$ to $y y_u/y_{l'}$, cf.~(\ref{eq.hierarchy-analytical}),~(\ref{eq.Evv-hierarchy-analytical-x}), and~(\ref{eq.Euv-hierarchy-analytical-x}).
\\[-0.2cm]
\item The contribution of the self-similar modes to the two-dimensional energy spectrum at the wall-normal location $y$ and wavelengths $\blambda$ can be obtained from a weighted integral of the energy densities of a the largest modes in the hierarchies that pass through the modes with $c_{l'} < c < c_u$ and $\blambda$, cf.~(\ref{eq.Euu-t-avg-analytical}),~(\ref{eq.Evv-t-avg-analytical}), and~(\ref{eq.Euv-t-avg-analytical}).
\\[-0.2cm]
\item The interaction coefficient for any triadically-consistent set of modes with $c_{l'} < c < c_u$ and $\lambda_x/\lambda_z > \gamma$ can be obtained from the largest mode in the corresponding self-similar hierarchy, i.e. the mode with $c = c_u$, cf.~(\ref{eq.M-N}).

\ei    

\vskip0.3cm

We provided an analytical support for Townsend's attached-eddy model by showing that the NSE admit self-similar resolvent modes. The wall-normal and spanwise scaling of the modes is the same as the attached eddies. It was also argued that the difference in the streamwise scaling of the self-similar modes and the attached eddies does not contradict Townsend's hypothesis. In addition, it was shown that the attached hairpin-like structures arise naturally in the log region since they are the dominant type of motion in this region. The proposed model-based formulation of the self-similar dynamics is expected to enable development of dynamically significant reduced-order models in the logarithmic region. It can also furnish a better understanding of the self-similar vortex clusters in the logarithmic region~\citep{deljimzanmos06}, the cascade of turbulent flux in the inertial subrange~\citep{hoyjim08}, and the self-similar mean dynamics~\citep{fifweiklemcm05, klefifwei09,kle13}.

\section*{Acknowledgments}
    \label{sec.Ack}

The support of Air Force Office of Scientific Research under grants FA 9550-09-1-0701 (P.M. Rengasamy Ponnappan) and FA 9550-12-1-0469 (P.M. Doug Smith) is gratefully acknowledged.

\appendix

\section{Edge effects}
\label{sec.edge}

In this section, we discuss the modes with speeds close to the edges of the logarithmic region, $c_l$ and $c_u$, and the modes that do not satisfy the aspect-ratio constraint. Owing to locality of singular functions around the critical layer, the modes with speeds $c_l \ll c \ll c_u$ are only affected by the logarithmic part of the mean velocity. However, the modes with speeds close to the edges of the logarithmic region may be affected by non-logarithmic parts of the mean velocity. 

\begin{figure}
    \begin{center}
    \begin{tabular}{ccc}
    \subfigure{\includegraphics[width=0.325\columnwidth]
    {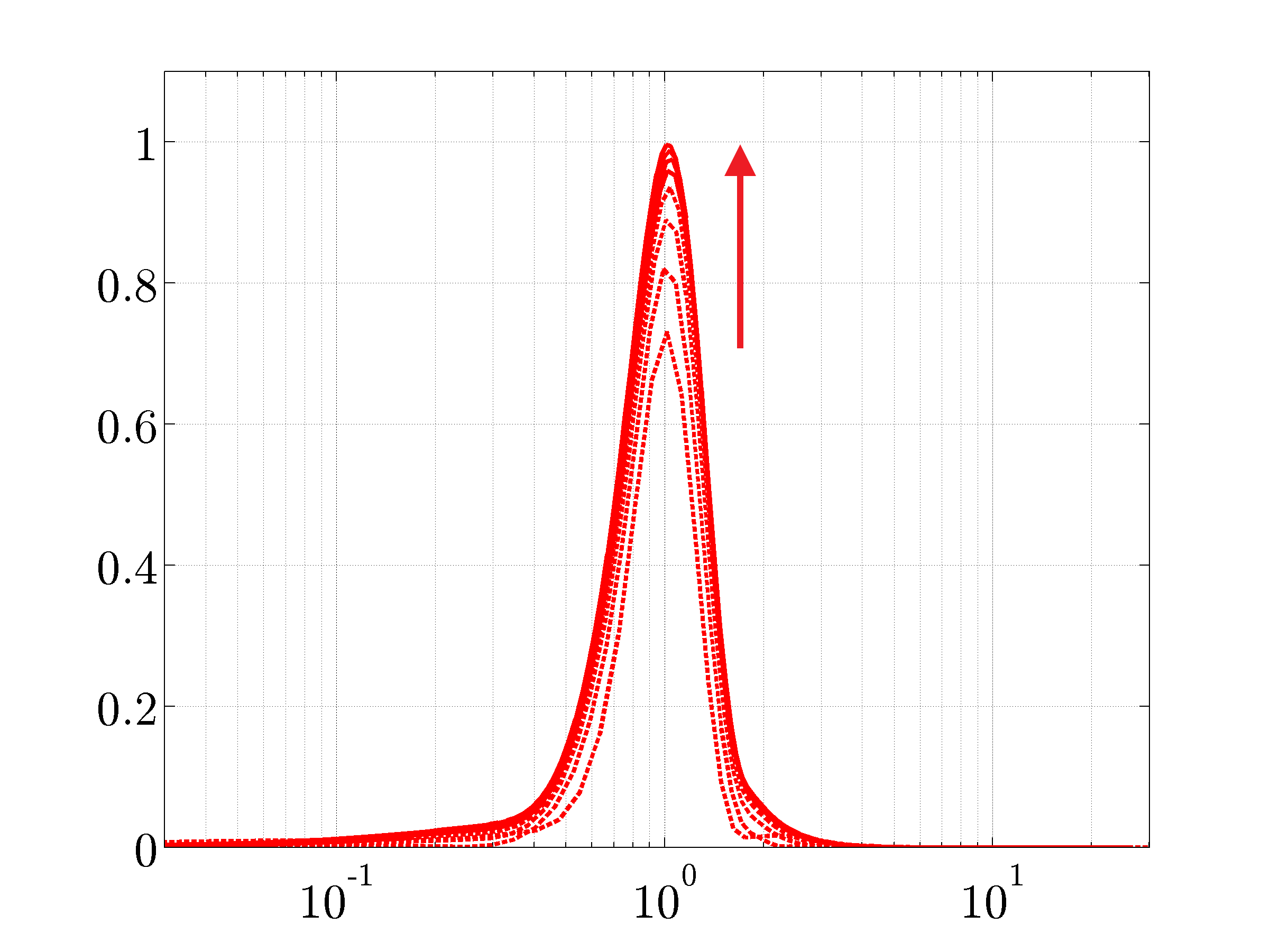}
    \label{fig.log-region-hierarchy-u-similarity-vs-y-yp0p1r-kx1kz10-R1e4}}
    &
    \hskip-0.3cm
    \subfigure{\includegraphics[width=0.325\columnwidth]
    {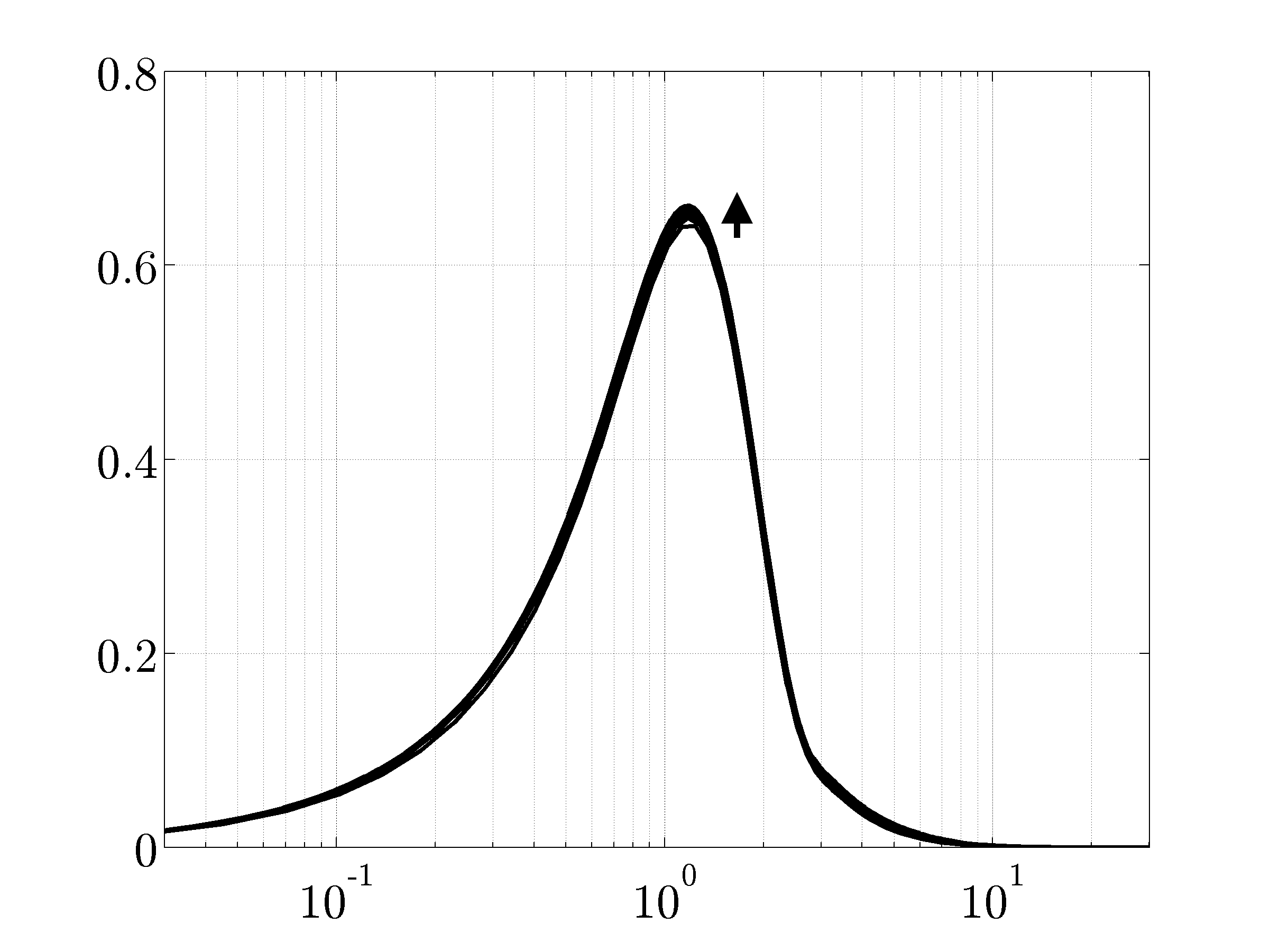}
    \label{fig.log-region-hierarchy-u-similarity-vs-y-Up2_3Uck-kx1kz10-R1e4}}
    &
    \hskip-0.3cm
    \subfigure{\includegraphics[width=0.325\columnwidth]
    {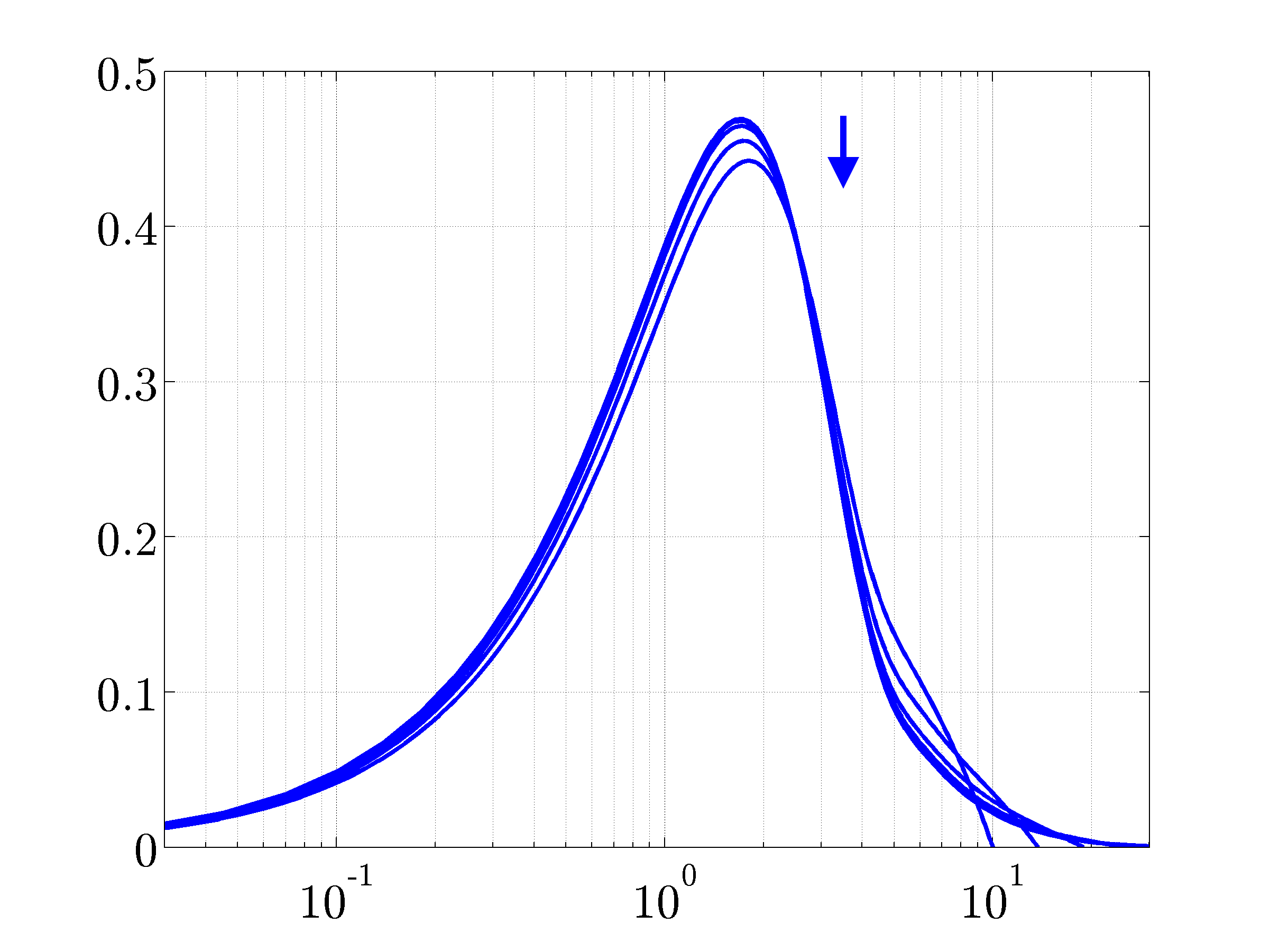}
    \label{fig.log-region-hierarchy-u-similarity-vs-y-yp100_Rb-kx1kz10-R1e4}}
    \\[0.2cm]
    $(a)$
    &
    \hskip-0.3cm
    $(b)$
    &
    \hskip-0.3cm
    $(c)$
    \end{tabular}
    \begin{tabular}{c}
    \\[-3.7cm]
    \begin{tabular}{c}
    \hskip-6.85cm
    \begin{turn}{90}
    \tc{black}{$~~~~\sqrt{y_c} \; \abs{\hat{u}_1}$}
    \end{turn}
    \end{tabular}
    \\[1.3cm]
    \begin{tabular}{c}
    \hskip0.2cm
    \tc{black}{$y/y_c$}
    \hskip3.7cm
    \tc{black}{$y/y_c$}
    \hskip3.7cm
    \tc{black}{$y/y_c$}
    \end{tabular}
    \end{tabular}
    \\[-0.2cm]
    \begin{tabular}{ccc}
    \subfigure{\includegraphics[width=0.325\columnwidth]
    {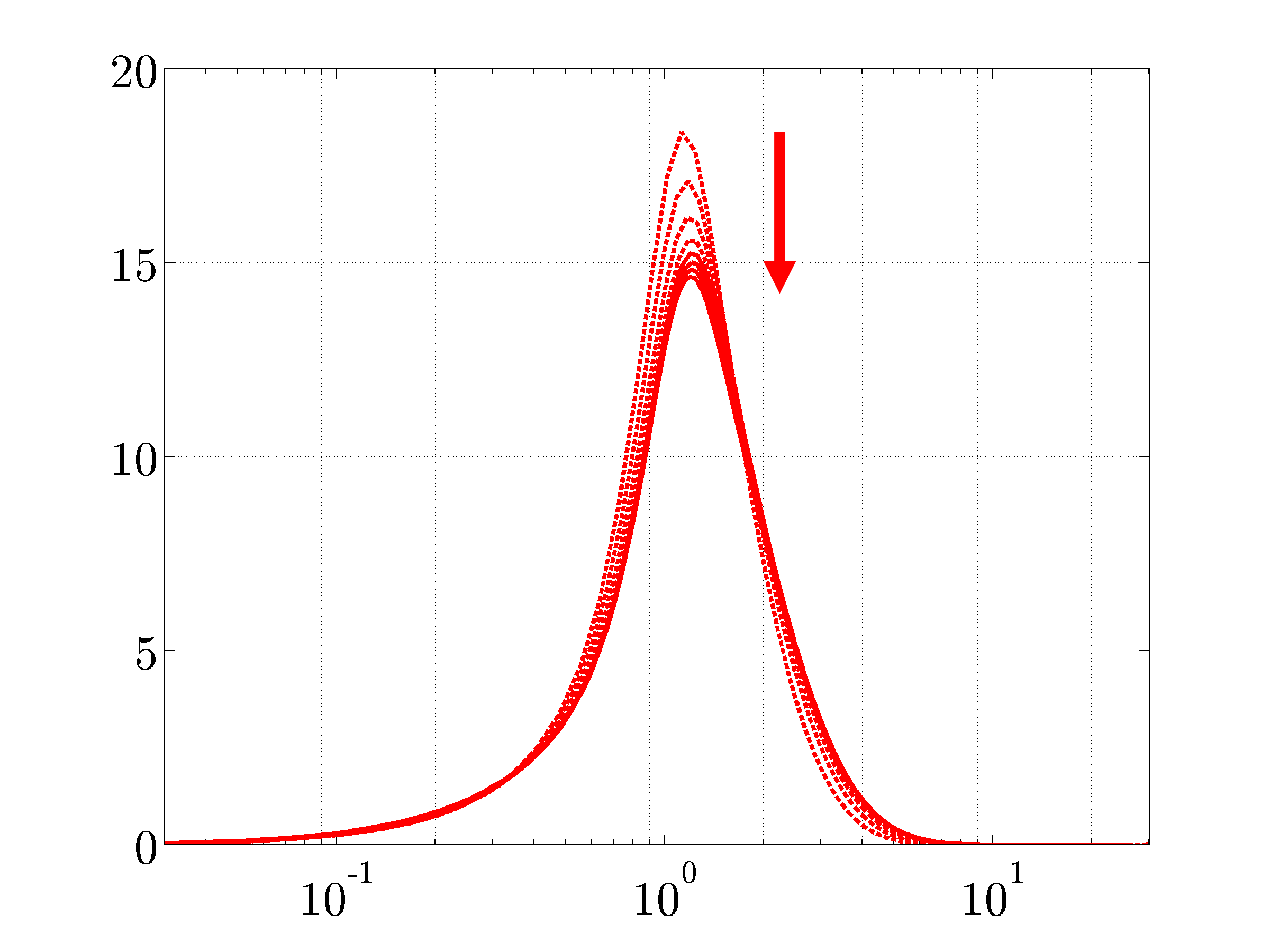}
    \label{fig.log-region-hierarchy-v-similarity-vs-y-yp0p1r-kx1kz10-R1e4}}
    &
    \hskip-0.3cm
    \subfigure{\includegraphics[width=0.325\columnwidth]
    {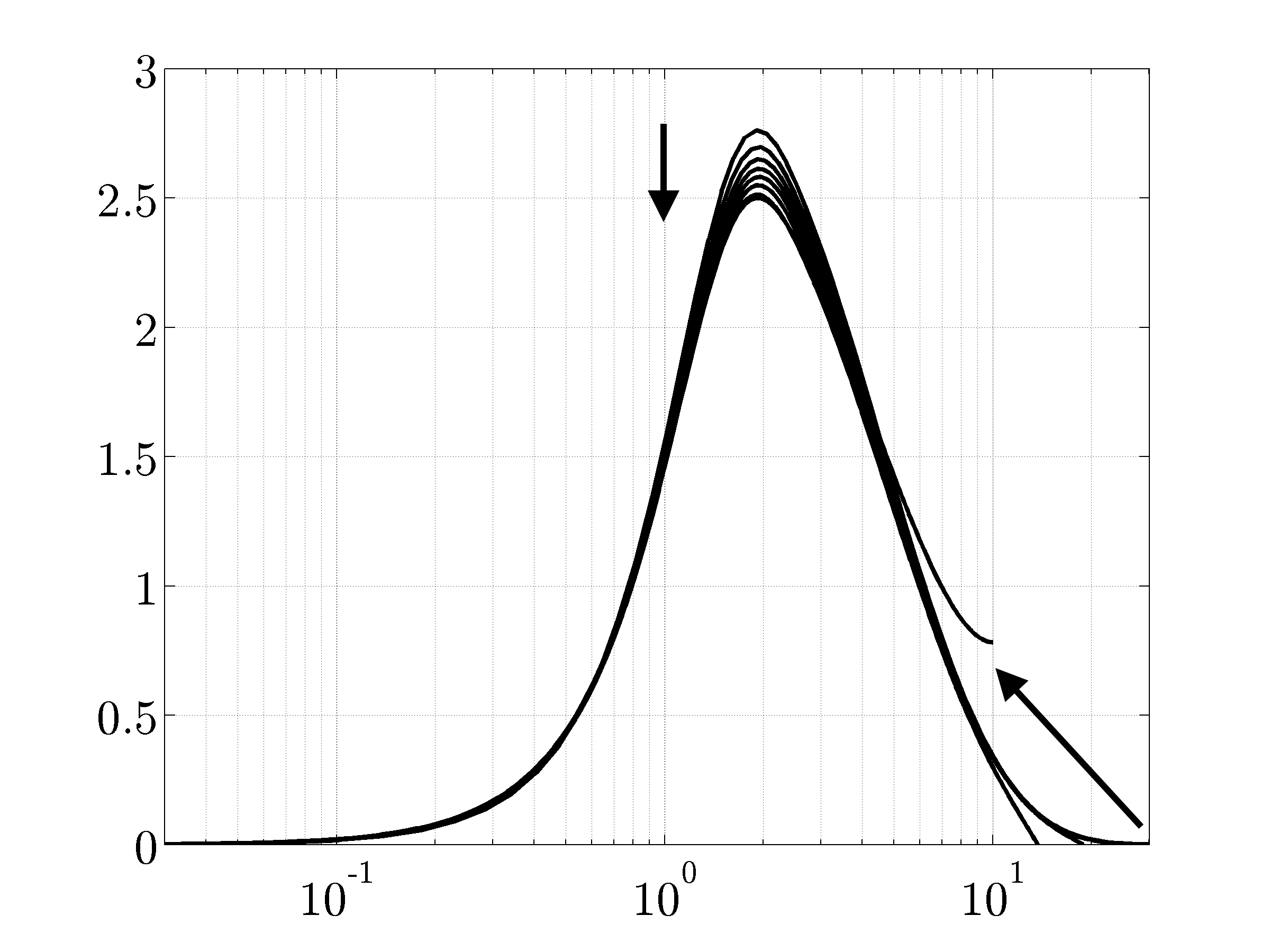}
    \label{fig.log-region-hierarchy-v-similarity-vs-y-Up2_3Uck-kx1kz10-R1e4}}
    &
    \hskip-0.3cm
    \subfigure{\includegraphics[width=0.325\columnwidth]
    {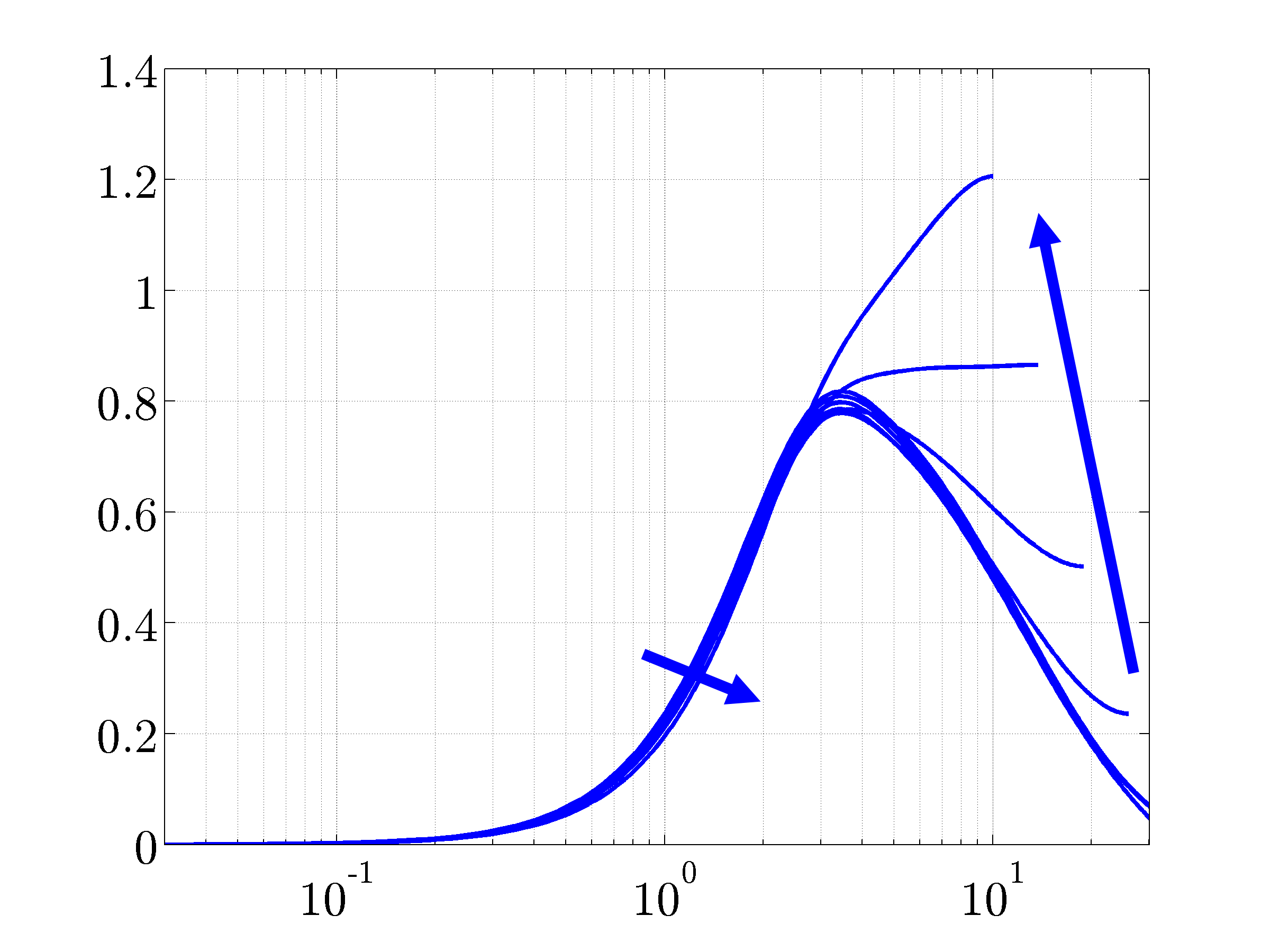}
    \label{fig.log-region-hierarchy-v-similarity-vs-y-yp100_Rb-kx1kz10-R1e4}}
    \\[0.2cm]
    $(d)$
    &
    \hskip-0.3cm
    $(e)$
    &
    \hskip-0.3cm
    $(f)$
    \end{tabular}
    \begin{tabular}{c}
    \\[-3.7cm]
    \begin{tabular}{c}
    \hskip-6.85cm
    \begin{turn}{90}
    \tc{black}{$~~~~y_c^+ \sqrt{y_c} \; \abs{\hat{v}_1}$}
    \end{turn}
    \end{tabular}
    \\[1.1cm]
    \begin{tabular}{c}
    \hskip0.2cm
    \tc{black}{$y/y_c$}
    \hskip3.7cm
    \tc{black}{$y/y_c$}
    \hskip3.7cm
    \tc{black}{$y/y_c$}
    \end{tabular}
    \end{tabular}
    \\[-0.2cm]
    \begin{tabular}{ccc}
    \subfigure{\includegraphics[width=0.325\columnwidth]
    {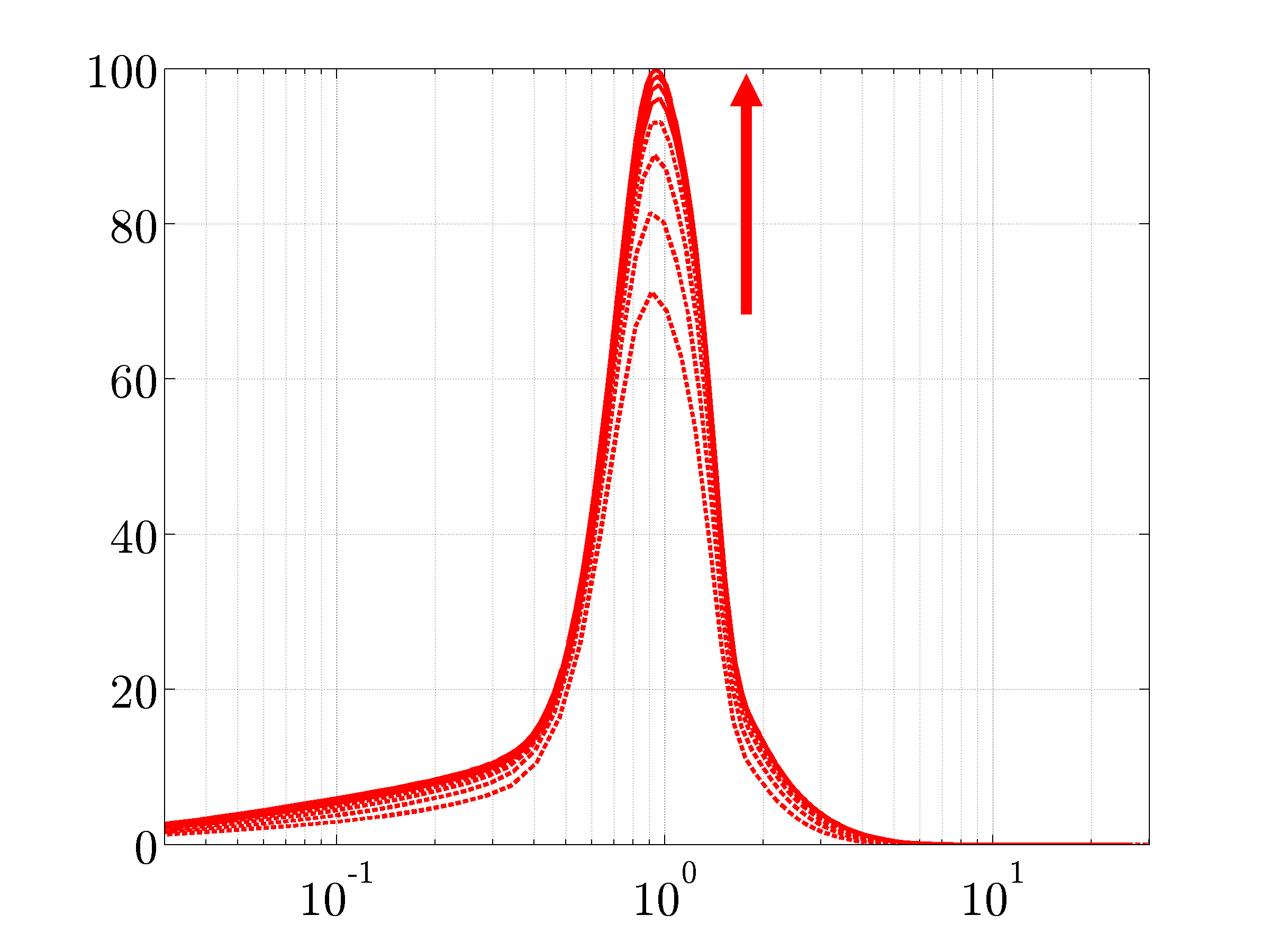}
    \label{fig.log-region-hierarchy-w-similarity-vs-y-yp0p1r-kx1kz10-R1e4}}
    &
    \hskip-0.3cm
    \subfigure{\includegraphics[width=0.325\columnwidth]
    {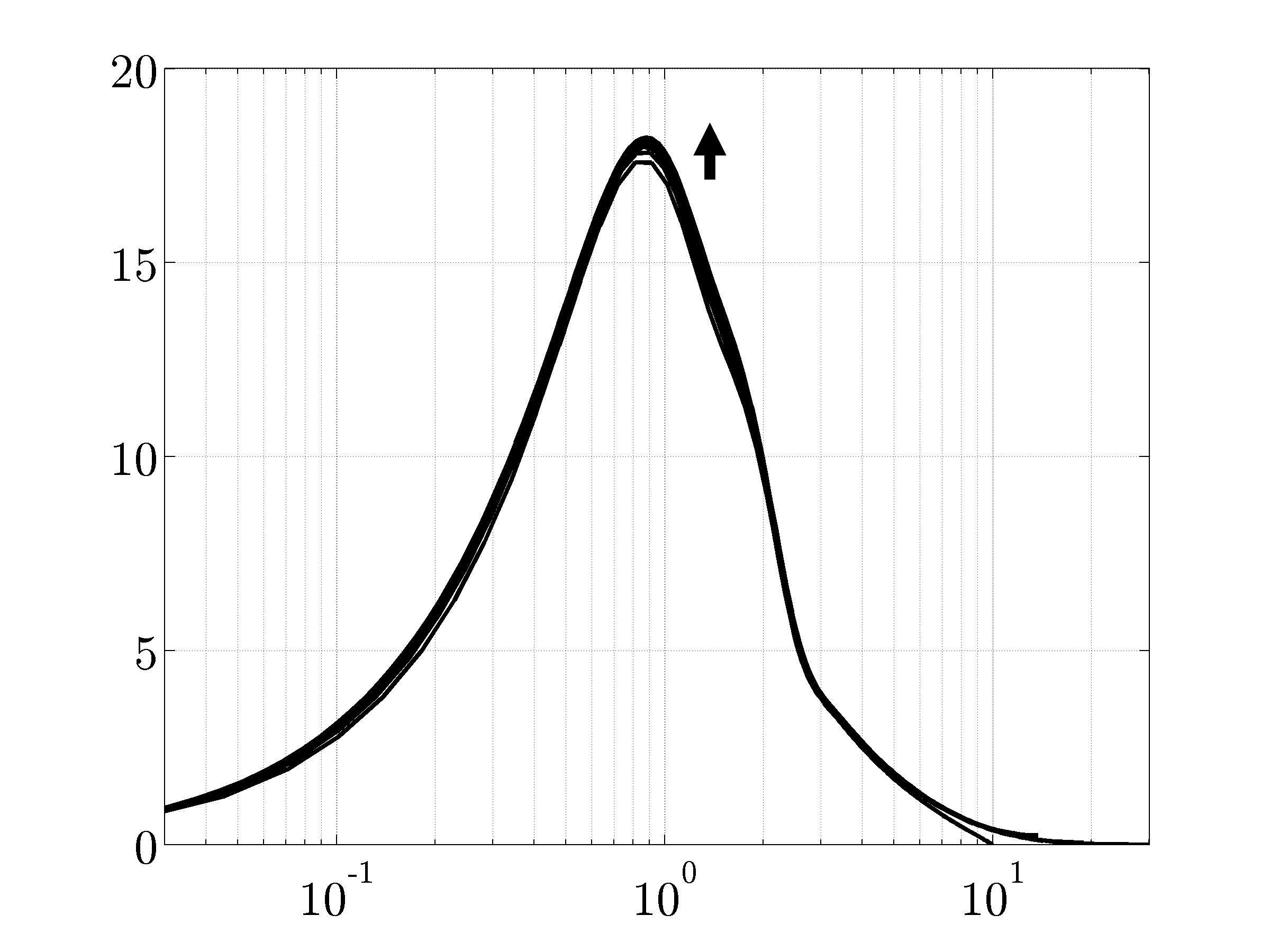}
    \label{fig.log-region-hierarchy-w-similarity-vs-y-Up2_3Uck-kx1kz10-R1e4}}
    &
    \hskip-0.3cm
    \subfigure{\includegraphics[width=0.325\columnwidth]
    {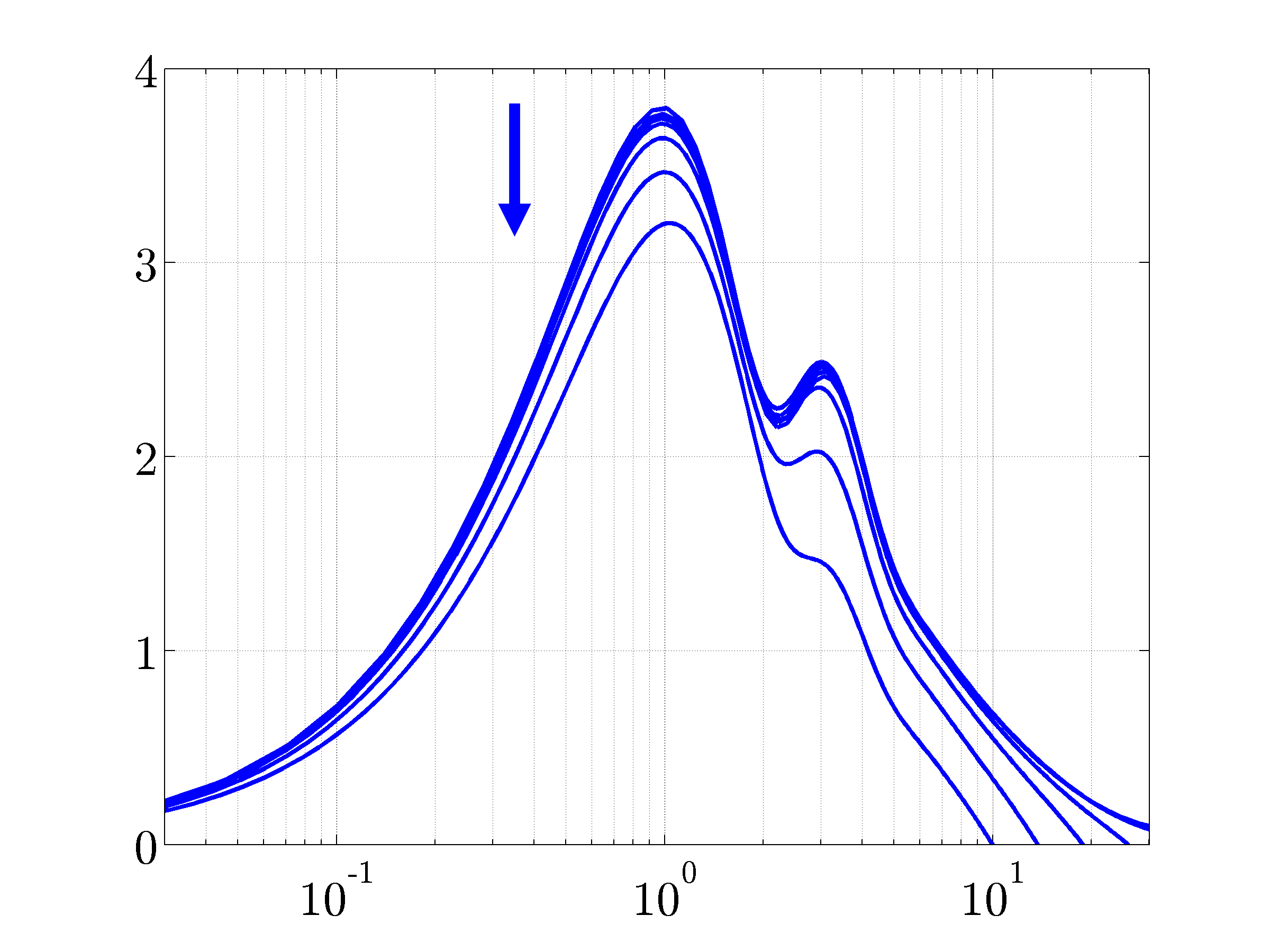}
    \label{fig.log-region-hierarchy-w-similarity-vs-y-yp100_Rb-kx1kz10-R1e4}}
    \\[0.2cm]
    $(g)$
    &
    \hskip-0.3cm
    $(h)$
    &
    \hskip-0.3cm
    $(i)$
    \end{tabular}
    \begin{tabular}{c}
    \\[-3.7cm]
    \begin{tabular}{c}
    \hskip-6.85cm
    \begin{turn}{90}
    \tc{black}{$~~~~y_c^+ \sqrt{y_c} \; \abs{\hat{w}_1}$}
    \end{turn}
    \end{tabular}
    \\[1.1cm]
    \begin{tabular}{c}
    \hskip0.2cm
    \tc{black}{$y/y_c$}
    \hskip3.7cm
    \tc{black}{$y/y_c$}
    \hskip3.7cm
    \tc{black}{$y/y_c$}
    \end{tabular}
    \end{tabular}
    \end{center}
    \caption{
    The normalized and scaled (according to table~\ref{table.scalings}) principal streamwise, (a)-(c), wall-normal, (d)-(f), and spanwise, (g)-(i), responses for the modes that belong to the hierarchies $h_1$, (a,d,g), $h_2$, (b,e,h), and $h_3$, (c,f,i) in figure~\ref{fig.hierarchies}. The arrows show the direction of increasing $y_c$ between $y_c^+ = 100$ and $y_c = 0.1$.
    }
    \label{fig.self-similar-edge}
    \end{figure}

Figure~\ref{fig.self-similar-edge} shows the principal response modes in the hierarchies $h_1$, $h_2$, and $h_3$ for speeds $c_l \leq c \leq c_u$, i.e. $y_l \leq y_c \leq y_u$, and $Re_\tau = 10^4$, cf.~figure~\ref{fig.hierarchies}. The modes are normalized and scaled according to the scalings of the self-similar class in table~\ref{table.scalings}. The arrows show the direction of increasing $y_c$. Figures~\ref{fig.log-region-hierarchy-u-similarity-vs-y-Up2_3Uck-kx1kz10-R1e4},~\ref{fig.log-region-hierarchy-v-similarity-vs-y-Up2_3Uck-kx1kz10-R1e4}, and~\ref{fig.log-region-hierarchy-w-similarity-vs-y-Up2_3Uck-kx1kz10-R1e4} show the modes corresponding to the hierarchy $h_2$. We see that the scaled streamwise and spanwise velocities lie on the top of each other and the scaled wall-normal velocities are approximately equal. The wall-normal velocities are wider in the wall-normal direction and as $y_c$ increases, they are affected by the symmetry constraints at the center of the channel. 

Figures~\ref{fig.log-region-hierarchy-u-similarity-vs-y-yp100_Rb-kx1kz10-R1e4},~\ref{fig.log-region-hierarchy-v-similarity-vs-y-yp100_Rb-kx1kz10-R1e4}, and~\ref{fig.log-region-hierarchy-w-similarity-vs-y-yp100_Rb-kx1kz10-R1e4} show the modes for the hierarchy $h_3$. The streamwise velocities approximately lie on the top of each other and the edge effects for large $y_c$ are small. On the other hand, the wall-normal and spanwise velocities are less localized in the wall-normal direction; for example, the bulk of the wall-normal velocity moves outside the upper edge of the logarithmic region for large values of $y_c$. Therefore, the self-similarity is weaker for the wall-normal and spanwise velocities. 

Figures~\ref{fig.log-region-hierarchy-u-similarity-vs-y-yp0p1r-kx1kz10-R1e4},~\ref{fig.log-region-hierarchy-v-similarity-vs-y-yp0p1r-kx1kz10-R1e4}, and~\ref{fig.log-region-hierarchy-w-similarity-vs-y-yp0p1r-kx1kz10-R1e4} show the modes corresponding to the hierarchy $h_1$. The streamwise wavelengths of the modes in this hierarchy are smaller than the other three hierarchies. Therefore, they are more localized in the wall-normal direction and are less affected by the edge effects. We see that the modes with large $y_c$ collapse on each other in contrast to the modes with small $y_c$. This is because the aspect ratio of the latter modes falls below the threshold $\gamma$, cf.~figure~\ref{fig.hierarchies}, and the self-similarity does not hold. 
    
\section{Derivation of the energy budget}
\label{sec.energy-budget-der}

Any resolvent mode with parameters $\kappa_x$, $\kappa_z$, $c$, and $\omega = c \kappa_x$ satisfies the NSE given in~(\ref{eq.NS-lin})
	\be
	\ba{rcl}
	-\mri \omega \hat{u}
	\, + \,
	\mri \kappa_x U \hat{u}
	\, + \,
	U' \hat{v}
	\, + \,
	\mri \kappa_x \hat{p}
	\, + \,
	(1/{Re}_\tau) (\kappa^2 \hat{u} - \hat{u}'')
	&\!\! = \!\!&
	\hat{f}_1,
	\\[0.2cm]
	-\mri \omega \hat{v}
	\, + \,
	\mri \kappa_x U \hat{v}
	\, + \,
	\hat{p}'
	\, + \,
	(1/{Re}_\tau) (\kappa^2 \hat{v} - \hat{v}'')
	&\!\! = \!\!&
	\hat{f}_2,
	\\[0.2cm]
	-\mri \omega \hat{w}
	\, + \,
	\mri \kappa_x U \hat{w}
	\, + \,
	\mri \kappa_z \hat{p}
	\, + \,
	(1/{Re}_\tau) (\kappa^2 \hat{w} - \hat{w}'')
	&\!\! = \!\!&
	\hat{f}_3,
	\\[0.2cm]
	\nabla \cdot \hat{\bu}
	&\!\! = \!\!&
	0.
	\ea
	\label{eq.NS-lin-expand}
	\ee
The inner product of both sides of~(\ref{eq.NS-lin-expand}) with the complex conjugate resolvent mode $\bu^*$ with parameters $-\kappa_x$, $-\kappa_z$, $c$, and $\omega = c \kappa_x$ yields
	\be
	\ba{c}
	\mri \kappa_x (U - c)
	(\overline{\hat{u}} \hat{u} 
	\, + \,
	\overline{\hat{v}} \hat{v}
	\, + \, 
	\overline{\hat{w}} \hat{w})
	\, + \,
	U' \overline{\hat{u}} \hat{v}
	\, + \,
	(\mri \kappa_x \overline{\hat{u}} \hat{p} 
	\, + \, 
	\overline{\hat{v}} \hat{p}' 
	\, + \,
	\mri \kappa_z \overline{\hat{w}} \hat{p})
	\, + \,
	\\[0.2cm]
	(1/{Re}_\tau) 
	\big(
	\kappa^2 
	(\overline{\hat{u}} \hat{u} 
	\, + \, 
	\overline{\hat{v}} \hat{v} 
	\, + \, 
	\overline{\hat{w}} \hat{w})
	\, - \,
	(\overline{\hat{u}} \hat{u}'' 
	\, + \, 
	\overline{\hat{v}} \hat{v}'' 
	\, + \,
	\overline{\hat{w}} \hat{w}'')
	\big)
	\, = \,
	\overline{\hat{u}} \hat{f}_1
	\, + \, 
	\overline{\hat{v}} \hat{f}_2
	\, + \,
	\overline{\hat{w}} \hat{f}_3.
	\ea
	\label{eq.NS-lin-inner-product}
	\ee
Since the mode $\bu^*$ satisfies the continuity equation, we have
	\be
	\mri \kappa_x \overline{\hat{u}} \hat{p} 
	\, + \,
	\overline{\hat{v}} \hat{p}' 
	\, + \, 
	\mri \kappa_z \overline{\hat{w}} \hat{p}
	\; = \;
	(\overline{\hat{v}} \hat{p})'
	\, - \,
	p 
	(-\mri \kappa_x \overline{\hat{u}}
	\, + \,
	\overline{\hat{v}}'
	\, - \, 
	\mri \kappa_z \overline{\hat{w}})
	\; = \;
	(\overline{\hat{v}} \hat{p})'.
	\label{eq.P-cont}
	\ee
In addition,
	\be
	\overline{\hat{u}} \hat{u}'' 
	\, + \, 
	\overline{\hat{v}} \hat{v}'' 
	\, + \,
	\overline{\hat{w}} \hat{w}''
	\; = \;
	(\overline{\hat{u}} \hat{u}'
	\, + \, 
	\overline{\hat{v}} \hat{v}'
	\, + \,
	\overline{\hat{w}} \hat{w}')'
	\, - \,
	(\overline{\hat{u}}' \hat{u}' 
	\, + \, 
	\overline{\hat{v}}' \hat{v}'
	\, + \,
	\overline{\hat{w}}' \hat{w}').
	\label{eq.D-derivative}
	\ee
Substituting~(\ref{eq.P-cont}) and~(\ref{eq.D-derivative}) in~(\ref{eq.NS-lin-inner-product}) yields
	\be
	\mri \kappa_x (U - c)
	\hat{\bu}^* \hat{\bu} 
	\, + \,
	U' \overline{\hat{u}} \hat{v}
	\, + \,
	(\overline{\hat{v}} \hat{p})' 
	\, + \,
	(1/{Re}_\tau) 
	\big(
	\kappa^2 
	\hat{\bu}^* \hat{\bu} 
	\, - \,
	(\hat{\bu}^* \hat{\bu}')'
	\, + \,
	\hat{\bu}'^* \hat{\bu}'
	\big)
	\, = \,
	\hat{\bu}^* \hat{\fvec}.
	\label{eq.NS-lin-inner-product-2}
	\ee
Integrating~(\ref{eq.NS-lin-inner-product-2}) from between the two walls and using the no-slip condition results in
	\be
	\mri \kappa_x 
	\ds{\int_0^2} \,
	(U - c)
	\hat{\bu}^* \hat{\bu} 
	\, \mrd y
	\, + \,
	\ds{\int_0^2} \,
	U' \overline{\hat{u}} \hat{v}
	\, \mrd y
	\, + \,
	(1/{Re}_\tau) 
	\ds{\int_0^2}
	\big(
	\kappa^2 
	\hat{\bu}^* \hat{\bu} 
	\, + \,
	\hat{\bu}'^* \hat{\bu}'
	\big)
	\mrd y
	\, = \,
	\ds{\int_0^2} \,
	\hat{\bu}^* \hat{\fvec}
	\, \mrd y.
	\label{eq.NS-lin-inner-product-integrate}
	\ee
Notice that the first term in~(\ref{eq.NS-lin-inner-product-integrate}) is purely imaginary and the third term is purely real. After including the modes across all four quadrants ($\pm\kappa_x$, $\pm\kappa_z$, and $c$), only the real part of~(\ref{eq.NS-lin-inner-product-integrate}) survives which yields the energy budget equation given in~(\ref{eq.budget-brief}) and~(\ref{eq.budget-def}). No physical effect could be attributed to the imaginary part of~(\ref{eq.NS-lin-inner-product-integrate}).

\section{Details of the interaction coefficient}
\label{sec.N-details}
 
Substituting the nonlinear forcing term from~(\ref{eq.f-convolution}) in~(\ref{eq.f-phi}) yields~(\ref{eq.f-convolution-expand}) where
	\be
	\ba{l}
	\cN_{lij} (\blambda, c, \blambda', c')
	\; = \;
	-
	\big(\dfrac{2\pi}{\lambda_x'}\big)^2 \dfrac{2\pi}{|\lambda_z'|}
	\,
	\sigma_i(\blambda', c') 
	\,
	\sigma_j(\blambda'', c'') 
	\;
	\ds{
	\int_{0}^{2}
	}
	\Big\{
	\\[0.2cm]
	\hskip0.6cm
	\hat{f}_{1l} (y, \blambda, c)
	\Big(
	\big(
	\hat{u}_i (y, \blambda', c')
	\,
	\overline{\hat{v}}_j (y, \blambda'', c'')
	\big)'
	\, + \,
	\\[0.2cm]
	\hskip1.8cm
	\mri 2\pi 
	\, 
	\hat{u}_i (y, \blambda', c')
	\,
	\big( 
	\overline{\hat{u}}_j (y, \blambda'', c'')/\lambda_x  
	\, + \,
	\overline{\hat{w}}_j (y, \blambda'', c'')/\lambda_z 
	\big)
	\Big)
	\, 
	+
	\\[0.2cm]
	\hskip0.6cm
	\hat{f}_{2l} (y, \blambda, c)
	\Big(
	\big(
	\hat{v}_i (y, \blambda', c')
	\,
	\overline{\hat{v}}_j (y, \blambda'', c'')
	\big)'
	\, + \,
	\\[0.2cm]
	\hskip1.8cm
	\mri 2\pi 
	\, 
	\hat{v}_i (y, \blambda', c')
	\,
	\big( 
	\overline{\hat{u}}_j (y, \blambda'', c'')/\lambda_x  
	\, + \,
	\overline{\hat{w}}_j (y, \blambda'', c'')/\lambda_z 
	\big)
	\Big)
	\, 
	+
	\\[0.2cm]
	\hskip0.6cm
	\hat{f}_{3l} (y, \blambda, c)
	\Big(
	\big(
	\hat{w}_i (y, \blambda', c')
	\,
	\overline{\hat{v}}_j (y, \blambda'', c'')
	\big)'
	\, + \,
	\\[0.2cm]
	\hskip1.8cm
	\mri 2\pi 
	\, 
	\hat{w}_i (y, \blambda', c')
	\,
	\big( 
	\overline{\hat{u}}_j (y, \blambda'', c'')/\lambda_x  
	\, + \,
	\overline{\hat{w}}_j (y, \blambda'', c'')/\lambda_z 
	\big)
	\Big)
	\,
	\Big\}
	\,
	\mrd y.
	\ea
	\label{eq.f-convolution-expand-ap}
	\ee

\section{Derivation of the scaling of the interaction coefficient}
\label{sec.N-details-scaling}

For a set of triadically-consistent modes in the self-similar hierarchies, notice that
\[
y_u/y_c \, = \, \mre^{\kappa (c_u - c)}, 
~~~
y_c/y_{c'} \, = \, \mre^{\kappa (c - c')}, 
~~~
y_c/y_{c''} \, = \, \mre^{\frac{\kappa \lambda_x}{\lambda_x + \lambda_x'} (c - c')}.
\]
Substituting the interaction coefficient from~(\ref{eq.f-convolution-expand-ap}) in~(\ref{eq.f-convolution-expand}) and defining
\[
\tilde{y} \, = \, y y_u/y_c,
\]
yields~(\ref{eq.f-convolution-scale}) where
	\be
	\ba{l}
	\cM_{lij} (\blambda_u, \blambda_u', c' - c)
	\; = \;
	\mre^{(3.5-1.5\frac{\lambda_x}{\lambda_x + \lambda_x'}) \kappa (c - c')}
	\,
	\big(\dfrac{2\pi}{\lambda_{x,u}'}\big)^2 \dfrac{2\pi}{|\lambda_{z,u}'|}
	\,
	\sigma_i(\blambda_{u}', c_u) \,
	\sigma_j(\blambda_{u}'', c_u) \,
	\;
	\ds{
      \int_{0}^{2}
	}
	\Big\{
	\\[0.2cm]
	\hskip0.35cm
	\Big(
	\mre^{\kappa (c' - c)} 
	\, 
	\hat{f}_{1l} (\tilde{y}, \blambda_u, c_u)
	\,
	\big(
	\hat{u}_i (\tilde{y} \mre^{\kappa (c - c')}, \blambda_{u}', c_u)
	\,
	\overline{\hat{v}}_{j} (\tilde{y} \mre^{\frac{\kappa \lambda_x}{\lambda_x + \lambda_x'} (c - c')}, \blambda_{u}'', c_u)
	\big)'
	\, + \,
	\\[0.2cm]
	\hskip1.76cm
	\hat{f}_{2l} (\tilde{y}, \blambda_u, c_u)
	\,
	\big(
	\hat{v}_i (\tilde{y} \mre^{\kappa (c - c')}, \blambda_{u}', c_u)
	\,
	\overline{\hat{v}}_{j} (\tilde{y} \mre^{\frac{\kappa \lambda_x}{\lambda_x + \lambda_x'} (c - c')}, \blambda_{u}'', c_u)
	\big)'
	\, + \,
	\\[0.2cm]
	\hskip1.76cm
	\hat{f}_{3l} (\tilde{y}, \blambda_u, c_u)
	\,
	\big(
	\hat{w}_i (\tilde{y} \mre^{\kappa (c - c')}, \blambda_{u}, c_u)
	\,
	\overline{\hat{v}}_{j} (\tilde{y} \mre^{\frac{\kappa \lambda_x}{\lambda_x + \lambda_x'} (c - c')}, \blambda_{u}'', c_u)
	\big)'
	\Big)
	\, + \,
	\\[0.2cm]
	\hskip0.35cm
	\mri 2\pi
	\,
	\Big(
	\,
	\mre^{\frac{\kappa \lambda_x}{\lambda_x + \lambda_x'} (c' - c)}
	\,
	\overline{\hat{u}}_j (\tilde{y} \mre^{\frac{\kappa \lambda_x}{\lambda_x + \lambda_x'} (c - c')}, \blambda_{u}'', c_u)/\lambda_{x,u}
	\, + \,
	\overline{\hat{w}}_j (\tilde{y} \mre^{\frac{\kappa \lambda_x}{\lambda_x + \lambda_x'} (c - c')}, \blambda_{u}'', c_u)/\lambda_{z,u}
	\Big)
	\Big(
	\\[0.2cm]
	\hskip1.2cm
	\mre^{\kappa (c' - c)}
	\,
	\hat{f}_{1l} (\tilde{y}, \blambda_u, c_u)
	\,
	\hat{u}_{i} (\tilde{y} \mre^{\kappa (c - c')}, \blambda_{u}', c_u)
	\, + \,
	\hat{f}_{2l} (\tilde{y}, \blambda_u, c_u)
	\,
	\hat{v}_{i} (\tilde{y} \mre^{\kappa (c - c')}, \blambda_{u}', c_u)
	\, + \,
	\\[0.2cm]
	\hskip1.2cm
	\hat{f}_{3l} (\tilde{y}, \blambda_u, c_u)
	\,
	\hat{w}_{i} (\tilde{y} \mre^{\kappa (c - c')}, \blambda_{u}', c_u)
	\Big)
	\,
	\Big\}
	\,
	\mrd \tilde{y}.
	\ea
	\label{eq.f-convolution-scale-ap}
	\ee
Notice that all the terms in~(\ref{eq.f-convolution-scale-ap}), including
\[
\dfrac{\lambda_x}{\lambda_x + \lambda_x'}
\, = \, 
\dfrac{1}{1 + \lambda_x'/\lambda_x}
\, = \,
\dfrac{1}{1 + (\lambda_{x,u}'/\lambda_{x,u}) \mre^{2 \kappa (c' - c)}},
\] 
can be expressed in terms of $\blambda_u$, $\blambda_u'$, and $c' - c$.

\bibliographystyle{jfm}
\bibliography{bib/couette,bib/mj-complete-bib,bib/periodic,bib/covariance,bib/control-pde,bib/ref-added-rm}

\begin{thebibliography}{29}
\expandafter\ifx\csname natexlab\endcsname\relax\def\natexlab#1{#1}\fi

\bibitem[Chakraborty {\em et~al.\/}(2005)Chakraborty, Balachandar \&
  Adrian]{chabaladr05}
{\sc Chakraborty, P., Balachandar, S. \& Adrian, R.~J.} 2005 On the
  relationships between local vortex identification schemes. {\em J. Fluid
  Mech.\/} {\bf 535}, 189--214.

\bibitem[Cheung \& Zaki(2014)]{chezak14}
{\sc Cheung, L.~C. \& Zaki, T.~A.} 2014 An exact representation of the
  nonlinear triad interaction terms in spectral space. {\em J. Fluid Mech.\/}
  {\bf 748}, 175--188.

\bibitem[Coles(1956)]{col56}
{\sc Coles, D.~E.} 1956 The law of the wake in the turbulent boundary layer.
  {\em J. Fluid Mech.\/} {\bf 1}, 191--226.

\bibitem[{del {\'A}lamo} {\em et~al.\/}(2006){del {\'A}lamo}, Jim{\'e}nez,
  Zandonade \& Moser]{deljimzanmos06}
{\sc {del {\'A}lamo}, J.~C., Jim{\'e}nez, J., Zandonade, P. \& Moser, R.~D.}
  2006 Self-similar vortex clusters in the turbulent logarithmic region. {\em
  J. Fluid Mech.\/} {\bf 561}, 329--358.

\bibitem[Fife {\em et~al.\/}(2005)Fife, Wei, Klewicki \&
  McMurtry]{fifweiklemcm05}
{\sc Fife, P., Wei, T., Klewicki, J. \& McMurtry, P.} 2005 Stress gradient
  balance layers and scale hierarchies in wall-bounded turbulent flows. {\em J.
  Fluid Mech.\/} {\bf 532}, 165--190.

\bibitem[Hoyas \& Jim{\'e}nez(2006)]{hoyjim06}
{\sc Hoyas, S. \& Jim{\'e}nez, J.} 2006 Scaling of the velocity fluctuations in
  turbulent channels up to ${R}e_{\tau} = 2003$. {\em Phys. Fluids\/} {\bf
  18}~(1), 011702.

\bibitem[Hoyas \& Jim{\'e}nez(2008)]{hoyjim08}
{\sc Hoyas, S. \& Jim{\'e}nez, J.} 2008 Reynolds number effects on the
  {R}eynolds-stress budgets in turbulent channels. {\em Phys. Fluids\/} {\bf
  20}, 101511.

\bibitem[Hwang \& Cossu(2010)]{hwacos10}
{\sc Hwang, Y. \& Cossu, C.} 2010 Linear non-normal energy amplification of
  harmonic and stochastic forcing in the turbulent channel flow. {\em J. Fluid
  Mech.\/} {\bf 664}, 51--73.

\bibitem[Hwang \& Cossu(2011)]{hwacos11}
{\sc Hwang, Y. \& Cossu, C.} 2011 Self-sustained processes in the logarithmic
  layer of turbulent channel flows. {\em Phys. Fluids\/} {\bf 23}, 061702.

\bibitem[Klewicki {\em et~al.\/}(2009)Klewicki, Fife \& Wei]{klefifwei09}
{\sc Klewicki, J., Fife, P. \& Wei, T.} 2009 On the logarithmic mean profile.
  {\em J. Fluid Mech.\/} {\bf 638}, 73--93.

\bibitem[Klewicki(2013)]{kle13}
{\sc Klewicki, J.~C.} 2013 Self-similar mean dynamics in turbulent wall flows.
  {\em J. Fluid Mech.\/} {\bf 718}, 596--621.

\bibitem[Kunkel \& Marusic(2006)]{kunmar06}
{\sc Kunkel, G.~J. \& Marusic, I.} 2006 Study of the near-wall-turbulent region
  of the high-{R}eynolds-number boundary layer using an atmospheric flow. {\em
  J. Fluid Mech.\/} {\bf 548}, 375--402.

\bibitem[Marusic \& Kunkel(2003)]{markun03}
{\sc Marusic, I. \& Kunkel, G.~J.} 2003 Streamwise turbulence intensity
  formulation for flat-plate boundary layers. {\em Phys. Fluids\/} {\bf
  15}~(8), 2461--2464.

\bibitem[Marusic {\em et~al.\/}(2013)Marusic, Monty, Hultmark \&
  Smits]{marmonhulsmi13}
{\sc Marusic, I., Monty, J.~P., Hultmark, M. \& Smits, A.~J.} 2013 On the
  logarithmic region in wall turbulence. {\em J. Fluid Mech.\/} {\bf 716},
  R3--1 -- 716 R3--11.

\bibitem[Marusic {\em et~al.\/}(1997)Marusic, Uddin \& Perry]{maruddper97}
{\sc Marusic, I., Uddin, A. K.~M. \& Perry, A.~E.} 1997 Similarity law for the
  streamwise turbulence intensity in zero-pressure-gradient turbulent boundary
  layers. {\em Phys. Fluids\/} {\bf 9}, 3718--3726.

\bibitem[Mathis {\em et~al.\/}(2009)Mathis, Hutchins \& Marusic]{mathutmar09}
{\sc Mathis, R., Hutchins, N. \& Marusic, I.} 2009 Large-scale amplitude
  modulation of the small-scale structures in turbulent boundary layers. {\em
  J. Fluid Mech.\/} {\bf 628}, 311--337.

\bibitem[McKeon \& Sharma(2010)]{mcksha10}
{\sc McKeon, B.~J. \& Sharma, A.~S.} 2010 A critical-layer framework for
  turbulent pipe flow. {\em J. Fluid Mech.\/} {\bf 658}, 336--382.

\bibitem[McKeon {\em et~al.\/}(2013)McKeon, Sharma \& Jacobi]{mckshajac13}
{\sc McKeon, B.~J., Sharma, A.~S. \& Jacobi, I.} 2013 {Experimental
  manipulation of wall turbulence: A systems approach}. {\em Phys. Fluids\/}
  {\bf 25}, 031301.

\bibitem[Moarref {\em et~al.\/}(2014)Moarref, Jovanovi\'c, Tropp, Sharma \&
  McKeon]{moajovtroshamckPOF14}
{\sc Moarref, R., Jovanovi\'c, M.~R., Tropp, J.~A., Sharma, A.~S. \& McKeon,
  B.~J.} 2014 A low-order decomposition of turbulent channel flow via resolvent
  analysis and convex optimization. {\em Phys. Fluids\/} {\bf 26}~(5), 051701.

\bibitem[Moarref {\em et~al.\/}(2013)Moarref, Sharma, Tropp \&
  McKeon]{moashatromckJFM13}
{\sc Moarref, R., Sharma, A.~S., Tropp, J.~A. \& McKeon, B.~J.} 2013
  Model-based scaling of the streamwise energy density in high-{R}eynolds
  number turbulent channels. {\em J. Fluid Mech.\/} {\bf 734}, 275--316.

\bibitem[Nickels {\em et~al.\/}(2007)Nickels, Marusic, Hafez, Hutchins \&
  Chong]{nicmarhafhutcho07}
{\sc Nickels, T.~B., Marusic, I., Hafez, S., Hutchins, N. \& Chong, M.~S.} 2007
  Some predictions of the attached eddy model for a high reynolds number
  boundary layer. {\em Phil. Trans. R. Soc. A\/} {\bf 365}, 807--822.

\bibitem[Perry \& Chong(1982)]{percho82}
{\sc Perry, A.~E. \& Chong, M.~S.} 1982 On the mechanism of wall turbulence.
  {\em J. Fluid Mech.\/} {\bf 119}~(173), 106--121.

\bibitem[Perry {\em et~al.\/}(1986)Perry, Henbest \& Chong]{perhencho86}
{\sc Perry, A.~E., Henbest, S. \& Chong, M.~S.} 1986 A theoretical and
  experimental study of wall turbulence. {\em J. Fluid Mech.\/} {\bf 165},
  163--199.

\bibitem[Perry \& Li(1990)]{perli90}
{\sc Perry, A.~E. \& Li, J.~D.} 1990 Experimental support for the attached-eddy
  hypothesis in zero-pressure-gradient turbulent boundary layers. {\em J. Fluid
  Mech.\/} {\bf 218}, 405--438.

\bibitem[Perry {\em et~al.\/}(1994)Perry, Marusic \& Li]{permarli94}
{\sc Perry, A.~E., Marusic, I. \& Li, J.~D.} 1994 Wall turbulence closure based
  on classical similarity laws and the attached eddy hypothesis. {\em Phys.
  Fluids\/} {\bf 6}, 1024--1035.

\bibitem[Sharma \& McKeon(2013)]{shamcK13}
{\sc Sharma, A.~S. \& McKeon, B.~J.} 2013 On coherent structure in wall
  turbulence. {\em J. Fluid Mech.\/} {\bf 728}, 196--238.

\bibitem[Smits {\em et~al.\/}(2011)Smits, McKeon \& Marusic]{smimckmar11}
{\sc Smits, A.~J., McKeon, B.~J. \& Marusic, I.} 2011 High-{R}eynolds number
  wall turbulence. {\em Annu. Rev. Fluid Mech.\/} {\bf 43}, 353--375.

\bibitem[Townsend(1961)]{tow61}
{\sc Townsend, A.~A.} 1961 Equilibrium layers and wall turbulence. {\em J.
  Fluid Mech.\/} {\bf 11}, 97--120.

\bibitem[Townsend(1976)]{tow76}
{\sc Townsend, A.~A.} 1976 {\em The structure of turbulent shear flow\/}.
  Cambridge University Press.

\end{thebibliography}

\end{document}